\documentstyle[12pt,epsf]{article}

\renewcommand{\thefootnote}{\fnsymbol{footnote}}
\newcommand{\newsection}{\setcounter{equation}{0}\section}

\def\appendix#1{\addtocounter{section}{1}\setcounter{equation}{0}
\renewcommand{\thesection}{\Alph{section}}
\section*{Appendix \thesection\protect\indent \parbox[t]{11.715cm} {#1}}
\addcontentsline{toc}{section}{Appendix \thesection\ \ \ #1} }

\newcommand{\weyl}{{\hat{\cal W}}}
\newcommand{\complex}{{\bb C}} 
\newcommand{\zed}{{\bb Z}} 
\newcommand{\real}{{\bb R}} 
\newcommand{\zeds}{{\bbs Z}} 
\newcommand{\mat}{{\bb M}} 
\newcommand{\mats}{{\bbs M}} 
\newcommand{\NO}{\,\mbox{$\circ\atop\circ$}\,} 
\newcommand{\id}{{1\!\!1}} 
\def\alg{{\cal A}}
\def\hil{{\cal H}}

\newif\ifold             \oldtrue

\font\mybb=msbm10 at 12pt
\def\bb#1{\hbox{\mybb#1}}
\font\mybbs=msbm10 at 9pt
\def\bbs#1{\hbox{\mybbs#1}}

\def\nn{\nonumber}
\newcommand{\tr}[1]{\:{\rm tr}\,#1}
\newcommand{\Tr}[1]{\:{\rm Tr}\,#1}

\def\e{{\,\rm e}\,}

\def\be{\begin{equation}}
\def\ee{\end{equation}}
\def\bea{\begin{eqnarray}}
\def\eea{\end{eqnarray}}
\def\bd{\begin{displaymath}}
\def\ed{\end{displaymath}}

\newcommand{\beq}{\begin{eqnarray}}
\newcommand{\eeq}{\end{eqnarray}}

\begin{document}
\begin{titlepage}
\begin{flushright}

\baselineskip=12pt

HWM--01--35\\
EMPG--01--14\\
hep--th/0109162\\
\hfill{ }\\
September 2001\\ Revised January 2003
\end{flushright}

\begin{center}

\baselineskip=24pt

\vspace{1cm}

{\Large\bf Quantum Field Theory on Noncommutative
Spaces\footnote{\baselineskip=12pt Based on invited lectures given at the
APCTP-KIAS Winter School on ``Strings and D-Branes 2000'', Seoul, Korea,
February 21--25 2000, at the Science Institute, University of Iceland,
Reykjavik, Iceland June 1--8 2000, and at the PIMS/APCTP/PITP Frontiers of
Mathematical Physics Workshop on ``Particles, Fields and Strings'', Simon
Fraser University, Vancouver, Canada, July 16--27 2001.}}

\baselineskip=12pt

\vspace{1cm}

{\bf Richard J. Szabo}
\\[6mm]
{\it Department of Mathematics\\ Heriot-Watt University\\ Riccarton, Edinburgh
EH14 4AS, U.K.}\\{\tt R.J.Szabo@ma.hw.ac.uk}
\\[30mm]

\end{center}

\vskip 1 cm

\begin{abstract}

\baselineskip=12pt

A pedagogical and self-contained introduction to noncommutative quantum field
theory is presented, with emphasis on those properties that are intimately tied
to string theory and gravity. Topics covered include the Weyl-Wigner
correspondence, noncommutative Feynman diagrams, UV/IR mixing, noncommutative
Yang-Mills theory on infinite space and on the torus, Morita equivalences of
noncommutative gauge theories, twisted reduced models, and an in-depth study of
the gauge group of noncommutative Yang-Mills theory. Some of the more
mathematical ideas and techniques of noncommutative geometry are also briefly
explained.

\end{abstract}

\vskip 1 cm

\begin{center}

{\sc To appear in Physics Reports}

\end{center}

\end{titlepage}
\setcounter{page}{2}

\tableofcontents
\newpage

\renewcommand{\thefootnote}{\arabic{footnote}} \setcounter{footnote}{0}

\newsection{Historical Introduction}

\subsection{Evidence for Spacetime Noncommutativity}

It was suggested very early on by the founding fathers of quantum
mechanics, most notably Heisenberg, in the pioneering days
of quantum field theory that one could use a noncommutative structure for
spacetime coordinates at very small length scales to introduce an effective
ultraviolet cutoff. It was Snyder~\cite{snyder} who first formalized
this idea in an article entirely devoted to the subject. This was
motivated by the need to control the divergences
which had plagued theories such as quantum electrodynamics from the very
beginning. It was purported to be superior to earlier suggestions of lattice
regularization in that it maintained Lorentz invariance. However, this
suggestion was largely ignored, but mostly because of its timing. At around the
same time, the renormalization program of quantum field theory finally proved
to be successful at accurately predicting numerical values for physical
observables in quantum electrodynamics.

The idea behind spacetime noncommutativity is very much inspired by quantum
mechanics. A quantum phase space is defined by replacing canonical position and
momentum variables $x^i,p_j$ with Hermitian operators $\hat x^i,\hat p_j$ which
obey the Heisenberg commutation relations $[\hat x_j,\hat
p^i]=i\,\hbar\,\delta^i_{~j}$. The phase space becomes smeared out and the
notion of a point is replaced with that of a Planck cell. In the classical
limit $\hbar\to0$, one recovers an ordinary space. It was von Neumann who first
attempted to rigorously describe such a quantum ``space'' and he dubbed this
study ``pointless geometry'', refering to the fact that the notion of a point
in a quantum phase space is meaningless because of the Heisenberg uncertainty
principle of quantum mechanics. This led to the theory of von Neumann algebras
and was essentially the birth of ``noncommutative geometry'', refering to the
study of topological spaces whose commutative $C^*$-algebras of functions are
replaced by noncommutative algebras~\cite{connesbook}. In this setting, the
study of the properties of ``spaces'' is done in purely algebraic terms
(abandoning the notion of a ``point'') and thereby allows for rich
generalizations.

Just as in the quantization of a classical phase space, a noncommutative
spacetime is defined by replacing spacetime coordinates $x^i$ by the Hermitian
generators $\hat x^i$ of a noncommutative $C^*$-algebra of ``functions on
spacetime''~\cite{connesbook} which obey the commutation relations
\beq
\left[\hat x^i\,,\,\hat x^j\right]=i\,\theta^{ij} \ .
\label{noncommalg}\eeq
The simplest special case of (\ref{noncommalg}) is where $\theta^{ij}$
is a constant, real-valued antisymmetric $D\times D$ matrix
($D$ is the dimension of spacetime) with dimensions of length squared. Since
the coordinates no longer commute, they cannot be simultaneously diagonalized
and the underlying space disappears, i.e. the spacetime manifold gets
replaced by a Hilbert space of states. Because of the induced spacetime
uncertainty relation,
\beq
\Delta x^i\,\Delta x^j\geq\frac12\,\left|\theta^{ij}\right| \ ,
\label{spacetimeuncert}\eeq
a spacetime point is replaced by a Planck cell of dimension given by the Planck
area. In this way one may think of ordinary spacetime coordinates $x^i$ as
macroscopic order parameters obtained by coarse-graining over scales smaller
than the fundamental scale $\Lambda\sim\sqrt\theta$. To describe
physical phenomena on scales of the order of $\theta$, the $x^i$'s
break down and must be replaced by elements of some noncommutative
algebra. Snyder's idea was that if one could find a coherent
description for the structure of spacetime which is pointless on small
length scales, then the ultraviolet divergences of quantum field
theory could be eliminated. It would be equivalent to using an
ultraviolet cutoff $\Lambda$ on momentum space integrations to compute
Feynman diagrams, which implicitly leads to a fundamental length scale
$\Lambda^{-1}$ below which all phenomena are ignored. The old belief
was therefore that the simplest, and most elegant, Lorentz-invariant
way of introducing $\Lambda$ is through noncommuting spacetime
``coordinates'' $\hat x^i$.\footnote{\baselineskip=12pt However, as we
  will discuss later on, this old idea is too naive and spacetime
  noncommutativity, at least in the form (\ref{noncommalg}), does not
  serve as an ultraviolet regulator.}

The ideas of noncommutative geometry were revived in the 1980's by the
mathematicians Connes, and Woronowicz and Drinfel'd, who generalized
the notion of a differential structure to the noncommutative
setting~\cite{connes1}, i.e. to arbitrary $C^*$-algebras, and also to
quantum groups and matrix pseudo-groups. Along with the definition of
a generalized integration~\cite{connesrieffel}, this led to an
operator algebraic description of (noncommutative) spacetimes (based
entirely on algebras of ``functions'') and it enables one to define
Yang-Mills gauge theories on a large class of noncommutative spaces. A
concrete example of physics in noncommutative spacetime is Yang-Mills
theory on a noncommutative torus~\cite{connesrieffel}. For quite some
time, the physical applications were based on geometric
interpretations of the standard model and its various fields and
coupling constants (the so-called Connes-Lott
model)~\cite{conneslottrev}. Other quantum field theories were also
studied along these lines (see for example~\cite{NCFTeg}). Gravity was
also eventually introduced in a unifying way~\cite{NCgrav}. The
central idea behind these approaches was to use a modified form of the
Kaluza-Klein mechanism in which the hidden dimensions are replaced by
noncommutative structures~\cite{NCKK}. For instance, in this
interpretation of the standard model~\cite{conneslottrev} the Higgs
field is a discrete $\zed_2$ gauge field on a noncommutative space,
regarded as an internal Kaluza-Klein type excitation. This led to an
automatic proof of the Higgs mechanism, independently of the details
of the Higgs potential. The input parameters are the masses of all
quarks and leptons, while the Higgs mass is a prediction of the
model. However, this approach suffered many weaknesses and eventually
died out. Most glaring was the problem that quantum radiative
corrections could not be incorporated in order to give satisfactory
predictions. Nevertheless, the model led to a revival of Snyder's idea
that classical general relativity would break down at the Planck scale because
spacetime would no longer be described by a differentiable
manifold~\cite{planckgrav}. At these length scales quantum gravitational
fluctuations become large and cannot be ignored~\cite{Ahl}.

More concrete evidence for spacetime noncommutativity came from string theory,
at present the best candidate for a quantum theory of gravity, which in the
1980's raised precisely the same sort of expectations about the structure of
spacetime at short distances. Because strings have a finite intrinsic length
scale $\ell_s$, if one uses only string states as probes of short distance
structure, then it is not possible to observe distances smaller than $\ell_s$.
In fact, based on the analysis of very high-energy string scattering
amplitudes~\cite{stringuncert}, string-modified Heisenberg uncertainty
relations have been postulated in the form
\beq
\Delta x=\frac\hbar2\left(\frac1{\Delta p}+\ell_s^2\,\Delta p\right) \ .
\label{strHeisen}\eeq
When $\ell_s=0$, the relation (\ref{strHeisen}) gives the usual quantum
mechanical prediction that the spatial extent of an object decreases as its
momentum grows. However, from (\ref{strHeisen}) it follows that the size of a
string grows with its energy. Furthermore, minimizing (\ref{strHeisen}) with
respect to $\Delta p$ yields an absolute lower bound on the measurability of
lengths in the spacetime, $(\Delta x)_{\rm
  min}=\ell_s$.\footnote{\baselineskip=12pt This bound can in fact be
  lowered to the 11-dimensional Planck length when one uses D0-branes
  as probes of short distance spacetime structure. This will be
  explained further in the next subsection.} Thus string theory
gives an explicit realization of the notion of the smearing out of spacetime
coordinates as described above. More generally, {\it spacetime} uncertainty
relations have been postulated in the form~\cite{spacetimeuncert}
\beq
\Delta x^i\,\Delta x^j=\ell_p^2
\label{spacetimeuncert1}\eeq
where $\ell_p$ is the Planck length of the spacetime. Thus the spacetime
configurations are smeared out and the notion of a ``point'' becomes
meaningless. In the low-energy limit $\ell_p\to0$, one recovers the usual
classical spacetime with commuting coordinates at large distance scales.

The apparent need in string theory for a description of spacetime in terms of
noncommutative geometry is actually even stronger than at first sight. This is
because of the notion of {\it quantum geometry}, which may be defined as the
appropriate modification of classical general relativity implied by string
theory. One instance of this is the quantum $T$-duality symmetry of strings on
a toroidal compactification~\cite{Tduality}. Consider, for example, closed
strings compactified on a circle ${\bf S}^1$ of radius $R$. Then $T$-duality
maps this string theory onto one with target space the circle of dual radius
$\tilde R=\ell_s^2/R$, and at the same time interchanges the Kaluza-Klein
momenta of the strings with their winding numbers around the ${\bf S}^1$ in the
spectrum of the quantum string theory. Because of this stringy symmetry, the
moduli space of string theories with target space ${\bf S}^1$ is parametrized
by radii $R\geq\ell_s$ (rather than the classical $R\geq0$), and very small
circles are unobservable because the corresponding string theory can be mapped
onto a completely equivalent one living in an ${\bf S}^1$ of very large radius.
This has led to a mathematically rigorous study of duality symmetries
\cite{fg}--\cite{lls1} using the techniques of noncommutative geometry. The
phenomenon of mirror symmetry is also possible to capture in this formalism,
which is based primarily on the geometry of the underlying worldsheet
superconformal field theories~\cite{for}. The main goal of these analyses is
the construction of an infinite-dimensional noncommutative ``effective target
space'' on which duality is realized as a true symmetry, i.e. as an isometry of
an appropriate Riemannian geometry. In this framework, a duality transformation
has a simple and natural interpretation as a change of ``coordinates'' inducing
the appropriate change of metric. It is inspired in large part by Witten's old
observation~\cite{witten1} that the de~Rham complex of a manifold can be
reconstructed from the geometry of two-dimensional supersymmetric
$\sigma$-models with target space the given manifold. A crucial ingredient of
this construction is the properties possessed by the closed string vertex
operator algebra, which in a particular low energy limit has the structure of a
deformation algebra of functions on the target space~\cite{lls1}. This sort of
deformation is very similar to what appears in Witten's open string field
theory~\cite{witten2}, which constitutes the original appearence of
noncommutative geometry in string theory. The relationships between closed
string theory and noncommutative geometry are reviewed in~\cite{lsrev}. Other
early aspects of the noncommutative geometry of strings may be found
in~\cite{otherncgstrings}.

Despite these successes, up until recently there have remained two main gaps in
the understanding of the role of noncommutative geometry in string theory:
\begin{itemize}
\item{While most of the formalism deals with closed strings, the role of {\it
open} strings was previously not clear.}
\item{There is no natural {\it dynamical} origin for the occurence of
noncommutative generalizations of field theories, and in particular of
Yang-Mills theory on a noncommutative space.}
\end{itemize}

\subsection{Matrix Models}

The answers to the latter two points are explained by open string degrees of
freedom known as D-branes~\cite{Dbranes}, which are fixed hypersurfaces in
spacetime onto which the endpoints of strings can attach. It was realized very
early on in studies of the physics of D-branes that their low-energy effective
field theory has configuration space which is described in terms of
noncommuting, matrix-valued spacetime coordinate fields~\cite{witten3}. This
has led to the Matrix theory conjecture~\cite{bfss} and also the so-called IIB
matrix model~\cite{ikkt}, both of which propose nonperturbative approaches to
superstring theories. The latter matrix model is obtained by dimensionally
reducing ordinary Yang-Mills theory to a point and its bosonic part is given by
the D-instanton action
\beq
S_{\rm IIB}=-\frac1{4g^2}\,\sum_{i\neq j}\tr\left[X^i\,,\,X^j\right]^2
\label{IIBbosaction}\eeq
where $X^i$, $i=1,\dots,D$, are $N\times N$ Hermitian matrices whose
entries are c-numbers. The global minimum of the action
(\ref{IIBbosaction}) is given by the equation
$[X^i,X^j]=0$,\footnote{\baselineskip=12pt Other classical minima
  include solutions with non-vanishing but constant commutator. This
  observation will be used in section~7 to establish a correspondence
  between the matrix model (\ref{IIBbosaction}) and noncommutative
  Yang-Mills theory.} so that the matrices $X^i$ are simultaneously
diagonalizable in the ground state. Their eigenvalues represent the
collective coordinates of the individual D-branes, and so at
tree-level we obtain an ordinary spacetime. However, the quantum
fluctuations about the classical minima give a spacetime whose
coordinates are described by noncommuting matrices. The noncommutative
geometry that arises in this way is due to the short open strings which connect
the individual D-branes to one another~\cite{witten3}. Because of these
excitations, D-branes can probe Planckian distances in spacetime at which their
worldvolume field theories are drastically altered by quantum gravitational
effects~\cite{dkps}. Furthermore, the matrix noncommutativity of the target
space of multiple D-brane systems agrees with the forms of the string-modified
uncertainty relations~\cite{dbraneuncert}.

A more concrete connection to noncommutative geometry came from studying the
toroidal compactifications of the matrix model (\ref{IIBbosaction})~\cite{cds}.
It was shown that the most general solutions $X^i$ to the so-called quotient
conditions for toroidal compactification are given by gauge connections on a
noncommutative torus. Substituting these $X^i$'s back into the D-instanton
action gives rise to Yang-Mills theory on a dual noncommutative torus. Thus,
these matrix models naturally lead to noncommutative Yang-Mills theory as their
effective field theories, and noncommutative geometry is now believed to be an
important aspect of the nonperturbative dynamics of superstring theory (and
M-theory). The noncommutativity was interpreted as the effect of turning on the
light-like component $C_{-ij}$ of the background three-form field of
11-dimensional supergravity wrapped on cycles of a torus through the
identification~\cite{cds}
\beq
\left(\Theta^{-1}\right)_{ij}=R\,\oint dx^i\wedge dx^j~C_{-ij} \ ,
\label{Theta3form}\eeq
where $R=\oint dx^-\,1$ (Here $\Theta^{ij}$ denote the dimensionless
noncommutativity parameters). This identification holds in the scaling limit
that defines Matrix theory via discrete light-cone
quantization~\cite{dlcqmatrix}. In the usual reduction of M-theory to Type II
superstring theory~\cite{Mtheory}, the three-form field $C$ becomes the
Neveu-Schwarz two-form field $B$, with $\theta\sim B^{-1}$. This
noncommutativity has been subsequently understood directly in the context of
open string quantization~\cite{ncopenstring}--\cite{sw}, so that noncommutative
geometry plays a role in the quantum dynamics of open strings in background
fields and in the presence of D-branes. The relationship between the matrix
noncommutativity of D-brane field theory and the noncommutativity due to
background supergravity fields is clarified in~\cite{ncrels}. At present,
noncommutative Yang-Mills theory is believed to be a useful tool in the
classification of string backgrounds, the best examples being the discoveries
of noncommutative instantons for $D=4$~\cite{ncinstanton}, and of solitons in
2+1-dimensional noncommutative gauge theory~\cite{ncgtsoliton,gn}. Other
stringy type topological defects in this latter context may also be
constructed~\cite{ncgtother}.

\subsection{Strong Magnetic Fields}

To quantify some of the previous remarks, we will now illustrate how
noncommutativity emerges in a simple quantum mechanical example, the Landau
problem~\cite{landau}. Consider a charged particle of mass $m$ moving in the
plane $\vec x=(x^1,x^2)$ and in the presence of a constant, perpendicular
magnetic field of magnitude $B$. The Lagrangian is
\beq
{\cal L}_m=\frac m2\,\dot{\vec x}^{\,2}-\dot{\vec x}\cdot\vec A
\label{Landaum}\eeq
where $A_i=-\frac B2\,\epsilon_{ij}\,x^j$ is the corresponding vector
potential. The Hamiltonian is ${\cal H}_m=\frac1{2m}\,\vec\pi^{\,2}$, where
$\vec\pi=m\dot{\vec x}=\vec p+\vec A$ is the gauge invariant mechanical
momentum (which is a physical observable), while $\vec p$ is the (gauge
variant) canonical momentum. From the canonical commutation relations it
follows that the physical momentum operators have the non-vanishing quantum
commutators
\beq
\left[\hat\pi^i\,,\,\hat\pi^j\right]=i\,B\,\epsilon^{ij} \ ,
\label{picommrels}\eeq
and so the momentum space in the presence of a background magnetic field $\vec
B$ becomes noncommutative. The points in momentum space are replaced by Landau
cells of area $B$ which serves as an infrared cutoff, i.e. $\vec\pi^{\,2}\geq
B$. In this way the noncommutativity regularizes potentially divergent
integrals such as $\int d^2\pi/\vec\pi^{\,2}\sim\ln B$.

Spatial noncommutativity arises in the limit $m\to0$ whereby the Landau
Lagrangian becomes
\beq
{\cal L}_0=-\frac B2\,\dot x^i\,\epsilon_{ij}\,x^j \ .
\label{Landau0}\eeq
This is a first order Lagrangian which is already expressed in phase space with
the spatial coordinates $x^1,x^2$ being the canonically conjugate variables, so
that
\beq
\left[\hat x^i\,,\,\hat x^j\right]=\frac iB\,\epsilon^{ij} \ .
\label{Bnoncomm}\eeq
This limiting theory is topological, in that the corresponding Hamiltonian
vanishes and there are no propagating degrees of freedom. Note that the space
noncommutativity (\ref{Bnoncomm}) alternatively follows from the momentum
noncommutativity (\ref{picommrels}) by imposing the first class constraints
$\vec\pi\approx\vec0$. The limit $m\to0$ thereby reduces the four dimensional
phase space to a two dimensional one which coincides with the configuration
space of the model. Such a degeneracy is typical in topological quantum field
theories~\cite{tqft}. The limit $m\to0$ with fixed $B$ is actually the
projection of the quantum mechanical spectrum of this system onto the lowest
Landau level (The mass gap between Landau levels is $B/m$). The same projection
can be done in the limit $B\to\infty$ of strong magnetic field with fixed mass
$m$.

This simple example has a more or less direct analog in string
theory~\cite{acny}. Consider bosonic strings moving in flat Euclidean space
with metric $g_{ij}$, in the presence of a constant Neveu-Schwarz two-form
$B$-field and with D$p$-branes. The $B$-field is equivalent to a constant
magnetic field on the branes, and it can be gauged away in the directions
transverse to the D$p$-brane worldvolume. The (Euclidean) worldsheet action is
\beq
S_\Sigma=\frac1{4\pi\alpha'}\,\int\limits_\Sigma\left(g_{ij}\,\partial_a x^i
\,\partial_a x^j-2\pi i\alpha'B_{ij}\,\epsilon^{ab}\,\partial_ax^i\,
\partial_b x^j\right)
\label{SSigmabulk}\eeq
where $\alpha'=\ell_s^2$, $\Sigma$ is the string worldsheet, and $x^i$ is the
embedding function of the strings into flat space. The term involving the
$B$-field in (\ref{SSigmabulk}) is a total derivative and for open strings it
can be written as an integral over the boundary of the string worldsheet,
\beq
S_{\partial\Sigma}=-\frac
i2\,\oint\limits_{\partial\Sigma}B_{ij}\,x^i\,\partial_tx^j \ ,
\label{SSigmabdry}\eeq
where $t$ is the coordinate of $\partial\Sigma$. Consider now the correlated
low-energy limit $g_{ij}\sim(\alpha')^2\sim\varepsilon\to0$, with $B_{ij}$
fixed~\cite{sw}. Then the bulk kinetic terms for the $x^i$ in
(\ref{SSigmabulk}) vanish, and the worldsheet theory is topological. All that
remains are the boundary degrees of freedom of the open strings which are
governed by the action (\ref{SSigmabdry}). Then, ignoring the fact that
$x^i(t)$ is the boundary value of a string, the one-dimensional action
(\ref{SSigmabdry}) coincides with that of the Landau action describing the
motion of electrons in a strong magnetic field. From this we may infer the
noncommutativity $[\hat x^i,\hat x^j]=(i/B)^{ij}$ of the coordinates of the
endpoints of the open strings which live in the D$p$-brane worldvolume. The
correlated low energy limit $\alpha'\to0$ taken above effectively decouples the
closed string dynamics from the open string dynamics. It also decouples the
massive open string states, so that the string theory reduces to a field
theory. Only the endpoint degrees of freedom remain and describe a
noncommutative geometry.\footnote{\baselineskip=12ptThe situation is actually a
little more subtle than that described above, since in the present case the
coordinates $x^i(t)$ do not simply describe the motion of particles but are
rather constrained to lie at the ends of strings. However, the general picture
that $x^i(t)$ become noncommuting operators remains valid always~\cite{sw}.}

\subsection{Outline and Omissions}

When the open string $\sigma$-model (\ref{SSigmabulk}) is coupled to gauge
field degrees of freedom which live on the worldsheet boundary
$\partial\Sigma$, the low-energy effective field theory may be described by
noncommutative Yang-Mills theory (modulo a certain factorization equivalence
that we shall describe later on)~\cite{sw}. Furthermore, it has been shown
independently that the IIB matrix model with D-brane backgrounds gives a
natural regularization of noncommutative Yang-Mills theory to all orders of
perturbation theory, with momentum space noncommutativity as in
(\ref{picommrels})~\cite{aiikkt}. The fact that quantum field theory on a
noncommutative space arises naturally in string theory and Matrix theory
strongly suggests that spacetime noncommutativity is a general feature of a
unified theory of quantum gravity. The goal of these lecture notes is to
provide a self-contained, pedagogical introduction to the basic aspects of
noncommutative field theories and in particular noncommutative Yang-Mills
theory. We shall pay particular attention to those aspects of these quantum
field theories which may be regarded as ``stringy''. Noncommutative field
theories have many novel properties which are not exhibited by conventional
quantum field theories. They should be properly understood as lying somewhere
between ordinary field theory and string theory, and the hope is that from
these models we may learn something about string theory and the classification
of its backgrounds, using the somewhat simpler techniques of quantum field
theory. Our presentation will be in most part at the field theoretical level,
but we shall frequently indicate how the exotic properties of noncommutative
field theories are intimately tied to string theory.

The organization of the remainder of this paper is as follows. In section 2 we
shall introduce the procedure of Weyl quantization which is a useful technique
for translating an ordinary field theory into a noncommutative one. In section
3 we shall take a very basic look at the perturbative expansion of
noncommutative field theories, using a simple scalar model to illustrate the
exotic properties that one uncovers. In section 4 we introduce noncommutative
Yang-Mills theory, and discuss its observables and some of its perturbative
properties. In section 5 we will describe the classic and very important
example of the noncommutative torus and gauge theories defined thereon. In
section 6 we shall derive a very important geometrical equivalence between
noncommutative Yang-Mills theories known as Morita
equivalence,\footnote{\baselineskip=12pt Morita equivalence is
  actually an algebraic rather than geometric equivalence. Here we
  mean {\it gauge} Morita equivalence which also maps geometrical
  structures defined in the gauge theory.} which we will
see is the analog of the $T$-duality symmetry of toroidally compactified open
strings. In section 7 we shall take a look at the matrix model formulations of
noncommutative gauge theories and a nonperturbative lattice regularization of
these models. Finally, in section 8 we will describe in some detail the local
and global properties of the gauge group of noncommutative Yang-Mills theory.

We conclude this introductory section with a brief list of the major
omissions in the present review article, and places where the
interested reader may find these topics. Other general reviews on the
subject, with very different emphasis than the present article, may be
found in~\cite{ncrevs}. Solitons and instantons in noncommutative
field theory are reviewed in~\cite{solitonrevs}. More general
star-products than the ones described here can be found
in~\cite{genstars} and references therein. The Seiberg-Witten map was
introduced in~\cite{sw} and has been the focal point of many
works. See~\cite{OkOog} for the recent exact solution, and references
therein for previous analyses. The stringy extension of noncommutative
gauge theory, defined by the noncommutative Born-Infeld action, is
analysed in~\cite{sw,leereg,NCDBI}, for example. The relationship
between noncommutative field theory and string field theory is reviewed
in~\cite{StringQFT}. A recent review of the more phenomenological
aspects of noncommutative field theory may be found
in~\cite{NCphen}. Finally, aspects of the $\theta$-expanded approach
to noncommutative gauge field theory, which among other things enables
a construction of noncommutative Yang-Mills theory for arbitrary gauge
groups, may be found in~\cite{thetaexp}.

\newsection{Weyl Quantization and the Groenewold-Moyal Product}

As we mentioned in section 1.1, many of the general ideas behind noncommutative
geometry are inspired in large part by the foundations of quantum mechanics.
Within the framework of canonical quantization, Weyl introduced an elegant
prescription for associating a quantum operator to a classical function of the
phase space variables~\cite{weylbook}. This technique provides a systematic way
to describe noncommutative spaces in general and to study field theories
defined thereon. In this section we shall introduce this formalism which will
play a central role in most of our subsequent analysis. Although we
will focus solely on the commutators (\ref{noncommalg}) with
constant $\theta^{ij}$, Weyl quantization also works for more general
commutation relations.

\subsection{Weyl Operators}

Let us consider the commutative algebra of (possibly complex-valued) functions
on $D$ dimensional Euclidean space $\real^D$, with product defined by the usual
pointwise multiplication of functions. We will assume that all fields defined
on $\real^D$ live in an appropriate Schwartz space of functions of sufficiently
rapid decrease at infinity~\cite{rieffelschwartz}, i.e. those functions whose
derivatives to arbitrary order vanish at infinity in both position and momentum
space. This condition can be characterized, for example, by the requirements
\beq
\sup_x\,\Bigl(1+|x|^2\Bigr)^{k+n_1+\dots+n_D}\,\Bigl|\partial_1^{n_1}\cdots
\partial_D^{n_D}f(x)\Bigr|^2~<~\infty
\label{Schwartzcondn}\eeq
for every set of integers $k,n_i\in\zed_+$, where $\partial_i=\partial/\partial
x^i$. In that case, the algebra of functions may be given the structure of a
Banach space by defining the $L^\infty$-norm
\beq
\|f\|^{~}_\infty=\sup_x\,\Bigl|f(x)\Bigr| \ .
\label{Linftynorm}\eeq

The Schwartz condition also implies that any function $f(x)$ may be described
by its Fourier transform
\beq
\tilde f(k)=\int d^Dx~\e^{-ik_ix^i}\,f(x) \ ,
\label{Fourierf}\eeq
with $\tilde f(-k)=\tilde f(k)^*$ whenever $f(x)$ is real-valued. We define a
noncommutative space as described in section 1.1 by replacing the local
coordinates $x^i$ of $\real^D$ by Hermitian operators $\hat x^i$ obeying the
commutation relations (\ref{noncommalg}). The $\hat x^i$ then generate a
noncommutative algebra of operators. Weyl quantization provides a one-to-one
correspondence between the algebra of fields on $\real^D$ and this ring of
operators, and it may be thought of as an analog of the operator-state
correspondence of local quantum field theory. Given the function $f(x)$ and its
corresponding Fourier coefficients (\ref{Fourierf}), we introduce its {\it Weyl
symbol} by
\beq
\weyl[f]=\int\frac{d^Dk}{(2\pi)^D}~\tilde f(k)~\e^{ik_i\hat x^i} \ ,
\label{Weylopdef}\eeq
where we have chosen the symmetric Weyl operator ordering prescription. For
example, $\weyl[\e^{ik_ix^i}]=\e^{ik_i\hat x^i}$. The Weyl operator $\weyl[f]$
is Hermitian if $f(x)$ is real-valued.

We can write (\ref{Weylopdef}) in terms of an explicit map $\hat\Delta(x)$
between operators and fields by using (\ref{Fourierf}) to get
\beq
\weyl[f]=\int d^Dx~f(x)\,\hat\Delta(x)
\label{WeylDelta}\eeq
where
\beq
\hat\Delta(x)=\int\frac{d^Dk}{(2\pi)^D}~\e^{ik_i\hat x^i}~\e^{-ik_ix^i} \ .
\label{Deltadef}\eeq
The operator (\ref{Deltadef}) is Hermitian,
$\hat\Delta(x)^\dagger=\hat\Delta(x)$, and it describes a mixed basis for
operators and fields on spacetime. In this way we may interpret the field
$f(x)$ as the coordinate space representation of the Weyl operator $\weyl[f]$.
Note that in the commutative case $\theta^{ij}=0$, the map (\ref{Deltadef})
reduces trivially to a delta-function $\delta^D(\hat x-x)$ and
$\weyl[f]|_{\theta=0}=f(\hat x)$. But generally, by the
Baker-Campbell-Hausdorff formula, for $\theta^{ij}\neq0$ it is a highly
non-trivial field operator.

We may introduce ``derivatives'' of operators through an anti-Hermitian linear
derivation $\hat\partial_i$ which is defined by the commutation relations
\beq
\left[\hat\partial_i\,,\,\hat x^j\right]=\delta_i^{~j}~~~~~~,~~~~~~\left[\hat
\partial_i\,,\,\hat\partial_j\right]=0 \ .
\label{derivdef}\eeq
Then it is straightforward to show that
\beq
\left[\hat\partial_i\,,\,\hat\Delta(x)\right]=-\partial_i\,\hat\Delta(x) \ ,
\label{partialDelta}\eeq
which upon integration by parts in (\ref{WeylDelta}) leads to
\beq
\left[\hat\partial_i\,,\,\weyl[f]\right]=\int d^Dx~\partial_if(x)\,
\hat\Delta(x)=\weyl[\partial_if] \ .
\label{partialWeyl}\eeq
{}From (\ref{partialDelta}) it also follows that translation generators can be
represented by unitary operators $\e^{v^i\hat\partial_i}$, $v\in\real^D$, with
\beq
\e^{v^i\hat\partial_i}\,\hat\Delta(x)\,\e^{-v^i\hat\partial_i}
=\hat\Delta(x+v) \ .
\label{Deltatransl}\eeq
The property (\ref{Deltatransl}) implies that any cyclic trace Tr defined on
the algebra of Weyl operators has the feature that $\Tr\,\hat\Delta(x)$ is
independent of $x\in\real^D$. From (\ref{WeylDelta}) it follows that the trace
Tr is uniquely given by an integration over spacetime,
\beq
\Tr\,\weyl[f]=\int d^Dx~f(x) \ ,
\label{Tracedef}\eeq
where we have chosen the normalization $\Tr\,\hat\Delta(x)=1$. In this sense,
the operator trace Tr is equivalent to integration over the noncommuting
coordinates $\hat x^i$. Note that $\hat\Delta(x)$ is not an element of the
algebra of fields and so its trace is {\it not} defined by (\ref{Tracedef}). It
should be simply thought of as an object which interpolates between fields on
spacetime and Weyl operators, whose trace is fixed by the given normalization.

The products of operators $\hat\Delta(x)$ at distinct points may be computed as
follows. Using the Baker-Campbell-Hausdorff
formula,\footnote{\baselineskip=12ptGoing back to the quantum mechanical
example in section 1.3 of a particle in a constant magnetic field, the relation
(\ref{BCH}) defines the algebra of magnetic translation operators for the
Landau levels~\cite{magtransl}.}
\beq
\e^{ik_i\hat x^i}~\e^{ik_i'\hat x^i}=\e^{-\frac i2\,\theta^{ij}k_ik_j'}
{}~\e^{i(k+k')_i\hat x^i} \ ,
\label{BCH}\eeq
along with (\ref{WeylDelta}), one may easily derive
\bea
\hat\Delta(x)\,\hat\Delta(y)&=&\int\!\!\!\int\frac{d^Dk}{(2\pi)^D}~
\frac{d^Dk'}{(2\pi)^D}~\e^{i(k+k')_i\hat x^i}~\e^{-\frac i2\,
\theta^{ij}k_ik_j'}~\e^{-ik_ix^i-ik_i'y^i}\nn\\&=&\int\!\!\!\int
\frac{d^Dk}{(2\pi)^D}~\frac{d^Dk'}{(2\pi)^D}~\int
d^Dz~\e^{i(k+k')_iz^i}~\hat\Delta(z)~\e^{-\frac
  i2\,\theta^{ij}k_ik_j'}~
\e^{-ik_ix^i-ik_i'y^i} \ . \nn\\& &
\label{Deltasderiv}\eea
If $\theta$ is an invertible matrix (this necessarily requires that the
spacetime dimension $D$ be even), then one may explicitly carry out the
Gaussian integrations over the momenta $k$ and $k'$ in (\ref{Deltasderiv}) to
get
\beq
\hat\Delta(x)\,\hat\Delta(y)=\frac1{\pi^D|\det\theta|}\,\int
d^Dz~\hat\Delta(z)~\e^{-2i(\theta^{-1})_{ij}(x-z)^i(y-z)^j} \ .
\label{Delta2prod}\eeq
In particular, using the trace normalization and the antisymmetry of
$\theta^{-1}$, from (\ref{Delta2prod}) it follows that the operators
$\hat\Delta(x)$ for $x\in\real^D$ form an orthonormal set,
\beq
\Tr\Bigl(\hat\Delta(x)\,\hat\Delta(y)\Bigr)=\delta^D(x-y) \ .
\label{Deltaortho}\eeq
This, along with (\ref{WeylDelta}), implies that the transformation
$f(x)\stackrel{\hat\Delta(x)}{\longmapsto}\weyl[f]$ is invertible with inverse
given by
\beq
f(x)=\Tr\Bigl(\weyl[f]\,\hat\Delta(x)\Bigr) \ .
\label{inverseDelta}\eeq
The function $f(x)$ obtained in this way from a quantum operator is usually
called a {\it Wigner distribution function}~\cite{wigner}. Therefore, the map
$\hat\Delta(x)$ provides a one-to-one correspondence between Wigner fields and
Weyl operators. We shall refer to this as the {\it Weyl-Wigner
  correspondence}. For an explicit formula for (\ref{Deltadef}) in
terms of parity operators, see~\cite{JMGBContemp}.

\subsection{The Star-Product}

Let us now consider the product of two Weyl operators $\weyl[f]$ and $\weyl[g]$
corresponding to functions $f(x)$ and $g(x)$. From (\ref{WeylDelta}),
(\ref{Delta2prod}) and (\ref{Deltaortho}) it follows that the coordinate space
representation of their product can be written as (for $\theta$ invertible)
\beq
\Tr\Bigl(\weyl[f]\,\weyl[g]\,\hat\Delta(x)\Bigr)=\frac1{\pi^D|\det\theta|}\,
\int\!\!\!\int d^Dy~d^Dz~f(y)\,g(z)~\e^{-2i(\theta^{-1})_{ij}(x-y)^i(x-z)^j}
\ .
\label{Weylprodcoord}\eeq
Using (\ref{Weylopdef}), (\ref{Fourierf}), and (\ref{BCH}) we deduce that
\beq
\weyl[f]\,\weyl[g]=\weyl[f\star g] \ ,
\label{Weylstar}\eeq
where we have introduced the {\it Groenewold-Moyal
star-product}~\cite{starprod}
\bea
f(x)\star g(x)&=&\int\!\!\!\int\frac{d^Dk}{(2\pi)^D}~\frac{d^Dk'}{(2\pi)^D}~
\tilde f(k)\,\tilde g(k'-k)~\e^{-\frac i2\,\theta^{ij}k_ik_j'}~\e^{ik_i'x^i}
\nn\\&=&f(x)~\exp\left(\frac i2\,\overleftarrow{\partial_i}\,
\theta^{ij}\,\overrightarrow{\partial_j}\right)~g(x)\nn\\&=&
f(x)\,g(x)+\sum_{n=1}^\infty
\left(\frac i2\right)^n\frac1{n!}\,\theta^{i_1j_1}\cdots\theta^{i_nj_n}\,
\partial_{i_1}\cdots\partial_{i_n}f(x)\,\partial_{j_1}\cdots\partial_{j_n}g(x)
\ .\nn\\& &
\label{starproddef}\eea

The star-product (\ref{starproddef}) is associative but noncommutative, and is
defined for constant, possibly degenerate $\theta$. For $\theta=0$ it
reduces to the ordinary product of functions. It is a particular
example of a star product which is normally defined in deformation
quantization as follows~\cite{bffls}. If $\cal A$ is an associative
algebra over a field $\bb K$,\footnote{\baselineskip=12pt Associativity is not
  required here. In fact, the following construction applies to Lie
  algebras as well, with all products understood as Lie brackets.}
then a {\it deformation} of $\cal A$ is a set of formal power series
$\sum_nf_n\,\lambda^n$ which form an algebra ${\cal A}[[\lambda]]$
over the ring of formal power series ${\bb K}[[\lambda]]$ in a
variable $\lambda$. The deformed algebra has the property that ${\cal
  A}[[\lambda]]/{\cal A}\,\lambda\cong{\cal A}$, i.e. the order
$\lambda^0$ parts form the original undeformed algebra. One can then
define a new multiplication law for the deformed algebra ${\cal
  A}[[\lambda]]$. For $f,g\in{\cal A}$, this is given by the
associative ${\bb K}[[\lambda]]$-bilinear product
\beq
f\star_\lambda g=f\,g+\sum_{n=1}^\infty\lambda^n\,C_n(f,g)
\label{formalstarlambda}\eeq
which may be extended to the whole of ${\cal A}[[\lambda]]$ by
linearity. The $C_n$'s are known as Hochschild two-cochains of the
algebra $\cal A$. The particular star product (\ref{starproddef})
defines the essentially unique (modulo redefinitions of $f$
and $g$ that are local order by order in $\theta$) deformation of the algebra
of functions on $\real^D$ to a noncommutative associative algebra whose product
coincides with the Poisson bracket of functions (with respect to the symplectic
form $\theta$) to leading order, i.e. $f\star g=fg+\frac
i2\,\theta^{ij}\partial_if\partial_jg+O(\theta^2)$, and whose coefficients in a
power series expansion in $\theta$ are local differential expressions which are
bilinear in $f$ and $g$~\cite{bffls}.

Note that the Moyal commutator bracket with the local coordinates $x^i$ can be
used to generate derivatives as
\beq
x^i\star f(x)-f(x)\star x^i=i\,\theta^{ij}\,\partial_jf(x) \ .
\label{xderiv}\eeq
In general, the star-commutator of two functions can be represented in a
compact form by using a bi-differential operator as in (\ref{starproddef}),
\beq
f(x)\star g(x)-g(x)\star f(x)=2i\,f(x)~\sin\left(\frac12\,
\overleftarrow{\partial_i}\,
\theta^{ij}\,\overrightarrow{\partial_j}\right)~g(x) \ ,
\label{starcomm}\eeq
while the star-anticommutator may be written as
\beq
f(x)\star g(x)+g(x)\star f(x)=2\,f(x)~\cos\left(\frac12\,
\overleftarrow{\partial_i}\,
\theta^{ij}\,\overrightarrow{\partial_j}\right)~g(x) \ .
\label{staranticomm}\eeq
A useful extension of the formula (\ref{starproddef}) is
\beq
f_1(x_1)\star\cdots\star f_n(x_n)=\prod_{a<b}\exp\left(\frac i2\,\theta^{ij}\,
\frac\partial{\partial x_a^i}\,\frac\partial{\partial x_b^j}\right)~f_1(x_1)
\cdots f_n(x_n) \ .
\label{starprodext}\eeq

Therefore, the spacetime noncommutativity may be encoded through ordinary
products in the noncommutative $C^*$-algebra of Weyl operators, or equivalently
through the deformation of the product of the commutative $C^*$-algebra of
functions on spacetime to the noncommutative star-product. Note that by
cyclicity of the operator trace, the integral
\beq
\Tr\Bigl(\weyl[f_1]\cdots\weyl[f_n]\Bigr)=\int d^Dx~f_1(x)\star\cdots\star
f_n(x)
\label{intstars}\eeq
is invariant under cyclic (but not arbitrary) permutations of the functions
$f_a$. In particular,
\beq
\int d^Dx~f(x)\star g(x)=\int d^Dx~f(x)\,g(x) \ ,
\label{fgstarint}\eeq
which follows for Schwartz functions upon integrating by parts over $\real^D$.

The above quantization method can be generalized to more complicated situations
whereby the commutators $[\hat x^i,\hat x^j]$ are not simply
c-numbers~\cite{mssw}. The generic situation is whereby both the coordinate and
conjugate momentum spaces are noncommutative in a correlated way. Then the
commutators $[\hat x^i,\hat x^j]$, $[\hat x^i,\hat p_j]$ and $[\hat p_i,\hat
p_j]$ are functions of $\hat x^i$ and $\hat p_i$, rather than just of $\hat
x^i$, and thereby define an algebra of pseudo-differential operators on the
noncommutative space. Such a situation arises in string theory when quantizing
open strings in the presence of a non-constant $B$-field~\cite{nonconstB}, and
it was the kind of noncommutative space that was considered originally in the
Snyder construction~\cite{snyder}. If $B$ is a closed two-form,
$dB=0$, then the associative star-product in these instances is given
by the Kontsevich formula~\cite{kontsevich} for the deformation
quantization associated with general Poisson structures, i.e. Poisson tensors
$\theta$ which are in general non-constant, obey the Jacobi identity,
and may be degenerate. This formula admits an elegant representation in
terms of the perturbative expansion of the Feynman path integral for a simple
topological open string theory~\cite{kontpath}. If $B$ is not closed,
then the straight usage of the Kontsevich formula leads to a
non-associative bidifferential operator, the non-associativity being
controlled by $dB$. However, one can still use associative
star-products within the framework of (noncommutative) gerbes. We
shall not deal with these generalizations in this paper, but only the
simplest deformation described above which utilizes a noncommutative
coordinate space and an independent, commutative momentum space.

In the case of a constant and non-degenerate $\theta$, the functional integral
representation of the Kontsevich formula takes the simple form of that of a
one-dimensional topological quantum field theory and the star-product
(\ref{starproddef}) may be written as
\bea
f(x)\star g(x)&=&\left\langle f\Bigl(\eta(1)\Bigr)\,g\Bigl(\eta(0)\Bigr)\,
\delta^D\Bigl(\eta(\pm\infty)-x\Bigr)\right\rangle_\eta\nn\\&=&
\int D\eta~\delta^D\Bigl(\eta(\pm\infty)-x\Bigr)\,
f\Bigl(\eta(1)\Bigr)\,g\Bigl(\eta(0)\Bigr)\nn\\&&\times\,
\exp\frac i2\,\int\limits_{-\infty}^\infty dt~
\eta^i(t)\,\left(\theta^{-1}\right)_{ij}\,\frac{d\eta^j(t)}{dt} \ .
\label{starpathrep}\eea
Here the integral runs over paths $\eta:\real\to\real^D$ and it is understood
as an expansion about the classical trajectories $\eta(t)=x$, which are
time-independent because the Hamiltonian of the theory (\ref{starpathrep})
vanishes. Notice that the underlying Lagrangian of (\ref{starpathrep})
coincides with that of the model of section~1.3 projected onto the lowest
Landau level. The beauty of this formula is that it involves ordinary products
of the fields and is thereby more amenable to practical computations. It also
lends a physical interpretation to the star-product. It does, however, require
an appropriate regularization in order to make sense of its perturbation
expansion~\cite{leereg}.

In the present case the technique described in this section has proven to be an
invaluable method for the study of noncommutative field theory. For instance,
stable noncommutative solitons, which have no counterparts in ordinary field
theory, have been constructed by representing the Weyl operator algebra on a
multi-particle quantum mechanical Hilbert space~\cite{gms,ncsoliton}. The
noncommutative soliton field equations may then be solved by any projection
operator on this Hilbert space. We note, however, that the general construction
presented above makes no reference to any particular representation of the Weyl
operator algebra. Later on we shall work with explicit representations of this
ring.

\newsection{Noncommutative Perturbation Theory}

In this section we will take a very basic look at the perturbative expansion of
noncommutative quantum field theory. To illustrate the general ideas, we shall
consider a simple, massive Euclidean $\phi^4$ scalar field theory in $D$
dimensions. To transform an ordinary scalar field theory into a noncommutative
one, we may use the Weyl quantization procedure of the previous section.
Written in terms of the Hermitian Weyl operator $\weyl[\phi]$ corresponding to
a real scalar field $\phi(x)$ on $\real^D$, the action is
\beq
S_{(4)}[\phi]=\Tr\left(\frac12\,\left[\hat\partial_i\,,\,\weyl[\phi]\right]^2+
\frac{m^2}2\,\weyl[\phi]^2+\frac{g^2}{4!}\,\weyl[\phi]^4\right) \ ,
\label{phi4actionWeyl}\eeq
and the path integral measure is taken to be the ordinary Feynman measure for
the field $\phi(x)$ (This choice is dictated by the string theory
applications). We may rewrite this action in coordinate space by using the map
(\ref{WeylDelta}) and the property (\ref{Weylstar}) to get
\beq
S_{(4)}[\phi]=\int d^Dx~\left[\frac12\,\Bigl(\partial_i\phi(x)\Bigr)^2+
\frac{m^2}2\,\phi(x)^2+\frac{g^2}{4!}\,\phi(x)\star\phi(x)\star\phi(x)\star
\phi(x)\right] \ .
\label{phi4actioncoord}\eeq
We have used the property (\ref{fgstarint}) which implies that noncommutative
field theory and ordinary field theory are identical at the level of free
fields. In particular, the bare propagators are unchanged in the noncommutative
case. The changes come in the interaction terms, which in the present case can
be written as
\beq
\Tr\Bigl(\weyl[\phi]^4\Bigr)=\prod_{a=1}^4\,\int\frac{d^Dk_a}{(2\pi)^D}~
\tilde\phi(k_a)~(2\pi)^D\,\delta^D\left(\sum_{a=1}^4k_a\right)~
V(k_1,k_2,k_3,k_4) \ ,
\label{phi4int}\eeq
where the interaction vertex in momentum space is
\beq
V(k_1,k_2,k_3,k_4)=\prod_{a<b}\e^{-\frac i2\,k_a\wedge k_b}
\label{phi4V}\eeq
and we have introduced the antisymmetric bilinear form
\beq
k_a\wedge k_b=k_{ai}\,\theta^{ij}\,k_{bj}=-k_b\wedge k_a
\label{wedgedef}\eeq
corresponding to the tensor $\theta$. We
will assume, for simplicity, throughout this section that $\theta$ is an
invertible matrix (so that $D$ is even). By using global Euclidean invariance
of the underlying quantum field theory, the antisymmetric matrix $\theta$ may
then be rotated into a canonical skew-diagonal form with skew-eigenvalues
$\vartheta_\alpha$, $\alpha=1,\dots,\frac D2$,
\beq
\theta=\pmatrix{0&\vartheta_1& & & \cr-\vartheta_1&0& & & \cr
& &\ddots& & \cr & & &0&\vartheta_{D/2}&\cr & & &-\vartheta_{D/2}&0\cr} \ ,
\label{thetacan}\eeq
corresponding to the choice of Darboux coordinates on $\real^D$. We denote by
$\|\theta\|$ the corresponding operator norm of $\theta$,
\beq
\|\theta\|=\max_{1\leq\alpha\leq\frac D2}\,|\vartheta_\alpha| \ .
\label{varthetamax}\eeq

{}From (\ref{phi4V}) we see that the interaction vertex in noncommutative field
theory contains a momentum dependent phase factor, and the interaction is
therefore non-local. It is, however, local to each fixed order in
$\theta$. Indeed, because of the star-product, noncommutative
quantum field theories are defined by a non-polynomial derivative interaction
which will be responsible for the novel effects that we shall uncover. Given
the uniqueness property of the Groenewold-Moyal deformation, noncommutative
field theory involves the non-polynomial derivative interaction which is
multi-linear in the interacting fields and which classically reduces smoothly
to an ordinary interacting field theory (but which is at most unique
up to equivalence). Notice that since the noncommutative
interaction vertex is a phase, it does not alter the convergence properties of
the perturbation series. When $\theta=0$, we recover the standard $\phi^4$
field theory in $D$ dimensions. Naively, we would expect that this non-locality
becomes negligible for energies much smaller than the noncommutativity scale
$\|\theta\|^{-1/2}$ (Recall the discussion of section 1.1). However, as we
shall see in this section, this is not true at the quantum level. This stems
from the fact that a quantum field theory on a noncommutative spacetime is
neither Lorentz covariant nor causal with respect to a {\it fixed}
$\theta$-tensor. However, as we have discussed,
noncommutative field theories can be embedded into string theory where the
non-covariance arises from the expectation value of the background $B$-field.
We will see in this section that the novel effects induced in these quantum
field theories can be dealt with in a systematic way, suggesting that these
models do exist as consistent quantum theories which may improve our
understanding of quantum gravity at very high energies where the notion of
spacetime is drastically altered.

In fact, even before plunging into detailed perturbative calculations, one can
see the effects of non-locality directly from the Fourier integral kernel
representation (\ref{Weylprodcoord}) of the star-product of two fields. The
oscillations in the phase of the integration kernel there suppress parts of the
integration region. Precisely, if the fields $f$ and $g$ are supported over a
small region of size $\delta\ll\sqrt{\|\theta\|}$, then $f\star g$ is
non-vanishing over a much larger region of size
$\|\theta\|/\delta$~\cite{mvrs}. This is exemplified in the star product of two
Dirac delta-functions,
\beq
\delta^D(x)\star\delta^D(x)=\frac1{\pi^D|\det\theta|} \ ,
\label{deltastarprod}\eeq
so that star product of two point sources becomes infinitely non-local. At the
field theoretical level, this means that very small pulses instantaneously
spread out very far upon interacting through the Groenewold-Moyal product, so
that very high energy processes can have important long-distance consequences.
As we will see, in the quantum field theory even very low-energy processes can
receive contributions from high-energy virtual particles. In particular, due to
this non-locality, the imposition of an ultraviolet cutoff $\Lambda$ will
effectively impose an infrared cutoff $1/\|\theta\|\Lambda$.

\subsection{Planar Feynman Diagrams}

By momentum conservation, the interaction vertex (\ref{phi4V}) is only
invariant up to {\it cyclic} permutations of the momenta $k_a$. Because of this
property, one needs to carefully keep track of the cyclic order in which lines
emanate from vertices in a given Feynman diagram. This is completely analogous
to the situation in the large $N$ expansion of a $U(N)$ gauge field theory or
an $N\times N$ matrix model~\cite{thooft}. {\it Noncommutative Feynman
diagrams} are therefore ribbon graphs that can be drawn on a Riemann surface of
particular genus~\cite{iikk}. This immediately hints at a connection with
string theory. In this subsection we will consider the structure of the planar
graphs, i.e. those which can be drawn on the surface of the plane or the
sphere, in a generic scalar field theory, using the $\phi^4$ model above as
illustration.

\begin{figure}[htb]
\epsfxsize=3in
\bigskip
\centerline{\epsffile{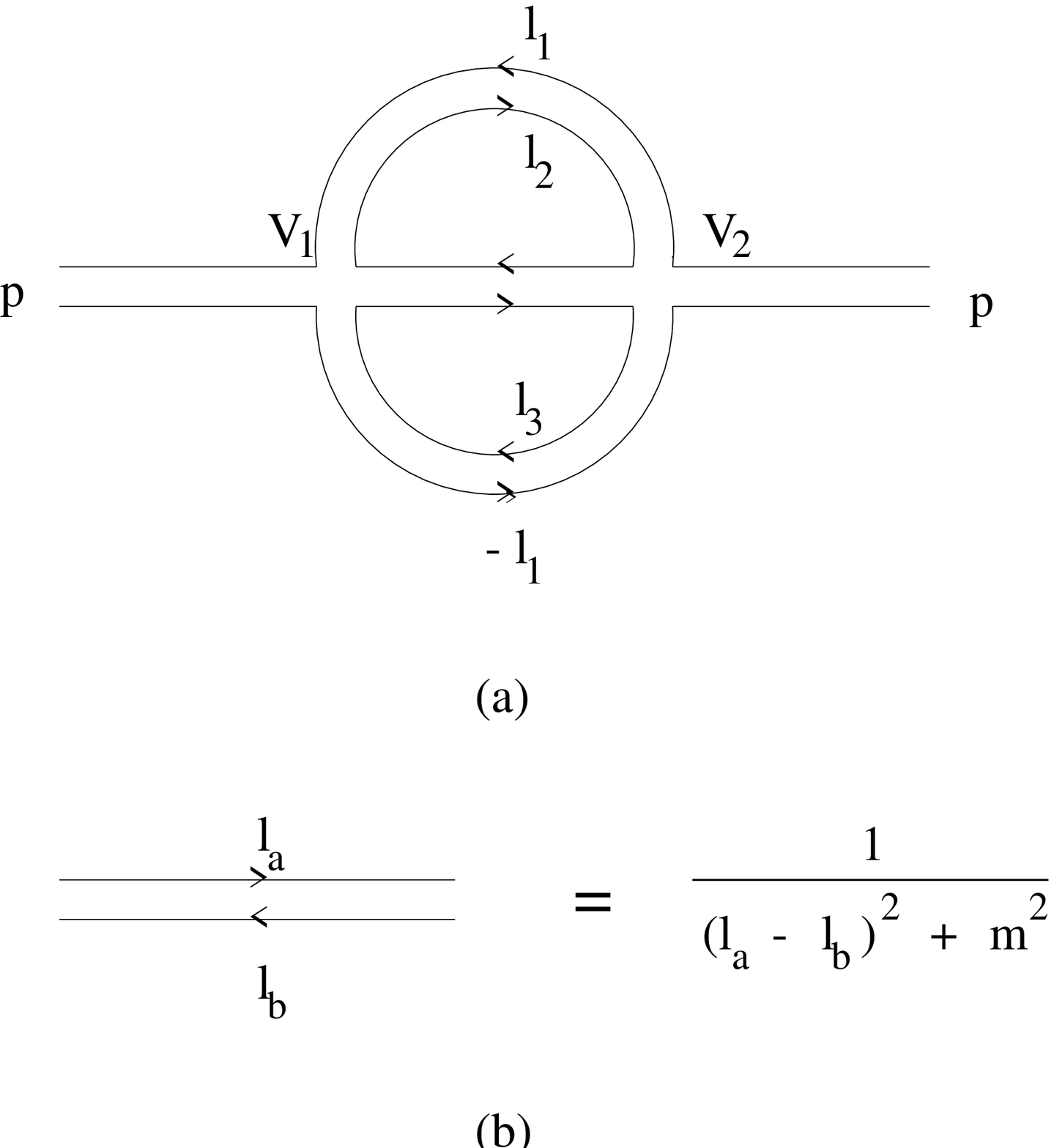}}
\caption{\baselineskip=12pt (a) {\it Example of a two-loop planar Feynman
diagram of external momentum $p$ in noncommutative $\phi^4$ theory. The
noncommutative phase factor at the first vertex is $V_1=\e^{-\frac
i2\,(l_2\wedge l_3+l_1\wedge l_2+l_3\wedge l_1)}$ while that at the second
vertex is $V_2=\e^{-\frac i2\,(l_2\wedge l_1+l_1\wedge l_3-l_2\wedge
l_3)}=(V_1)^{-1}$.} (b) {\it The massive scalar propagator in ribbon
notation.}}
\bigskip
\label{ribbonex}\end{figure}

Consider an $L$-loop planar graph, and let $k_1,\dots,k_n$ be the cyclically
ordered momenta which enter a given vertex $V$ of the graph through $n$
propagators. By introducing an oriented ribbon structure to the propagators of
the diagram, we label the index lines of the ribbons by the ``momenta''
$l_1,\dots,l_{L+1}$ such that $k_a=l_{m_a}-l_{m_{a+1}}$, where
$m_a\in\{1,\dots,L+1\}$ with $l_{m_{n+1}}=l_{m_1}$ (see fig.~\ref{ribbonex}).
Because adjacent edges in a ribbon propagator are given oppositely flowing
momenta, this construction automatically enforces momentum conservation at each
of the vertices. Given these decompositions, a noncommutative vertex $V$ such
as (\ref{phi4V}) will decompose as
\beq
V=\prod_{a=1}^n\e^{-\frac i2\,l_{m_a}\wedge l_{m_{a+1}}}
\label{Vplanargen}\eeq
into a product of phases, one for each incoming propagator. However, the
momenta associated to a given line will flow in the opposite direction at the
other end of the propagator (fig.~\ref{ribbonex}), so that the phase associated
with any internal propagator is equal in magnitude and opposite in sign at its
two ends. Therefore, the overall phase factor associated with any planar
Feynman diagram is~\cite{filk}
\beq
V_{\rm p}(p_1,\dots,p_n)=\prod_{a<b}\e^{-\frac i2\,p_a\wedge p_b}
\label{planarphase}\eeq
where $p_1,\dots,p_n$ are the cyclically ordered external momenta of the graph.
The phase factor (\ref{planarphase}) is completely independent of the details
of the internal structure of the planar graph.

We see therefore that the contribution of a planar graph to the noncommutative
perturbation series is just the corresponding $\theta=0$ contribution
multiplied by the phase factor (\ref{planarphase}). This phase factor is
present in all interaction terms in the bare Lagrangian, and in all tree-level
graphs computed with it. At $\theta=0$, divergent terms in the perturbation
expansion are determined by products of local fields, and the phase
(\ref{planarphase}) modifies these terms to the star-product of local fields.
We conclude that planar divergences at $\theta\neq0$ may be absorbed into
redefinitions of the bare parameters if and only if the corresponding
commutative quantum field theory is renormalizable~\cite{mvrs}. This dispells
the naive expectation that the Feynman graphs of noncommutative quantum field
theory would have better ultraviolet behaviour than the commutative ones (at
least for the present class of noncommutative spaces)~\cite{dispell}. Note that
here the renormalization procedure is not obtained by adding local
counterterms, but rather the counterterms are of an identical non-local form as
those of the bare Lagrangian. In any case, at the level of planar graphs for
scalar fields, noncommutative quantum field theory has precisely the same
renormalization properties as its commutative counterparts.

\subsubsection{String Theoretical Interpretation}

The factorization of the noncommutativity parameters in planar amplitudes
brings us to our first analogy to string theory. Consider the string
$\sigma$-model that was described in section 1.3. The open string propagator on
the boundary of a disk $\Sigma$ in a constant background $B$ field is given
by~\cite{schomerus,sw,acny}
\beq
\left\langle x^i(t)\,x^j(t')\right\rangle=-\alpha'G^{ij}\ln(t-t')^2
+\frac i2\,\theta^{ij}~{\rm sgn}(t-t') \ ,
\label{boundaryprop}\eeq
where
\beq
\theta^{ij}=-(2\pi\alpha')^2\left(\frac1{g+2\pi\alpha'B}\,B\,
\frac1{g-2\pi\alpha'B}\right)^{ij} \ ,
\label{thetaBrel}\eeq
and
\beq
G_{ij}=g_{ij}-(2\pi\alpha')^2\left(Bg^{-1}B\right)_{ij}
\label{openstringmetric}\eeq
is the metric seen by the {\it open} strings ($g_{ij}$ is the metric seen by
the closed strings). Consider an operator on $\partial\Sigma$ of the general
form $P[\partial x,\partial^2x,\dots]\,\e^{ip_ix^i}$, where $P$ is a polynomial
in derivatives of the coordinates $x^i$ along the D$p$-brane worldvolume. The
sign term in (\ref{boundaryprop}), which is responsible for the worldvolume
noncommutativity, does not contribute to contractions of the operators
$\partial^nx^i$ when we evaluate quantum correlation functions using the Wick
expansion. It follows then that the correlation functions in the background
fields $G,\theta$ may be computed as~\cite{sw}
\bea
& &\left\langle\prod_nP_n\Bigl[\partial
x(t_n),\partial^2x(t_n),\dots\Bigr]~\e^{ip_{ni}x^i(t_n)}\right
\rangle_{G,\theta}\nn\\& &~~~~~~=\prod_{n>m}\e^{-\frac i2\,p_n\wedge p_m
\,{\rm sgn}(t_n-t_m)}\,\left\langle\prod_n
P_n\Bigl[\partial x(t_n),\partial^2x(t_n),\dots\Bigr]~\e^{ip_{ni}x^i(t_n)}
\right\rangle_{G,\theta=0} \ . \nn\\& &
\label{stringcorrfact}\eea
This result holds for generic values of the string slope $\alpha'$. It
implies that $\sigma$-model correlation functions in a background
$B$-field may be computed by simply replacing ordinary products of fields by
star-products and the closed string metric $g$ by the open string metric $G$.
Therefore, the $\theta$-dependence of disk amplitudes when written in terms of
the open string variables $G$ and $\theta$ (rather than the closed string ones
$g$ and $B$) is very simple. These two tensors represent the metric and
noncommutativity parameters of the underlying noncommutative space. This
implies that the tree-level, low-energy effective action for open strings in a
$B$-field is obtained from that at $B=0$ by simply replacing ordinary products
of fields by star-products. By adding gauge fields to the D$p$-brane
worldvolume, this is essentially how noncommutative Yang-Mills theory arises as
the low-energy effective field theory for open strings in background
Neveu-Schwarz two-form fields~\cite{sw}. This phenomenon corresponds exactly to
the factorization of planar diagrams that we derived above. The
one-loop, annulus diagram corrections to these results are derived
in~\cite{Laidlaw}.

\subsection{Non-Planar Feynman Diagrams}

\begin{figure}[htb]
\epsfxsize=3in
\bigskip
\centerline{\epsffile{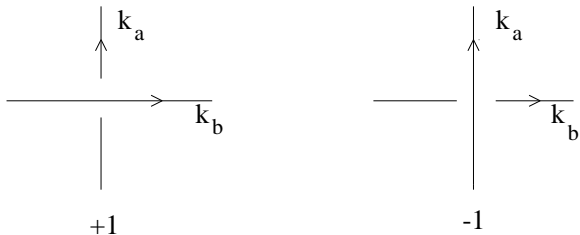}}
\caption{\baselineskip=12pt {\it Positive and negative crossings in a
non-planar Feynman graph.}}
\bigskip
\label{nonplanar}\end{figure}

The construction of the previous subsection breaks down in the case of
non-planar Feynman diagrams, which have propagators that cross over each other
or over external lines (fig.~\ref{nonplanar}). It is straightforward to show
that the total noncommutative phase factor for a general graph which
generalizes the planar result (\ref{planarphase}) is given by~\cite{filk}
\beq
V_{\rm np}(p_1,\dots,p_n)=V_{\rm p}(p_1,\dots,p_n)~\prod_{a,b}\e^{-\frac
i2\,\cap_{ab}\,k_a\wedge k_b} \ ,
\label{nonplanarphase}\eeq
where $\cap_{ab}$ is the signed intersection matrix of the graph which counts
the number of times that the $a$-th (internal or external) line crosses over
the $b$-th line (fig.~\ref{nonplanar}). By momentum conservation it follows
that the matrix $\cap_{ab}$ is essentially unique. Therefore, the $\theta$
dependence of non-planar graphs is much more complicated and we expect them to
have a much different behaviour than their commutative counterparts. In
particular, because of the extra oscillatory phase factors which occur, we
expect these diagrams to have an improved ultraviolet behaviour. When internal
lines cross in an otherwise divergent graph, the phase oscillations provide an
effective cutoff $\Lambda_{\rm eff}=\|\theta\|^{-1/2}$ and render the diagram
finite. For instance, it turns out that all one-loop non-planar diagrams are
finite, as we shall see in the next subsection. However, it is {\it not} the
case that all non-planar graphs (without divergent planar subgraphs) are
finite~\cite{mvrs}. At $\theta\neq0$, it is possible to demonstrate the
convergence of the Feynman integral associated with a diagram $\cal G$,
provided that $\cal G$ has no divergent planar subgraphs and all subgraphs of
$\cal G$ have non-positive degree of divergence. The general concensus
at present seems to be that these noncommutative scalar field theories
{\it are} renormalizable to all orders of perturbation
theory~\cite{scalarren}, although there are dangerous counterexamples
at two-loop order and at present such renormalizability statements are
merely conjectures. An explicit example of a field theory
which is renormalizable is provided by the noncommutative Wess-Zumino
model~\cite{GGRdaS,MorSchap}. In general some non-planar graphs are divergent,
but, as we will see in the next subsection, these divergences should be viewed
as infrared divergences.

Non-planar diagrams can also be seen to exhibit an interesting stringy
phenomenon. Consider the limit of maximal noncommutativity, $\theta\to\infty$,
or equivalently the ``short-distance'' limit of large momenta and fixed
$\theta$. The planar graphs have no internal noncommutative phase factors,
while non-planar graphs contain at least one. In the limit $\theta\to\infty$,
the latter diagrams therefore vanish because of the rapid oscillations of their
Feynman integrands. It can be shown~\cite{mvrs} that a noncommutative Feynman
diagram of genus $h$ is suppressed relative to a planar graph by the factor
$\prod_\alpha1/(E^2\,\vartheta_\alpha)^{2h}$, where $E$ is the total energy of
the amplitude. Therefore, if $G_{\rm conn}(p_1,\dots,p_n;\theta)$ is any
connected $n$-point Green's function in momentum space, then
\beq
\lim_{\|\theta\|\to\infty}\,\prod_{a<b}\e^{\frac i2\,p_a\wedge p_b}\,
G_{\rm conn}(p_1,\dots,p_n;\theta)=G_{\rm conn}^{\rm planar}(p_1,\dots,p_n)
\label{maxplanar}\eeq
for each $n$, and the maximally noncommutative quantum field theory is given
entirely by planar diagrams. But this is exactly the characteristic feature of
high-energy string scattering amplitudes, and thus in the high momentum or
maximal noncommutativity limit the field theory resembles a string theory. Note
that in this regard it is the largest skew-eigenvalue $\|\theta\|$ of $\theta$
which plays the role of the topological expansion parameter, i.e. $\|\theta\|$
is the analog of the rank $N$ in the large $N$ 't~Hooft genus expansion of
multi-colour field theories~\cite{thooft}.

\subsection{UV/IR Mixing}

\begin{figure}[htb]
\epsfxsize=3in
\bigskip
\centerline{\epsffile{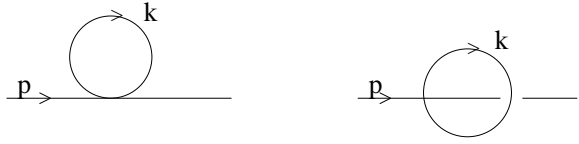}}
\caption{\baselineskip=12pt {\it The one-loop planar and non-planar irreducible
Feynman diagrams contributing to the two-point function in noncommutative
$\phi^4$ theory.}}
\bigskip
\label{oneloop}\end{figure}

In this subsection we will illustrate some of the above points with an explicit
computation, which will also reveal another exotic property of noncommutative
field theories. The example we will consider is mass renormalization in the
noncommutative $\phi^4$ theory (\ref{phi4actioncoord}) in four dimensions. For
this, we will evaluate the one-particle irreducible two-point function
\beq
\Pi(p)=\left\langle\tilde\phi(p)\,\tilde\phi(-p)\right\rangle_{\rm 1PI}=
\sum_{n=0}^\infty g^{2n}\,\Pi^{(n)}(p)
\label{1PItotal}\eeq
to one-loop order. The bare two-point function is $\Pi^{(0)}(p)=p^2+m^2$, and
at one-loop order there is (topologically) one planar and one non-planar
Feynman graph which are depicted in fig.~\ref{oneloop}. The symmetry factor for
the planar graph is twice that of the non-planar graph, and they lead to the
respective Feynman integrals
\bea
\Pi_{\rm p}^{(1)}(p)&=&\frac13\,\int\frac{d^Dk}{(2\pi)^D}\,
\frac1{k^2+m^2} \ ,\label{Pi1loopintp}\\\Pi_{\rm np}^{(1)}(p)&=&\frac16\,
\int\frac{d^Dk}{(2\pi)^D}\,\frac{\e^{ik\wedge p}}{k^2+m^2} \ .
\label{Pi1loopintnp}\eea
The planar contribution (\ref{Pi1loopintp}) is proportional to the standard
one-loop mass correction of commutative $\phi^4$ theory, which for $D=4$ is
quadratically ultraviolet divergent. The non-planar contribution is expected to
be generically convergent, because of the rapid oscillations of the phase
factor $\e^{ik\wedge p}$ at high energies. However, $k\wedge p=0$ when
$p_i\,\theta^{ij}=0$, i.e. whenever $\theta=0$ or, if $\theta$ is invertible,
whenever the external momentum $p$ vanishes. In that case the phase factor in
(\ref{Pi1loopintnp}) becomes ineffective at damping the large momentum
singularities of the integral, and the usual ultraviolet divergences of the
planar counterpart (\ref{Pi1loopintp}) creep back in through the relation
\beq
\Pi_{\rm p}^{(1)}=2\,\Pi_{\rm np}^{(1)}(p=0) \ .
\label{pnprel}\eeq
The non-planar graph is therefore singular at small $p_i\,\theta^{ij}$, and the
effective cutoff for a one-loop graph in momentum space is $1/\sqrt{|p\bullet
p|}$, where we have introduced the positive-definite inner product
\beq
p\bullet q=-p_i\,\left(\theta^2\right)^{ij}\,q_j=q\bullet p
\label{symproddef}\eeq
with $(\theta^2)^{ij}=\delta_{kl}\,\theta^{ik}\,\theta^{lj}$. Thus, at
small momenta the noncommutative phase factor is irrelevant and the
non-planar graph inherits the usual ultraviolet singularities, but now in the
form of a long-distance divergence. Turning on the noncommutativity parameters
$\theta^{ij}$ thereby replaces the standard ultraviolet divergence with a
singular infrared behaviour. This exotic mixing of the ultraviolet and infrared
scales in noncommutative field theory is called {\it UV/IR mixing}~\cite{mvrs}.

Let us quantify this phenomenon somewhat. To evaluate the Feynman integrals
(\ref{Pi1loopintp}) and (\ref{Pi1loopintnp}), we introduce the standard
Schwinger parametrization
\beq
\frac1{k^2+m^2}=\int\limits_0^\infty d\alpha~\e^{-\alpha(k^2+m^2)} \ .
\label{Schwinger}\eeq
By substituting (\ref{Schwinger}) into (\ref{Pi1loopintp},\ref{Pi1loopintnp})
and doing the Gaussian momentum integration, we arrive at
\beq
\Pi_{\rm np}^{(1)}(p)=\frac1{6(4\pi)^{D/2}}\,\int\limits_0^\infty
\frac{d\alpha}{\alpha^{D/2}}~\e^{-\alpha m^2-\frac{p\bullet p}{4\alpha}-
\frac1{\Lambda^2\alpha}} \ ,
\label{Pi1loopSchwinger}\eeq
where the momentum space ultraviolet divergence has now become a small $\alpha$
divergence in the Schwinger parameter, which we have regulated by
$\Lambda\to\infty$. The integral (\ref{Pi1loopSchwinger}) is elementary to do
and the result is
\beq
\Pi_{\rm np}^{(1)}(p)=\frac{m^{\frac{D-2}2}}{6(2\pi)^{D/2}}\,
\left(p\bullet p+\frac4{\Lambda^2}\right)^{\frac{2-D}4}\,K_{\frac{D-2}2}
\left(m\,\sqrt{p\bullet p+\frac4{\Lambda^2}}\,\right) \ ,
\label{Pi1loopexpl}\eeq
where $K_\nu(x)$ is the irregular modified Bessel function of order $\nu$. The
complete renormalized propagator up to one-loop order is then given by
\beq
\Pi(p)=p^2+m^2+2g^2\,\Pi_{\rm np}^{(1)}(0)+g^2\,\Pi_{\rm np}^{(1)}(p)+O(g^4)
\label{Pi1looptotal}\eeq
where we have used (\ref{pnprel}).

Let us now consider the leading divergences of the function
(\ref{Pi1looptotal}) in the case $D=4$. From the asymptotic behaviour
$K_\nu(x)\simeq2^{\nu-1}\,\Gamma(\nu)\,x^{-\nu}+\dots$ for $x\to0$ and
$\nu\neq0$, the expansion of (\ref{Pi1loopexpl}) in powers of
$\frac1{\Lambda^2}$ produces the leading singular behaviour
\beq
\Pi_{\rm np}^{(1)}(p)=\frac1{96\pi^2}\,\left(\Lambda_{\rm eff}^2-m^2
\ln\frac{\Lambda_{\rm eff}^2}{m^2}\right)+O(1) \ ,
\label{Pi1loopexpand}\eeq
where the effective ultraviolet cutoff is given by
\beq
\Lambda_{\rm eff}^2=\frac1{\frac1{\Lambda^2}+p\bullet p} \ .
\label{Lambdaeff}\eeq
Note that in the limit $\Lambda\to\infty$, the non-planar one-loop graph
(\ref{Pi1loopexpand}) remains finite, being effectively regulated by the
noncommutativity of spacetime, i.e. $\Lambda_{\rm eff}^2\to\frac1{p\bullet p}$
for $\Lambda\to\infty$. However, the ultraviolet divergence is restored in
either the commutative limit $\theta\to0$ or the infrared limit $p\to0$. In the
zero momentum limit $p\to0$, we have $\Lambda_{\rm eff}\simeq\Lambda$, and we
recover the standard mass renormalization of $\phi^4$ theory in four
dimensions,
\beq
m_{\rm ren}^2=m^2+\frac1{32}\,\frac{g^2\Lambda^2}{\pi^2}-\frac1{32}\,
\frac{g^2m^2}{\pi^2}\,\ln\frac{\Lambda^2}{m^2}+O(g^4) \ ,
\label{massren}\eeq
which diverges as $\Lambda\to\infty$. On the other hand, in the ultraviolet
limit $\Lambda\to\infty$, we have $\Lambda_{\rm eff}^2\simeq\frac1{p\bullet
p}$, and the corrected propagator assumes a complicated, non-local form that
cannot be attributed to any (mass) renormalization. Notice, in particular, that
the renormalized propagator contains both a zero momentum pole and a
logarithmic singularity $\ln p\bullet p$. From this analysis we conclude that
the limit $\Lambda\to\infty$ and the low momentum limit $p\to0$ do not commute,
and noncommutative quantum field theory exhibits an intriguing mixing of the
ultraviolet ($\Lambda\to\infty$) and infrared ($p\to0$) regimes. The
noncommutativity leads to unfamiliar effects of the ultraviolet modes on the
infrared behaviour which have no analogs in conventional quantum field theory.

This UV/IR mixing is one of the most fascinating aspects of noncommutative
quantum field theory. To recapitulate, we have seen that a divergent diagram in
the $\theta=0$ theory is typically regulated by the noncommutativity at
$\theta\neq0$ which renders it finite, but as $p\to0$ the phases become
ineffective and the diagram diverges at vanishing momentum. The pole at $p=0$
that arises in the propagator for the $\phi$ field comes from the {\it high}
momentum region of integration (i.e. $\Lambda\to\infty$), and it is thereby a
consequence of very high energy dynamics. This contribution to the self-energy
has a huge effect on the propagation of long-wavelength particles. In position
space, it leads to long-ranged correlations, since the correlation functions of
the noncommutative field theory will decay algebraically for small
$g$~\cite{mvrs}, in contrast to normal correlation functions which decay
exponentially for $m\neq0$. Indeed, it is rather surprising to have found
infrared divergences in a massive field theory. Roughly speaking, when a
particle of momentum $p_j$ circulates in a loop of a Feynman graph, it can
induce an effect at distance $|\theta^{ij}p_j|$, and so the high momentum end
of Feynman integrals give rise to power law long-range forces which are
entirely absent in the classical field theory. We may conclude from the
analysis of this subsection that noncommutative quantum field theory below the
noncommutativity scale is nothing like conventional, commutative quantum field
theory.

The strange mixing of ultraviolet and infrared effects in noncommutative field
theory can be understood heuristically by going back to the quantum mechanical
example of section 1.3. Indeed, the field quanta in the present field theory
can be thought of as pairs of opposite charges, i.e. electron-hole bound
states, moving in a strong magnetic field~\cite{bigatti,mms-j1}. Recall from
section 1.3 that in this limit the position and momentum coordinates of such a
charge are related by $x^i=\theta^{ij}p_j$, with
$\theta^{ij}=B^{-1}\,\epsilon^{ij}$. Thus a particle with momentum $p$ along,
say, the $x^1$-axis will have a spatial extension of size $|\theta p|$ in the
$x^2$-direction, and the size of the particle grows with its momentum. In other
words, the low-energy spectrum of a noncommutative field theory includes, in
addition to the usual point-like, particle degrees of freedom, electric
dipole-like excitations. More generally, this can be understood by combining
the
induced spacetime uncertainty relation (\ref{spacetimeuncert}) that arises in
the noncommutative theory with the standard Heisenberg uncertainty relation.
The resulting uncertainties then coincide with the string-modified uncertainty
relations (\ref{strHeisen}). Therefore, this UV/IR mixing phenomenon may be
regarded as another stringy aspect of noncommutative quantum field theory. It
can also be understood in terms of noncommutative Gaussian
wavepackets~\cite{gms,mvrs}.

\subsubsection{String Theoretical Interpretation}

\begin{figure}[htb]
\epsfxsize=5in
\bigskip
\centerline{\epsffile{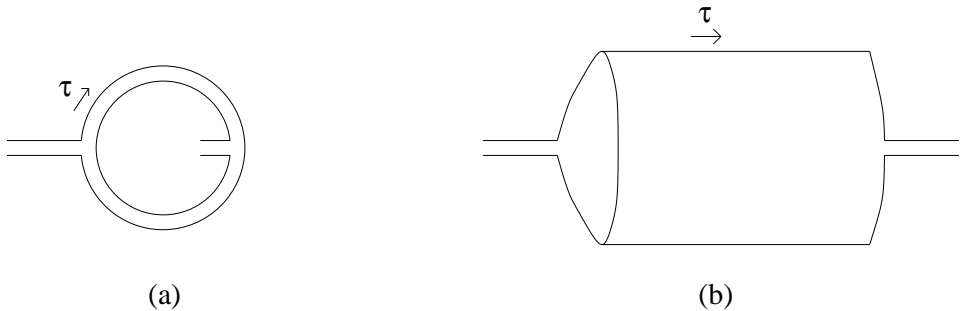}}
\caption{\baselineskip=12pt {\it The double twist diagram in} (a) {\it the open
string channel and} (b) {\it the closed string channel.}}
\bigskip
\label{twist}\end{figure}

As we have alluded to above, the unusual properties of noncommutative quantum
field theories are not due to inconsistencies in their definitions, but rather
unexpected consequences of the non-locality of the star-product interaction
which gives the field theory a stringy nature and is therefore well-suited to
be an effective theory of strings. The UV/IR mixing has a more precise analog
in string theory in the context of a particular open string amplitude known as
the double twist diagram~\cite{mvrs}. This non-planar, non-orientable diagram
is depicted in the open string channel in fig.~\ref{twist}(a). Note that
symbolically it coincides with the ribbon graph for the one-loop non-planar
mass renormalization in noncommutative $\phi^3$ theory. By applying the modular
transformation $\tau\mapsto-1/\tau$ to the Teichm\"uller parameter of the
annular one-loop open string diagram, it gets transformed into the cylindrical
closed string diagram of fig.~\ref{twist}(b). The latter amplitude behaves like
$1/p_i\,g^{ij}\,p_j$ for small momenta~\cite{mvrs}. In string perturbation
theory, one integrates over the moduli of string diagrams, and the region of
moduli space corresponding to high energies in the open string loop describes
the tree-level exchange of a light closed string state. Therefore, an
ultraviolet phenomenon in the open string channel corresponds to an infrared
singularity in the closed string channel. This is precisely the same behaviour
that was observed at the field theoretical level above, if we identify the
closed string metric with the noncommutativity parameter through
$g^{ij}\sim-(\theta^2)^{ij}$. In the correlated decoupling limit $\alpha'\to0$
described in section~1.3, this is exactly what is found from
(\ref{openstringmetric}) when the open string metric is taken to be
$G^{ij}=\delta^{ij}$, as it is in the present case. Thus the exotic properties
unveiled above may indeed be attributed to stringy behaviours of noncommutative
quantum field theories.

The occurence of infrared singularities in massive field theories suggests the
presence of new light degrees of freedom~\cite{mvrs,vrs}. From our analysis of
the one-loop renormalization of the scalar propagator, we have seen that, in
addition to the original pole at $p^2=-m^2$, there is a pole at $p^2=O(g^2)$
which arises from the high loop momentum modes of the scalar field $\phi$. In
order to write down a Wilsonian effective action which correctly describes the
low momentum behaviour of the theory, it is necessary to add new light fields
to the action. For instance, the quadratic infrared singularity obtained above
can be reproduced by a Feynman diagram in which $\phi$ turns into a new field
$\chi$ and then back into $\phi$, where the field $\chi$ couples to $\phi$
through an action of the form
\beq
S_\chi=\int d^Dx\left[g\,\chi(x)\,
\phi(x)+\frac12\,\partial\chi(x)\bullet\partial\chi(x)+
\frac{\Lambda^2}2\,\Bigl(\partial\bullet\partial\chi(x)\Bigr)^2\right] \ .
\label{chiaction}\eeq
This process is completely analogous to the string channel duality discussed
above, with the field $\phi$ identified with the open string modes and $\chi$
with the closed string mode. Other stringy aspects of UV/IR mixing can be
observed by studying the noncommutative quantum field theory at finite
temperature~\cite{finiteT}. Then, at the level of non-planar graphs, one finds
stringy winding modes corresponding to states which wrap around the compact
thermal direction. This gives an alternative picture to the field theoretical
analog of the open-closed string channel duality discussed in this section.
Perturbative string calculations also confirm explicitly the UV/IR
mixing~\cite{stringUVIR}. A similar analysis can be done for the linear and
logarithmic infrared singularities~\cite{mvrs}, and also for the corrections to
vertex functions~\cite{mvrs,adbs}. At higher loop orders, however, the momentum
dependences become increasingly complicated and are far more difficult to
interpret~\cite{vrs}. Other aspects of this phenomenon may be found
in~\cite{otherUVIR}. Even field theories which do not exhibit the UV/IR mixing
phenomenon, such as the noncommutative Wess-Zumino model~\cite{GGRdaS}, show
exotic effects like the dipole picture~\cite{GGRdaSlow}. The perturbative
properties of the corresponding supersymmetric model are studied
in~\cite{BGPR}.

In Minkowski spacetime with noncommuting time direction, i.e.
$\theta^{0i}\neq0$, one encounters severe acausal effects, such as events which
precede their causes and objects which grow instead of Lorentz contract as they
are boosted~\cite{acausal}. Such a quantum field theory is neither causal nor
unitary in certain instances~\cite{nonunitary}. In a theory with space-like
noncommutativity, one can perform a boost and induce a time-like component for
$\theta$. The resulting theory is still unitary~\cite{CaiOhta}. The Lorentz
invariant condition for unitarity is $p\bullet p>0$, which has two solutions
corresponding to space-like and light-like noncommutativity. For space-like
$\theta$ one can always boost to a frame in which $\theta^{0i}=0$. However, for
light-like noncommutativity, one cannot eliminate $\theta^{0i}$ by any finite
boost.

In string theory with a background electric field,
however, stringy effects conspire to cancel such acausal effects~\cite{ncos}.
There is no low-energy limit in this case in which both $\theta^{ij}$ and
$G^{ij}$ can be kept fixed when $\alpha'\to0$, because, unlike the case of
magnetic fields, electric fields in string theory have a limiting critical
value above which the vacuum becomes unstable~\cite{burgess}, and one cannot
take the external field to be arbitrarily large. There is no low-energy limit
in which one is left only with a noncommutative field theory. Instead, such a
theory of open strings should be considered in a somewhat different decoupling
limit whose effective theory is not a noncommutative field theory but rather a
theory of open strings in noncommutative spacetime~\cite{ncos}. The closed
string dynamics are still decoupled from the open string sector, so that the
theory represents a new sort of non-critical string theory which does not
require closed strings for its consistency. The effective string scale of this
theory is of the order of the noncommutativity scale, so that stringy effects
do not decouple from noncommutative effects and an open string theory emerges,
rather than a field theory. This new model is known as {\it noncommutative open
string theory}~\cite{ncos}. Other such open string theories have been found
in~\cite{OhtaTomino}. One can also get a light-like noncommutative quantum
field theory from a consistent field theory limit of string theory in the
presence of electromagnetic fields satisfying $E^2=B^2=0$ and $E\cdot
B=0$~\cite{AGM}.

\newsection{Noncommutative Yang-Mills Theory}

Having now become acquainted with some of the generic properties of
noncommutative quantum field theory, we shall focus most of our attention in
the remainder of this paper to gauge theories on a noncommutative space, which
are the relevant field theories for the low-energy dynamics of open strings in
background supergravity fields and on D-branes~\cite{cds,sw}. The Weyl
quantization procedure of section 2 generalizes straightforwardly to the
algebra of $N\times N$ matrix-valued functions on $\real^D$. The star-product
then becomes the tensor product of matrix multiplication with the
Groenewold-Moyal product (\ref{starproddef}) of functions. This extended
star-product is still associative. We can therefore use this method to
systematically construct noncommutative gauge theories on
$\real^D$~\cite{mssw}.

\subsection{Star-Gauge Symmetry}

Let $A_i(x)$ be a Hermitian $U(N)$ gauge field on $\real^D$ which may be
expanded in terms of the Lie algebra generators $t_a$ of $U(N)$ as
$A_i=A_i^a\,t_a$, with $\tr^{~}_N(t_a\,t_b)=\delta_{ab}$, $a,b=1,\dots,N^2$,
and $[t_a,t_b]=if_{ab}^{~~c}\,t_c$. Here the $t_a$ live in the
fundamental representation of the $U(N)$ gauge group and $\tr^{~}_N$
denotes the ordinary matrix trace. In fact, many of the expressions in
the following do not close in the $U(N)$ Lie algebra, as they will
involve products rather than commutators of the generators. We
introduce a Hermitian Weyl operator corresponding to $A_i(x)$ by
\beq
\weyl[A]_i=\int d^Dx~\hat\Delta(x)\otimes A_i(x) \ ,
\label{Weylgaugeop}\eeq
where $\hat\Delta(x)$ is the map (\ref{Deltadef}) and the tensor product
between the coordinate and matrix representations is written explicitly for
emphasis. We may then write down the appropriate noncommutative version of the
Yang-Mills action as
\beq
S_{\rm YM}=-\frac1{4g^2}\,\Tr\otimes\tr^{~}_N\left(\left[\hat\partial_i\,,\,
\weyl[A]_j\right]-\left[\hat\partial_j\,,\,\weyl[A]_i\right]-i
\left[\weyl[A]_i\,,\,\weyl[A]_j\right]\right)^2
\label{WeylYMaction}\eeq
where Tr is the operator trace (\ref{Tracedef}) over the spacetime coordinate
indices. Using (\ref{Weylgaugeop}), (\ref{partialWeyl}), (\ref{Deltaortho}) and
(\ref{Weylstar}), the action (\ref{WeylYMaction}) can be written as
\beq
S_{\rm YM}=-\frac1{4g^2}\,\int d^Dx~\tr^{~}_N\Bigl(F_{ij}(x)\star
F^{ij}(x)\Bigr)
\ ,
\label{starYMaction}\eeq
where
\bea
F_{ij}&=&\partial_iA_j-\partial_jA_i-i\left(A_i\star A_j-A_j\star A_i\right)
\nn\\&=&\partial_iA_j-\partial_jA_i-i\left[A_i,A_j\right]+\frac12\,
\theta^{kl}\,\Bigl(\partial_kA_i\,\partial_lA_j-\partial_kA_j\,
\partial_lA_i\Bigr)+O(\theta^2)
\label{starfieldstrength}\eea
is the noncommutative field strength of the gauge field $A_i(x)$. Thus the
gauge field belongs to the tensor product of the Groenewold-Moyal deformed
algebra of functions on $\real^D$ with the algebra of $N\times N$ matrices.
Note that the action (\ref{starYMaction}) defines a non-trivial interacting
theory even for the simplest case of rank $N=1$, which for $\theta=0$ is just
pure electrodynamics.

Let us consider the symmetries of the action (\ref{WeylYMaction}). It is
straightforward to see that it is invariant under any inhomogeneous
transformation of the form
\beq
\weyl[A]_i~\longmapsto~\weyl[g]\,\weyl[A]_i\,\weyl[g]^\dagger-i\,\weyl[g]
\left[\hat\partial_i\,,\,\weyl[g]^\dagger\right] \ ,
\label{Weylgaugetr}\eeq
with $\weyl[g]$ an arbitrary unitary element of the unital $C^*$-algebra of
matrix-valued Weyl operators,\footnote{\baselineskip=12pt Actually, this
algebra does not contain an identity element because we are restricting to the
space of Schwartz fields. It can, however, be easily extended to a unital
algebra. We will elaborate on this point in section~8.} i.e.
\beq
\weyl[g]\,\weyl[g]^\dagger=\weyl[g]^\dagger\,\weyl[g]=\hat\id\otimes\id_N \ ,
\label{Weylunitary}\eeq
where $\hat\id$ is the identity on the ordinary Weyl operator algebra and
$\id_N$ is the $N\times N$ unit matrix. Given the one-to-one correspondence
between Weyl operators and fields, we may expand the unitary operator
$\weyl[g]$ in terms of an $N\times N$ matrix field $g(x)$ on $\real^D$ as
\beq
\weyl[g]=\int d^Dx~\hat\Delta(x)\otimes g(x) \ .
\label{Weylgexp}\eeq
The unitarity condition (\ref{Weylunitary}) is then equivalent to
\beq
g(x)\star g(x)^\dagger=g(x)^\dagger\star g(x)=\id_N \ .
\label{starunitary}\eeq
In this case we say that the matrix field $g(x)$ is {\it star-unitary}. Note
that (\ref{starunitary}) implies that the adjoint $g^\dagger$ of $g$ is equal
to the inverse of $g$ with respect to the {\it star-product} on the deformed
algebra of functions on spacetime, but for $\theta\neq0$ we generally have that
$g^\dagger\neq g^{-1}$. In other words, generally
$\weyl[g^{-1}]\neq\weyl[g]^{-1}$. The explicit relationship between $g^\dagger$
and $g^{-1}$ can worked out order by order in $\theta$ by using the infinite
series representation of the star-product in (\ref{starproddef}). To leading
orders we have (for $\theta$ invertible)
\beq
g^\dagger=g^{-1}+\frac i2\,\theta^{ij}\,g^{-1}\,\Bigl(\partial_i\,g\Bigr)
\,g^{-1}\,\Bigl(\partial_j\,g\Bigr)\,g^{-1}+O(\theta^2) \ .
\label{gdagginv}\eeq

{}From the Weyl-Wigner correspondence it follows that the function $g(x)$
parametrizes the local {\it star-gauge transformation}
\beq
A_i(x)~\longmapsto~g(x)\star A_i(x)\star g(x)^\dagger-i\,g(x)\star
\partial_i\,g(x)^\dagger \ .
\label{stargaugetr}\eeq
The invariance of the noncommutative Yang-Mills gauge theory action
(\ref{starYMaction}) under (\ref{stargaugetr}) follows from the cyclicity of
both the operator and matrix traces, and the corresponding covariant
transformation rule for the noncommutative field strength,
\beq
F_{ij}(x)~\longmapsto~g(x)\star F_{ij}(x)\star g(x)^\dagger \ .
\label{Fstargaugetr}\eeq
The noncommutative gauge theory obtained in this way reduces to conventional
$U(N)$ Yang-Mills theory in the commutative limit $\theta=0$.

However, because of the way that the theory is constructed above from
associative algebras, there is no direct way to get other gauge
groups~\cite{mssw,matsubara}. The important point here is that
expressions in noncommutative gauge theory in general involve the
enveloping algebra of the underlying Lie group. Because of
the property
\beq
(g\star h)^\dagger=h^\dagger\star g^\dagger \ ,
\label{Moyaldagger}\eeq
the Groenewold-Moyal product $g\star h$ of two unitary matrix fields is always
unitary and the group $U(N)$ (in the fundamental representation) is
closed under the star-product. However, the
special unitary group $SU(N)$ does not give rise to any gauge group on
noncommutative $\real^D$, because in general $\det(g\star
h)\neq\det(g)\star\det(h)$. In contrast to the commutative case, the $U(1)$ and
$SU(N)$ sectors of the decomposition
\beq
U(N)=U(1)\times SU(N)/\zed_N
\label{UNdecomp}\eeq
do not decouple because the $U(1)$ ``photon'' interacts with the $SU(N)$
gluons~\cite{armoni}. Physically, this $U(1)$ corresponds to the center of mass
coordinate of a system of $N$ D-branes and it represents the interactions of
the short open string excitations on the D-branes with the bulk supergravity
fields. In the case of a vanishing background $B$-field, the closed and open
string dynamics decouple and one is effectively left with an $SU(N)$ gauge
theory, but this is no longer true when $B\neq0$. It has been argued, however,
that one can still define orthogonal and symplectic star-gauge groups by using
anti-linear anti-unitary automorphisms of the Weyl operator
algebra~\cite{sospstar}. We shall see in section~8 that these automorphisms are
related to some standard operators in noncommutative geometry which can be
thought of as generating charge conjugation symmetries of the field theory.
Physically, these cases correspond to the stability of orientifold
constructions with background $B$-fields and D$p$-branes~\cite{sospstar}.
Notice also that, in contrast to the case of noncommutative scalar field
theory, the corresponding quantum measure for path integration is {\it not}
simply the ordinary gauge-fixed Feynman measure for the $U(N)$ gauge field
$A_i(x)$, because it must be defined by gauge-fixing the star-unitary gauge
group, i.e. the group of unitary elements of the matrix-valued Weyl operator
algebra. We shall return to this point in section~4.3. The noncommutative gauge
symmetry group will be described in some detail in section~8.

\subsection{Noncommutative Wilson Lines}

We now turn to a description of star-gauge invariant observables in
noncommutative Yang-Mills theory~\cite{iikk,amns1,amns2}. Let $C_v$ be an
arbitrary oriented smooth contour in spacetime $\real^D$. The line $C_v$ is
parametrized by the smooth embedding functions $\xi( t):[0,1]\to\real^D$ with
endpoints $\xi(0)=0$ and $\xi(1)=v$ in $\real^D$. The holonomy of a
noncommutative gauge field over such a contour is described by the
noncommutative parallel transport operator
\bea
{\cal U}(x;C_v)&=&{\rm P}~\exp_\star\,i\int\limits_{C_v}d\xi^i~A_i(x+\xi)
\nn\\&=&1+\sum_{n=1}^\infty i^n\int\limits_0^1d t_1~\int
\limits_{ t_1}^1d t_2~\cdots\int\limits_{ t_{n-1}}^1d t_n~
\frac{d\xi^{i_1}( t_1)}{d t_1}\,\cdots\,\frac{d\xi^{i_n}
( t_n)}{d t_n}\nn\\& &\times\,A_{i_1}\Bigl(x+\xi( t_1)\Bigr)
\star\cdots\star A_{i_n}\Bigl(x+\xi( t_n)\Bigr) \ ,
\label{partransport}\eea
where P denotes path ordering and we have used the extended star-product
(\ref{starprodext}). The operator (\ref{partransport}) is an $N\times N$
star-unitary matrix field depending on the line $C_v$. Under the star-gauge
transformation (\ref{stargaugetr}), it transforms as
\beq
{\cal U}(x;C_v)~\longmapsto~g(x)\star{\cal U}(x;C_v)\star g(x+v)^\dagger \ .
\label{calUgaugetr}\eeq
The noncommutative holonomy can be alternatively represented
as~\cite{alekbytsko}
\beq
{\cal U}(x;C_v)={\cal G}(x)\star{\cal G}(x+v)^\dagger \ ,
\label{calUcalG}\eeq
where ${\cal G}(x)$ is a solution of the noncommutative parallel transport
equation
\beq
\partial_i\,{\cal G}(x)=i\,A_i(x)\star{\cal G}(x)
\label{paralleleqn}\eeq
which in general depends on the choice of integration path.

Observables of noncommutative gauge theory must be star-gauge invariant. Using
the holonomy operators (\ref{partransport}) and assuming that $\theta$ is
invertible, it is straightforward to associate a star-gauge invariant
observable to {\it every} contour $C_v$ by~\cite{iikk,amns1,amns2}
\beq
{\cal O}(C_v)=\int d^Dx~\tr^{~}_N\Bigl({\cal
U}(x;C_v)\Bigr)\star\e^{ik_i(v)x^i}
\label{stargaugeobs}\eeq
where the line parameter
\beq
k_i(v)=\left(\theta^{-1}\right)_{ij}\,v^j
\label{linemomentum}\eeq
can be thought of as the total momentum of $C_v$. The star-gauge invariance of
(\ref{stargaugeobs}) follows from the fact that the plane wave
$\e^{ik_i(v)x^i}$ for any $v\in\real^D$ is the unique function with the
property that
\beq
\e^{ik_i(v)x^i}\star g(x)\star\e^{-ik_i(v)x^i}=g(x+v)
\label{startransl}\eeq
for arbitrary functions $g(x)$ on $\real^D$. Using (\ref{calUgaugetr}),
(\ref{startransl}), and the cyclicity of the traces Tr and $\tr^{~}_N$, the
star-gauge invariance of the operator (\ref{stargaugeobs}) follows.

To establish the property (\ref{startransl}), via Fourier transformation it
suffices to prove it for arbitrary plane waves $\e^{ip_ix^i}$. Then, using the
coordinate space representation of the Baker-Campbell-Hausdorff formula
(\ref{BCH}) and the star-unitarity of any plane wave, we have the identity
\beq
\e^{ik_ix^i}\star\e^{ip_ix^i}\star\e^{-ik_ix^i}=\e^{ik_ix^i}\star
\e^{-ik_ix^i}\star\e^{ip_ix^i}~\e^{ip_i\theta^{ij}k_j}=\e^{ip_i(x^i+
\theta^{ij}k_j)} \ ,
\label{startranslproof}\eeq
from which (\ref{startransl},\ref{linemomentum}) follows. This means that, in
noncommutative gauge theory, the spacetime translation group is a subgroup of
the star-gauge group. In fact, the same is true of the rotation group of
$\real^D$ (c.f. (\ref{xderiv}))~\cite{lsz}. The fact that the Euclidean group
is contained in the star-gauge symmetry implies that the local dynamics of
gauge invariant observables is far more restricted in noncommutative Yang-Mills
theory as compared to the commutative case. We shall describe such spacetime
symmetries in more detail in section~8.

The most striking fact about the construction (\ref{stargaugeobs}) is that in
the noncommutative case there are gauge invariant observables associated with
{\it open} contours $C_v$, in contrast to the commutative case where only
closed loops $C_0$ would be allowed. The translational symmetry generated by
the star-product leads to a larger class of observables in noncommutative gauge
theory. Let us make a few further remarks concerning the above construction:
\begin{itemize}
\item{For an open line $C_v$ with relative separation vector $v$ between its
two endpoints, the parameter (\ref{linemomentum}) has a natural interpretation
as its total momentum (by the Fourier form of the integral
(\ref{stargaugeobs})). It follows that the longer the curve is, the larger its
momentum is. This is simply the characteristic UV/IR mixing phenomenon that we
encountered in the previous section. If one increases the momentum $k_j$ in a
given direction, then the contour will extend in the other spacetime directions
proportionally to $\theta^{ij}k_j$. In the electric dipole interpretation of
section~3.3, the relationship (\ref{linemomentum},\ref{startransl}) follows if
we demand that the dipole quanta of the field theory interact by joining at
their ends. We will see some more manifestations of this exotic property later
on.}
\item{In the commutative limit $\theta=0$ we have $v=0$, which is the
well-known property that there are no gauge-invariant quantities associated
with open lines in ordinary Yang-Mills theory.}
\item{When $\theta=0$, the quantity (\ref{stargaugeobs}) can be defined for
closed contours by replacing the plane wave $\e^{ik_i(v)x^i}$ by an arbitrary
function $f(x)$, since in that case the total momentum of a closed loop is
unrestricted. In particular, we can take $f(x)$ to be delta-function supported
about some fixed spacetime point and recover the standard gauge-invariant
Wilson loops of Yang-Mills theory. However, for $\theta\neq0$, closed loops
have vanishing momentum, and only the unit function $\e^{ik_i(v)x^i}=1$ is
permitted in (\ref{stargaugeobs}). Thus, although there is a larger class of
observables in noncommutative Yang-Mills theory, the dynamics of closed Wilson
loops is severely restricted as compared to the commutative case. Indeed, the
requirement of star-gauge invariance is an extremely stringent restriction on
the quantum field theory. It means that there is no local star-gauge invariant
dynamics, because everything must be smeared out by the Weyl operator trace Tr.
The fact that there are no local operators such as the gluon operator
$\tr^{~}_NF_{ij}(x)^2$ suggests that the gauge dynamics below the
noncommutativity scale can be quite different from the commutative case. This
is evident in the dual supergravity computations of noncommutative Wilson
loops~\cite{wilsonsugra}, which show that while the standard area law behaviour
may be observed at very large distance scales, below the noncommutativity scale
it breaks down and is replaced by some unconventional behaviour. This makes it
unclear how to interpret quantities such as a static quark potential in
noncommutative gauge theory.}
\item{The gauge-invariant Wilson line operators have been shown to constitute
an overcomplete set of observables for noncommutative gauge
theory~\cite{amns2}, just like in the commutative case. This is due to the fact
that fluctuations in the shape of $C_v$ leave the corresponding holonomy
invariant. They may be used to construct gauge invariant operators which carry
definite momentum and which reduce to the usual local gauge invariant operators
of ordinary gauge field theory in the commutative limit as
follows~\cite{ghi,dharwadia}. For this, we let
\beq
C_k^{(0)}~:~\xi^j(t)=k_i\,\theta^{ij}\,t~~~~~~,~~~~~~0\leq t\leq1
\label{straightline}\eeq
be the straight line path from the origin to the point $v^j=k_i\,\theta^{ij}$,
and let ${\cal O}(x)$ be any local operator of {\it ordinary} Yang-Mills theory
which transforms in the adjoint representation of the gauge group. Then a
natural star-gauge invariant operator is obtained by attaching the operator
${\cal O}(x)$ at one end of a Wilson line of non-vanishing momentum,
\beq
\tilde{\cal O}(k)=\tr_N^{~}\int d^Dx~{\cal O}(x)\star{\cal U}\left(x;C_k^{(0)}
\right)\star\e^{ik_ix^i} \ .
\label{calOk}\eeq
The collection of operators of the form (\ref{calOk}) generate a convenient set
of gauge-invariant operators which are the natural generalizations of the
standard local gauge theory operators in the commutative limit. For small $k$
or $\|\theta\|$, the seperation $v$ of the open Wilson lines becomes small, and
(\ref{calOk}) reduces to the usual Yang-Mills operator in momentum space. In
this sense, it is possible to generate operators which are local in momentum in
noncommutative gauge theory.}
\item{Correlation functions of the operators (\ref{calOk}) exhibit many of the
stringy features of noncommutative gauge
theory~\cite{ghi,rvr}. They can also be used to construct the appropriate gauge
invariant operators that couple noncommutative gauge fields on a D-brane to
massless closed string modes in flat space~\cite{sugracoupling}, and thereby
yield explicit expressions for the gauge theory operators dual to bulk
supergravity fields in this case. We will return to this point in section~8.}
\end{itemize}

The observables (\ref{stargaugeobs}) may also be expressed straightforwardly in
terms of Weyl operators~\cite{amns1,amns2}, though we shall not do so here.
Here we will simply point out an elegant path integral representation of the
noncommutative holonomy operator (\ref{partransport}) in the case of a $U(1)$
gauge group~\cite{okuyama}. Let us introduce, as in the Kontsevich formula
(\ref{starpathrep}), auxilliary bosonic fields $\eta^i(t)$ which live on the
contour $C_v$ and which have the free propagator
\beq
\left\langle\eta^i(t)\,\eta^j(t')\right\rangle_\eta=\left[\left(
-i\,\theta^{-1}\otimes\partial_t\right)^{-1}\right]^{ij}(t,t')=
\frac i2\,\theta^{ij}~{\rm sgn}(t-t') \ .
\label{etaprop}\eeq
It is then straightforward to see that the parallel transport operator
(\ref{partransport}) can be expressed in terms of the path integral expectation
value
\bea
{\cal U}(x;C_v)&=&\left\langle\exp i\int\limits_{C_v}d\xi^i~A_i(x+\xi+\eta)
\right\rangle_\eta\nn\\&=&\int D\eta~\exp i\int\limits_{C_v}dt~
\left[\frac12\,\eta^i(t)\,\left(\theta^{-1}\right)_{ij}\,
\frac{d\eta^j(t)}{dt}
+A_i\Bigl(x+\xi(t)+\eta(t)\Bigr)\,\frac{d\xi^i(t)}{dt}\right] \ .\nn\\& &
\label{holonomypath}\eea
The equivalence between the two representations follows from expanding the
gauge field $A_i(x+\xi+\eta)$ as a formal power series in $\eta^i(t)$ and
applying Wick's theorem. Because of the $\theta$ dependence of the propagator
(\ref{etaprop}), the Wick contractions produce the appropriate series
representation of the extended star-product in (\ref{starprodext}), while the
${\rm sgn}(t-t')$ term produces the required path ordering operation P in the
Wick expansion. Again, the beauty of the formula (\ref{holonomypath}) is that
it uses ordinary products of fields and is therefore much more amenable to
practical, perturbative computations involving noncommutative Wilson lines.
Other descriptions of the noncommutative holonomy may be found
in~\cite{leereg,wilsonother}.

\subsection{One-Loop Renormalization}

In order to analyse the perturbative properties of noncommutative Yang-Mills
theory, one needs to first of all gauge-fix the star-gauge invariance of the
model. This can be done in a straightforward way by adapting the standard
Faddeev-Popov technique to the noncommutative
case~\cite{armoni,krajwulk,martinruiz}. The gauge fixed noncommutative
Yang-Mills action assumes the form
\bea
S_{\rm YM}&=&\int d^Dx~\tr^{~}_N\Biggl(-\frac1{4g^2}\,F_{ij}(x)\star F^{ij}(x)
+\frac2\xi\,\left(\partial^iA_i(x)\right)^2\Biggr.\nn\\& &-\Biggl.\,
2\,\bar c(x)\star\partial^i\,\nabla_i\,c(x)+2\,\partial^i\,\nabla_i\,c(x)\star
\bar c(x)\Biggr)
\label{YMgaugefix}\eea
where $c=c^a\,t_a$ and $\bar c=\bar c^a\,t_a$ are noncommutative fermionic
Faddeev-Popov ghost fields which transform in the adjoint representation of the
local star-gauge group,
\beq
c(x)~\longmapsto~g(x)\star c(x)\star g(x)^\dagger~~~~~~,~~~~~~
\bar c(x)~\longmapsto~g(x)\star\bar c(x)\star g(x)^\dagger \ .
\label{FPtransf}\eeq
The constant $\xi$ is the covariant gauge-fixing parameter, and $\nabla_i$
denotes the star-gauge covariant derivative which is defined by
\beq
\nabla_i\,c=\partial_i\,c-i\left(A_i\star c-c\star A_i\right) \ .
\label{covderiv}\eeq

Feynman rules for noncommutative Yang-Mills theory may now be written
down~\cite{sheikhren}. Because of the noncommutative interaction
vertices analogous to (\ref{phi4V}), the effective ``structure
constants'' of the star-gauge group will involve oscillatory functions
of the momenta of the lines.\footnote{\baselineskip=12pt Explicit
  presentations of the genuine structure constants of the
  noncommutative gauge symmetry group may be found in~\cite{lsz}.} For
instance, in the case of a $U(1)$ gauge group, the Feynman rules are
easily read off from those of ordinary non-abelian gauge theory by
simply replacing Lie algebra structure constants $f_{ab}^{~~c}$ with
the momentum dependent functions
\beq
f_{kp}^{~~q}=2\sin(k\wedge p)\,(2\pi)^D\,\delta^D(k+p+q) \ ,
\label{fkpqdef}\eeq
and sums over Lie algebraic indices by integrations over momenta. In the
generic case of rank $N\geq1$, the only subtleties which arise are that the
Feynman rules will involve not only commutators but also anticommutators of the
$U(N)$ generators $t_a$. They will therefore depend on both the antisymmetric
structure constants $f_{ab}^{~~c}$ and the symmetric tensors $d_{ab}^{~~c}$,
where
\beq
t_a\,t_b=\frac i2\,f_{ab}^{~~c}\,t_c+\frac12\,d_{ab}^{~~c}\,t_c \ .
\label{fddef}\eeq
In the Feynman-'t~Hooft gauge $\xi=1$, the gluon propagator is
$-(i/p^2)\,G_{ij}\,\delta^{ab}$ and the ghost propagator is
$-(i/p^2)\,\delta^{ab}$. The three-gluon vertex is given by
\bea
V_{(3)}(k,p,q)^{~~c}_{ab~ijl}&=&-2g\left[\left(d_{ab}^{~~c}
\sin k\wedge p-if_{ab}^{~~c}
\cos k\wedge p\right)\Bigl(k-q\Bigr)_l\,G_{ij}\right.\nn\\& &+\Bigl.
\,{\rm permutations}\Bigr]\,(2\pi)^D\,\delta^D(k+p+q) \ ,
\label{3gluon}\eea
the four-gluon vertex by
\bea
& &V_{(4)}(k,p,q,r)_{abe~ijlm}^{~~~g}\nn\\& &~~~~~~=
-4ig^2\left[\left(d_{ab}^{~~c}\sin k\wedge p-i
f_{ab}^{~~c}\cos k\wedge p\right)\Bigl(d_{ce}^{~~g}\sin q\wedge r-if_{ce}^{~~g}
\cos q\wedge r\Bigr)\right.\nn\\& &~~~~~~~~~~\times\left.\Bigl(G_{il}\,G_{jm}
-G_{im}\,G_{jl}\Bigr)+\,{\rm permutations}\right]
\,(2\pi)^D\,\delta^D(k+p+q+r) \ , \nn\\& &
\label{4gluon}\eea
and the ghost-ghost-gluon vertex by
\beq
V_{\rm gh}(k,p,q)_{ab~i}^{~~c}=-2g\,k_i\left(d_{ab}^{~~c}
\sin k\wedge p-if_{ab}^{~~c}\cos
k\wedge p\right)\,(2\pi)^D\,\delta^D(k+p+q) \ .
\label{ghostgluon}\eeq
The Feynman rules for the $U(1)$ case follow from substituting $f_{ab}^{~~c}=0$
and $d_{ab}^{~~c}=1$ in the above.

With these rules, it is relatively straightforward to do perturbative
calculations in noncommutative Yang-Mills theory in parallel to the commutative
case. Essentially the only tricks involved are the usages of various
trigonometric identities to simplify the momentum integrations over the
oscillatory functions involved. We shall not go into any details here, but
simply quote a few of the many interesting results that have been obtained.
First, let us consider the one-loop renormalization of the gluon
propagator~\cite{mvrs}. Since star-products and matrix products always appear
together, the notion of planarity in the sense of the large $N$ expansion is
the same as that for the noncommutative interactions which was discussed in
section 3. Therefore, the large $\theta$ genus expansion of the theory will
produce a similar sort of string expansion as in ordinary large $N$ gauge
theory. Moreover, non-planar one-loop $U(N)$ diagrams will contribute only to
the $U(1)$ part of the theory. Using the noncommutative version of the standard
background field gauge, the divergent part of the one-loop effective action to
quadratic order in momentum space for $D=4$ is found to be~\cite{mvrs,mst,MvR}
\bea
\Gamma^{(1)}_{\rm eff}[A]&=&-\frac14\,\int\frac{d^4k}{(2\pi)^4}
{}~\left[\left(\frac1{g^2}-\frac{11N^2}{24\pi^2}\,\ln\frac{\Lambda^2}{k^2}
\right)\tr^{~}_N\Bigl(\widetilde{\partial A}\Bigr)^2(k)\right.
\nn\\& &+\,\frac{4N}{\pi^2}\,\frac{\theta^{ij}\,\theta^{kl}}{(k\bullet k)^2}\,
\left(\tr^{~}_N\left(\widetilde{\partial_iA_j}\right)\,
\tr^{~}_N\left(\widetilde{\partial_kA_l}\right)\right)(k)
\nn\\&&+\left.\,\frac{11N}{24\pi^2}\,\ln\left(\frac1{k^2\,(k\bullet k)}\right)
\Bigl(\tr^{~}_N\widetilde{\partial A}\Bigr)^2(k)\right] \ .
\label{Gammaquad}\eea
After renormalization, the one-loop Gell~Mann-Low beta-function may be computed
from (\ref{Gammaquad}) as~\cite{martinruiz,hayakawa}
\beq
\beta(g^2)=\frac{\partial\,g^2}
{\partial\ln\Lambda}=-\frac{22}3\,\frac{g^4N^2}{8\pi^2} \ .
\label{betafn}\eeq
Note that this formula holds even for $N=1$, and it follows that noncommutative
$U(1)$ gauge theory is {\it asymptotically free}. The effective coupling
constant grows at large distance scales and leads to interesting strong
coupling effects. This is also true of course for all $N$. In fact, apart from
the non-planar terms which are generically finite at $\theta\neq0$, the
effective action (\ref{Gammaquad}) is the same as that of ordinary commutative
$SU(N)$
Yang-Mills theory in the large $N$ limit (in which only planar 't~Hooft
diagrams survive). Therefore, the perturbative beta-function for noncommutative
$U(N)$ Yang-Mills theory for any $N$ can be simply found from that of the
ordinary $SU(N)$ theory. This remarkable coincidence will be explained in
section 6 when we discuss Morita equivalence of noncommutative gauge theories.
Note that, in contrast to the ordinary commutative case whereby the dynamics in
the center $U(1)$ of the $U(N)$ gauge group is always decoupled and free, in
the noncommutative case it runs with the same beta-function as the rest of the
$SU(N)$ gauge theory~\cite{armoni}.

The full, gauge-invariant noncommutative effective action for pure Yang-Mills
theory involves open Wilson lines~\cite{armonilopez}. Notice also that the
non-planar $U(1)$ part of the effective action (\ref{Gammaquad}) has a
logarithmic infrared singularity, similarly to the case of
noncommutative scalar field theory. Here, unlike the power-like UV/IR
mixing which seems to imply the alarming feature that perturbation
theory is no longer reliable, logarithmic UV/IR mixing may be put to
good use. This has been pointed out in~\cite{CKTUVIR} where it was
suggested that the $U(1)$ UV/IR mixed degrees of freedom of a $U(N)$
gauge field theory have a direct physical interpretation. There are
examples of supersymmetric theories in which they decouple from the
$SU(N)$ degrees of freedom, and eventually become weakly-coupled in
the infrared, playing the role of the hidden sector which breaks
supersymmetry. In this way the unfamiliar behaviour of the $U(1)$
running coupling constant in the extreme infrared is not interpreted
as an artifact of perturbation theory, but is instead turned into a
useful mechanism to break supersymmetry. Physical interpretations of
UV/IR mixing from the D-brane perspective may also be found
in~\cite{MvR,armonilopez}.

Unlike the standard high momentum divergences of ordinary quantum
field theories, which can be typically removed by a choice of
regularization scheme, here the noncommutative momenta play the role
of the regulators and lead to new infrared singularities which cannot
be straightforwardly removed. These effects can also be characterized as
non-analytic behaviour in the noncommutativity parameter $\theta$, so that the
noncommutative field theory does not recover ordinary field theory at the
quantum level in the limit $\theta\to0$. However, it is difficult to analyse
the renormalizability properties of noncommutative Yang-Mills theory along the
lines that we discussed in section 3. Part of the problem stems from the fact,
discussed in the previous subsection, that noncommutative gauge theories appear
to have no local gauge invariant operators, and so it is difficult to deduce
what (infrared) effects will be induced by the noncommutativity. Naively, one
would expect that the theory would have at worst logarithmic divergences
(unlike the scalar field theory studied in section 3 which also contained
quadratic divergences), but from (\ref{Gammaquad}) we see that both linear and
logarithmic infrared singularities arise~\cite{mst}. Because noncommutative
Yang-Mills theory already contains massless fields, it is difficult to
disentangle the usual infrared effects from the new ones induced by the
noncommutativity. It is not clear in this case what the new light degrees of
freedom look like. It has been shown, however, that noncommutative quantum
electrodynamics, i.e.
noncommutative $U(1)$ gauge theory minimally coupled (with respect to the
star-gauge invariance) to noncommutative fermion fields, is free from the
infrared poles in $\theta$ but still contains the anticipated logarithmic
non-analyticity~\cite{mst}. An exception is noncommutative supersymmetric
Yang-Mills theory with 16 supercharges in which UV/IR mixing appears to be
absent altogether~\cite{uvirsym}. A lot of effort has also been expelled into
analysing the ultraviolet structure of the quantum field theory, and it is
believed that in lower spacetime dimensions noncommutative Yang-Mills theory is
renormalizable in precisely the same way that its commutative counterpart
is~\cite{sheikhren,ncymren}. It also appears to be
gauge-invariant~\cite{uvirsym} and unitary~\cite{bgnv} in perturbation theory,
consistent with the fact that these models may be naturally embedded into
string theory. Other aspects of perturbative noncommutative gauge theories are
discussed in~\cite{1loopother}.

\newsection{Gauge Theory on the Noncommutative Torus}

The study of massless field theories on a torus is of great interest in the
noncommutative case because the compactness of the spacetime gives a natural
infrared regularization of the theory. One may therefore analyse more carefully
the ultraviolet behaviour and also the new light degrees of freedom which are
responsible for the UV/IR mixing. From a more mathematical point of view, the
noncommutative torus constitutes one of the original examples in noncommutative
geometry~\cite{connesbook} which captures the essential topological changes
which occur when one deforms a compact space. It is perhaps the most basic
example which still contains a rich geometrical structure. In this section we
shall describe some basic aspects of the noncommutative torus with particular
emphasis on the properties of vector bundles defined over them. From the study
of the global properties of gauge theories defined on this space, we will pave
the way for our discussion of Morita equivalence in the next section.

\subsection{The Noncommutative Torus}

Most of what we have said about noncommutative quantum field theory is true
when $\real^D$ is replaced by a $D$ dimensional torus ${\bf T}^D$, with only
subtle changes that we shall now explain. Let $\Sigma^i_{~a}$ be the $D\times
D$ period matrix of ${\bf T}^D$, which is a vielbein for its metric, i.e.
$\Sigma^i_{~a}\,\delta^{ab}\,\Sigma^j_{~b}=G^{ij}$. Here and in the following
the indices $i,j,\dots$ will label spacetime directions while $a,b,\dots$ will
denote indices in the frame bundle of ${\bf T}^D$. The matrices $\Sigma^i_{~a}$
parametrize the moduli of $D$ dimensional tori and they may be regarded as maps
from the frame bundle to the tangent bundle of ${\bf T}^D$. They define the
periods of the directions of ${\bf T}^D$,
\beq
x^i\sim x^i+\Sigma^i_{~a}~~~~~~,~~~~~~a=1,\dots,D \ ,
\label{xSigma}\eeq
for each $i=1,\dots,D$. When $\Sigma^i_{~a}$ is not proportional to
$\delta^i_{~a}$, the identifications (\ref{xSigma}) for $a\neq i$ describe how
the torus is tilted in its parallelogram representation.

Smooth functions on the torus must be single-valued, which implies that the
corresponding Fourier momenta $\vec k$ are quantized as
\beq
k_i=2\pi\left(\Sigma^{-1}\right)_i^{~a}\,m_a~~~~~~,~~~~~~m_a\in\zed \ .
\label{momquant}\eeq
Therefore, to describe the deformation of the function algebra, one cannot use
the unbounded operators $\hat x^i$ obeying (\ref{noncommalg}). Instead, one
must restrict to the proper subalgebra of the algebra of noncommutative
$\real^D$ that is generated by the {\it Weyl basis} of unitary operators
\beq
\hat Z^a=\e^{2\pi i\,(\Sigma^{-1})_i^{~a}\,\hat x^i}
\label{Weylbasis}\eeq
which generate the algebra
\beq
\hat Z^a\,\hat Z^b=\e^{-2\pi i\,\Theta^{ab}}\,\hat Z^b\,\hat Z^a \ ,
\label{noncommtorus}\eeq
where
\beq
\Theta^{ab}=2\pi\left(\Sigma^{-1}\right)_i^{~a}\,\theta^{ij}\,\left(\Sigma^{-1}
\right)_j^{~b}
\label{Thetadim}\eeq
are the corresponding dimensionless noncommutativity parameters. The
commutation relations (\ref{noncommtorus}) define the ``algebra of functions''
on the {\it noncommutative torus}. Formally, if ${\cal L}\cong\zed^D$ is the
lattice of rank $D$ (with bilinear form $G_{ij}$) which generates the torus as
the quotient space ${\bf T}^D=\real^D/{\cal L}$, then the projective regular
representations ${\cal L}_{\Theta}$ in (\ref{noncommtorus}) of the lattice
group ${\cal L}$ are labelled by an element $\Theta^{ab}$ of the second
Hochschild cohomology group $H^2({\cal L},U(1))$. This latter characterization
can be generalized to describe other sorts of noncommutative compactifications
of $\real^D$~\cite{howumatrix}.

Any function on ${\bf T}^D$ can be expanded as a Fourier series
\beq
f(x)=\sum_{\vec m\in\zeds^D}f_{\vec m}~\e^{2\pi i\,(\Sigma^{-1})_i^{~a}
\,m_ax^i} \ .
\label{Fourierseries}\eeq
The corresponding Weyl algebra is generated by the operators (\ref{Weylbasis})
and Weyl quantization takes the form of the map
\beq
\weyl[f]=\int d^Dx~f(x)\,\hat\Delta(x) \ ,
\label{weylftorus}\eeq
where the integration is taken over ${\bf T}^D$ and
\beq
\hat\Delta(x)=\frac1{|\det\Sigma|}\,\sum_{\vec m\in\zeds^D}\,\prod_{a=1}^D
\left(\hat Z^a\right)^{m_a}~\prod_{a<b}\e^{-\pi i\,m_a\,\Theta^{ab}
\,m_b}~\e^{-2\pi i\,(\Sigma^{-1})_i^{~a}\,m_ax^i}
\label{Deltatorus}\eeq
is a periodic field operator,
\beq
\hat\Delta(x+\Sigma^i_{~a}\,\hat\imath)=\hat\Delta(x)~~~~~~,~~~~~~
a=1,\dots,D \ ,
\label{Deltaperiodic}\eeq
with $\hat\imath$ a unit vector in the $i$-th direction of spacetime. Like on
$\real^D$, we may introduce anti-Hermitian, commuting linear derivations
$\hat\partial_i$ which on the noncommutative torus are defined by their actions
on the Weyl basis,
\beq
\left[\hat\partial_i\,,\,\hat Z^a\right]=2\pi i\left(\Sigma^{-1}\right)_i^{~a}
\,\hat Z^a \ .
\label{derivWeyl}\eeq
The basis (\ref{Deltatorus}) then has the requisite property
\beq
\left[\hat\partial_i\,,\,\hat\Delta(x)\right]=-\partial_i\,\hat\Delta(x) \ .
\label{Deltaderivnctorus}\eeq

\subsection{Topological Quantum Numbers}

A $U(N)$ noncommutative Yang-Mills theory on the torus ${\bf T}^D$ can be
constructed in much the same way as we did in the previous section. If we
restrict to gauge field configurations which are single-valued functions on
${\bf T}^D$, then everything we have said goes through without a hitch, with
single-valued star-unitary functions $g(x)$ parametrizing the star-gauge
transformations (\ref{stargaugetr}). The only difference which arises is that,
like in the commutative case, there are extra observables associated with the
non-trivial homotopy of the torus. The most general star-gauge invariant
observable is still given by (\ref{stargaugeobs}), but now there is a larger
set of line momenta. Because the momenta are now quantized as in
(\ref{momquant}), the identification of the translation vector $v$ in
(\ref{startransl}) is ambiguous up to an integer translation of the periods of
${\bf T}^D$, and the relationship (\ref{linemomentum}) is now modified to
\beq
v^i=\theta^{ij}\,k_j(v,n)+\Sigma^i_{~a}\,n^a
\label{vtorus}\eeq
for arbitrary integer-valued vectors $n^a$. When $\theta=0$, the relationship
(\ref{vtorus}) reproduces the well-known result that the only open line
observables in ordinary Yang-Mills theory are those which are associated with
loops that wind $n^a$ times around the $a$-th non-contractible cycle of the
torus. Therefore, we obtain the analog of Polyakov lines in noncommutative
Yang-Mills theory associated with the different homotopy classes of the
torus~\cite{amns1,amns2}.

More interesting things happen, however, when we consider gauge field
configurations of non-vanishing topological charge on the noncommutative torus.
An elegant way to keep track of the quantum numbers associated with
topologically non-trivial gauge fields is through their Chern numbers. In the
commutative case, these would be represented by the integers
$\mu_{(n)}(E)=\oint\tr^{~}_NF^n/(2\pi)^n$ defined in terms of the curvature
two-form $F$ of some gauge connection of a $U(N)$ gauge bundle $E$ over ${\bf
T}^D$, and suitably integrated over cycles of the torus. For $n=0$ they produce
the rank $N$ of the vector bundle $E$, for $n=1$ they yield the fluxes $Q_{ab}$
of the gauge fields through the surface formed by the $a$-th and $b$-th cycles
of ${\bf T}^D$, and for $n=2$ they give the instanton number $k$ of the bundle
$E$ when $D=4$. We can collect these integers into the inhomogeneous Grassmann
form
\beq
{\rm ch}_0(E)=N+\sum_{n=1}^d\frac1{n!}\,\mu_{(n)}(E)_{a_1\cdots a_{2n}}\,
\rho^{a_1}\cdots\rho^{a_{2n}} \ ,
\label{muEdef}\eeq
where here and in the following we will assume that the spacetime torus has
even dimension $D=2d$. We have introduced a set $\rho^a$, $a=1,\dots,D$, of
anticommuting Grassmann variables,
\beq
\rho^a\rho^b=-\rho^b\rho^a \ ,
\label{Grassmann}\eeq
which can be thought of as local generators of the cotangent bundle of ${\bf
T}^D$. The quantity ${\rm ch}_0(E)$ then defines an integer cohomology class of
the ordinary torus ${\bf T}^D$. Given these integers which characterize the
given bundle $E$, there is an elegant formula for the {\it noncommutative Chern
character}
\beq
{\rm ch}_\Theta(E)=\Tr\otimes\tr^{~}_N\,\exp\frac{\weyl[F]}{2\pi}=\sum_{n\geq0}
\frac1{(2\pi)^n\,n!}\,\Tr\otimes\tr^{~}_N\left(\weyl[F]\right)^n
\label{Chernstar}\eeq
which characterizes the corresponding gauge bundle over the noncommutative
torus. Here $F$ is the noncommutative curvature two-form of the bundle with
local components $F_{ab}=\Sigma^i_{~a}\,F_{ij}\,\Sigma^j_{~b}$, where $F_{ij}$
is defined by (\ref{starfieldstrength}) for an arbitrary gauge connection $A$.
It can be regarded as an element of the ordinary cohomology ring $H^{\rm
even}({\bf T}^D,\real)$ of even degree differential forms on the torus. The
quantity (\ref{Chernstar}) can be written in terms of (\ref{muEdef}) through
the Elliott formula~\cite{elliott}
\beq
{\rm ch}_\Theta(E)=\exp\left(-\frac12\,\Theta^{ab}\,
\frac\partial{\partial\rho^a}\,\frac\partial{\partial\rho^b}\right)~
{\rm ch}_0(E) \ ,
\label{chThetamuE}\eeq
with $\Theta$ regarded as a two-cycle of the homology group $H_2({\bf
T}^D,\real)$~\cite{ans}. The coefficients of $\rho^{a_1}\cdots\rho^{a_{2n}}$ in
the expansion of (\ref{chThetamuE}) define the {\it $n$-th noncommutative Chern
numbers} of the given noncommutative gauge theory. They represent the
topological invariants of the corresponding deformation $E\mapsto E_\Theta$
from a commutative to a noncommutative gauge bundle. In the commutative limit
$\Theta=0$, ${\rm ch}_0(E)$ generates the ordinary integer-valued Chern
numbers. But for $\Theta\neq0$ they are non-integral in general.

For example, in two dimensions we find
\beq
{\rm ch}_\Theta(E)=(N-Q\,\Theta)+Q\,\rho^1\rho^2 \ ,
\label{ch2D}\eeq
where $Q$ is the magnetic flux through ${\bf T}^2$. We see here that in general
the rank of a bundle over the noncommutative torus is no longer necessarily an
integer or even a rational number. This is a common feature of vector bundles
over noncommutative spaces~\cite{connesbook}. The integral curvature
$\Tr\otimes\tr^{~}_N\weyl[F]/2\pi$, on the other hand, is always an integer,
because in noncommutative geometry the top Chern number $\int d^Dx~{\rm
ch}_\Theta(E)$ always computes the index of a Fredholm
operator~\cite{connesbook}, analogously to the commutative case. In
fact, in any dimension the topological numbers of $E$ are all integers
which can be obtained from the K-theory class of $E$. Similarly, in
four dimensions the noncommutative Chern character is
\beq
{\rm ch}_\Theta(E)=\left(N+\frac12\,\Theta^{ab}\,Q_{ab}+k
\,\tilde\Theta_{ab}\,\Theta^{ab}\right)+\frac12\,\Bigl(Q+k\,\tilde\Theta
\Bigr)_{ab}\,\rho^a\rho^b+k\,\rho^1\rho^2\rho^3\rho^4 \ ,
\label{ch4D}\eeq
where $\tilde\Theta_{ab}=\frac12\,\epsilon_{abcd}\,\Theta^{cd}$.

Note that (\ref{chThetamuE}) in general agrees with the formula for D-brane
charges in background supergravity fields as computed from a Wess-Zumino type
action~\cite{douglaswz}, in which the sum over all Ramond-Ramond form
potentials couples to the generalized Mukai vector $\nu(E)={\rm
ch}_0(E)\wedge\e^{B/2\pi}\in H^{\rm even}({\bf T}^D,\real)$ of the given vector
bundle $E\to{\bf T}^D$. The $2n$-th component of ${\rm ch}_0(E)$ in
(\ref{muEdef}) gives the number of D$(2n)$-branes which wrap the various
$2n$-cycles of ${\bf T}^D$. The Chern character (\ref{chThetamuE}) measures the
fact that D2-branes in background $B$-fields have an effective D0-brane charge,
and similarly for other branes. This is seen explicitly in (\ref{ch2D}), which
shows that the number of D2-branes is unaffected in two dimensions by the
presence of the $B$-field, but the number of D0-branes is shifted by the
product of the number of D2-branes and the Neveu-Schwarz two-form field along
the D2-branes.

\subsection{Large Star-Gauge Transformations}

Having described how to construct topological invariants of gauge theories on
the noncommutative torus, let us now turn to their local aspects. We will
consider the noncommutative gauge theory which is defined by the action
\beq
S_{\rm YM}=-\frac1{4g^2}\,\int d^Dx~\tr^{~}_N\Bigl(F_{ij}(x)-f_{ij}\,
\id_N\Bigr)_\star^2
\label{SYMmulti}\eeq
where $F_{ij}(x)$ is the noncommutative field strength tensor
(\ref{starfieldstrength}). The constant, antisymmetric background flux $f_{ij}$
will be fixed later on. At the classical level, the action (\ref{SYMmulti}) is
minimized by gauge field configurations of non-vanishing topological charge. On
a compact space, gauge fields of non-vanishing flux are not single-valued
functions and must be defined on the corresponding covering space. We therefore
regard the noncommutative gauge fields $A_i(x)$ as functions on $\real^D$ which
obey the twisted boundary conditions
\beq
A_i(x+\Sigma^j_{~a}\,\hat\jmath)=\Omega_a(x)\star A_i(x)\star
\Omega_a(x)^\dagger-i\,\Omega_a(x)\star\partial_i\,\Omega_a(x)^\dagger \ ,
\label{twistedbc}\eeq
where $\Sigma^j_{~a}$ are the periods of ${\bf T}^D$ and $\Omega_a(x)$ are the
transition functions of the bundle which are $N\times N$ star-unitary matrices.
Once we have taken the global gauge transformations (\ref{twistedbc}) of the
theory into account, we may use star-gauge invariance to write the action
(\ref{SYMmulti}) in terms of gauge fields on the torus.

By iterating (\ref{twistedbc}) we find a set of consistency conditions
\beq
\Omega_a(x+\Sigma^i_{~b}\,\hat\imath)\star\Omega_b(x)=
\Omega_b(x+\Sigma^i_{~a}\,\hat\imath)\star\Omega_a(x)
\label{cocycle}\eeq
which require that the transition functions define {\it cocycles} of the local
star-gauge group. We will make the gauge choice
\beq
\Omega_a(x)=\e^{i\alpha_{ai}x^i}\otimes\Gamma_a \ ,
\label{Omegagauge}\eeq
where $\alpha$ is a real-valued constant $D\times D$ matrix with the
antisymmetry property $(\alpha\Sigma)^\top=-\alpha\Sigma$ which ensures that
the transition function $\Omega_a(x)$ has the periodicity
$\Omega_a(x+\Sigma^i_{~a}\,\hat\imath)=\Omega_a(x)$. The matrix $\alpha$
appears as the $U(1)$ factor in the given gauge choice and it will essentially
account for the abelian fluxes of the gauge fields. The $\Gamma_a$ are constant
$SU(N)$ matrices. From (\ref{cocycle}) it follows that they must commute up to
some phases,
\beq
\Gamma_a\Gamma_b=\e^{2\pi iQ_{ab}/N}\,\Gamma_b\Gamma_a \ ,
\label{WeyltHooftalg}\eeq
where $Q$ is an antisymmetric $D\times D$ matrix. Taking the determinant of
both sides of (\ref{WeyltHooftalg}) shows that $Q_{ab}\in\zed$. The commutation
relations (\ref{WeyltHooftalg}) define the {\it Weyl-'t~Hooft algebra} in $D$
dimensions~\cite{weylbook,thooftflux}, with $Q$ the matrix of non-abelian
$SU(N)$ 't~Hooft fluxes through the various non-trivial two-cycles of the torus
(Recall that magnetic flux on compact spaces with non-contractible two-cycles
is always quantized). From (\ref{cocycle}) we find the matrix-valued
consistency condition
\beq
Q=\frac N{2\pi}\,\Bigl(2\alpha\,\Sigma-\alpha\,\theta\,\alpha^\top\Bigr) \ .
\label{matrixcons}\eeq

We will now rewrite the noncommutative gauge theory (\ref{SYMmulti}) in terms
of gauge fields whose vacuum configuration has vanishing magnetic flux (i.e.
$A_i(x)=0$ up to a star-gauge transformation). These new field configurations
will therefore be single-valued functions on the torus. For this, we introduce
a fixed, multi-valued background abelian gauge field $a_i(x)$ to absorb the
flux $f_{ij}$. A gauge choice which is compatible with (\ref{Omegagauge}) is
given by
\beq
a_i(x)=\frac12\,F_{ij}\,x^j\otimes\id_N
\label{aifixed}\eeq
where $F$ is a real-valued constant antisymmetric $D\times D$ matrix. From
(\ref{WeyltHooftalg}) and the identity (c.f. (\ref{xderiv}))
\beq
x^i\star\e^{i\alpha_{aj}x^j}-\e^{i\alpha_{aj}x^j}\star x^i=-\theta^{ik}\,
\alpha_{ak}~\e^{i\alpha_{aj}x^j}
\label{xalphaid}\eeq
it follows that the twisted boundary conditions (\ref{twistedbc}) for the gauge
field (\ref{aifixed}) are then equivalent to the matrix identities
\beq
\alpha=-\Sigma^\top F\,\frac1{2\,\id_D+\theta F}~~~~~~,~~~~~~
F=2\alpha^\top\,\frac1{\Sigma-\theta\alpha^\top} \ .
\label{alphaFids}\eeq
We decompose the gauge field configurations $A_i(x)$ of the theory
(\ref{SYMmulti}) into the particular solution (\ref{aifixed},\ref{alphaFids})
of the twisted boundary conditions and a fluctuating part around the fixed
background as
\beq
A_i(x)=a_i(x)+{\cal A}_i(x) \ ,
\label{Aixdecomp}\eeq
where the field ${\cal A}_i(x)$ satisfies the covariant twisted boundary
conditions
\beq
{\cal A}_i(x+\Sigma^j_{~a}\,\hat\jmath)=\Omega_a(x)\star{\cal A}_i(x)\star
\Omega_a(x)^\dagger \ .
\label{adjointsection}\eeq
The condition (\ref{adjointsection}) requires that the fluctuating field ${\cal
A}_i(x)$ be an {\it adjoint section} of the given gauge bundle over the
noncommutative torus. Substituting (\ref{adjointsection}) and (\ref{aifixed})
into the action (\ref{SYMmulti}) we arrive at
\beq
S_{\rm YM}=-\frac1{4g^2}\,\int d^Dx~\tr^{~}_N
\left({\cal F}_{ij}(x)+f^\star_{ij}-f_{ij}\,\id_N\right)_\star^2 \ ,
\label{SYMdecomp}\eeq
where
\beq
{\cal F}_{ij}=D_i{\cal A}_j-D_j{\cal A}_i-i\left({\cal A}_i\star
{\cal A}_j-{\cal A}_j\star{\cal A}_i\right)
\label{calFdef}\eeq
with
\beq
D_i=\partial_i+i\,a_i
\label{nabladef}\eeq
a fiducial connection of constant curvature $F_{ij}$, and
\beq
f_{ij}^\star=\partial_ia_j-\partial_ja_i-i\left(a_i\star a_j-a_j\star a_i
\right)=\left(F+\frac14\,F\,\theta\,F\right)_{ij}\otimes\id_N
\label{noncommcurva}\eeq
is the noncommutative field strength of the background gauge field
(\ref{aifixed}). Requiring that ${\cal A}_i(x)=0$ be the vacuum field
configuration of the theory up to a star-gauge transformation fixes
$f_{ij}\,\id_N=f_{ij}^\star$ in (\ref{SYMdecomp}), and the action becomes
\beq
S_{\rm YM}=-\frac1{4g^2}\,\int d^Dx~\tr^{~}_N\Bigl({\cal F}_{ij}(x)\star
{\cal F}^{ij}(x)\Bigr) \ .
\label{SYMsingle}\eeq

Since the classical gauge field configurations of the theory (\ref{SYMsingle})
have vanishing curvature ${\cal F}_{ij}(x)=0$, we would like to interpret them
as single-valued functions. This will be done in the next section, where we
shall find a suitable basis of the noncommutative $C^*$-algebra of functions in
which the covariant derivatives (\ref{nabladef}) act as ordinary derivative
operators. We will do so by finding the most general adjoint section obeying
(\ref{adjointsection}) and interpreting the resulting model as a new gauge
theory on a new noncommutative torus. For this, it will be convenient to solve
the covariant constraint (\ref{adjointsection}) using Weyl operators. Using the
map (\ref{Deltatorus}), we may associate to the adjoint section ${\cal A}_i(x)$
the Hermitian Weyl operator
\beq
\weyl[{\cal A}]_i=\int d^Dx~\hat\Delta(x)\otimes{\cal A}_i(x)
\label{WeylcalA}\eeq
in terms of which the action (\ref{SYMsingle}) becomes
\beq
S_{\rm YM}=-\frac1{4g^2}\,\Tr\otimes\tr^{~}_N\left(\left[\hat D_i\,,\,
\weyl[{\cal A}]_j\right]-\left[\hat D_j\,,\,\weyl[{\cal A}]_i\right]-i
\left[\weyl[{\cal A}]_i\,,\,\weyl[{\cal A}]_j\right]\right)^2 \ ,
\label{SYMop}\eeq
where
\beq
\hat D_i=\hat\partial_i+\frac i2\,F_{ij}\,\hat x^j
\label{hatnabla}\eeq
is a linear derivation on the Weyl operator algebra of constant curvature
\beq
\left[\hat D_i\,,\,\hat D_j\right]=i\left(F+\frac14\,F\,\theta\,F
\right)_{ij} \ .
\label{constcurv}\eeq
The twisted boundary conditions (\ref{adjointsection}) may then be written in
terms of Weyl operators as
\beq
\e^{\Sigma^j_{~a}\hat\partial_j}~\weyl[{\cal A}]_i~\e^{-\Sigma^j_{~a}
\hat\partial_j}=\weyl[\Omega]_a~\weyl[{\cal A}]_i\,\left(\weyl[\Omega]_a
\right)^\dagger \ ,
\label{adjointsectionop}\eeq
where
\beq
\weyl[\Omega]_a=\e^{i\alpha_{ai}\hat x^i}\otimes\Gamma_a
\label{WeylOmega}\eeq
are the unitary Weyl operators corresponding to the transition functions in the
gauge (\ref{Omegagauge}).

The background abelian flux $f_{ij}$ can be written in terms of the geometrical
parameters of the given constant curvature bundle by using (\ref{alphaFids})
and the identity
\beq
\left(\frac1{\id_D-\theta\alpha^\top\Sigma^{-1}}\right)^2\otimes\id_N=
\left(\id_D+\frac12\,\theta\,F\right)^2\otimes\id_N=\id_D\otimes\id_N+
\theta\,f^\star
\label{Fids}\eeq
to write (\ref{matrixcons}) in the form
\beq
\Sigma^\top\,f~\Sigma=2\pi\,\frac1{N\,\id_D-Q\,\Theta}\,Q \ ,
\label{fQrel}\eeq
where $\Theta$ is the dimensionless noncommutativity parameter
(\ref{Thetadim}). The identity (\ref{fQrel}) gives the relationship between the
central curvatures and the magnetic fluxes of the gauge field configurations.
Note that in the commutative case $\Theta=0$, the $SU(N)$ 't~Hooft flux $Q$ is
equivalent to the $U(1)$ flux $f=F$ of the bundle in (\ref{constcurv}). The
't~Hooft flux was originally introduced for ordinary $SU(N)$ gauge theory and
it is the only way in that case to twist the boundary conditions on the gauge
fields~\cite{thooftflux,lebpolros}. For this reason the matrices $\Gamma_a$
which generate the Weyl-'t~Hooft algebra (\ref{WeyltHooftalg}) are sometimes
refered to as {\it twist eaters}. In the commutative case, keeping the phase
$\alpha$ in (\ref{Omegagauge}) is redundant (see (\ref{matrixcons})), because
it can be cancelled by using the global decomposition (\ref{UNdecomp}) of the
$U(N)$ gauge group. The quotient there means that an element
$(\e^{i\alpha},g)\in U(1)\times SU(N)$ is identified with
$(\e^{i\alpha}\,\omega^{-1},\omega g)$ for any $N$-th root of unity
$\omega\in\zed_N$. The $U(1)$ twists can in this way be consistently cancelled
by the $SU(N)/\zed_N$ sector of the gauge theory and one can simply set
$\alpha=0$ without loss of generality. However, as is evident from the formulas
above, this is no longer true in the noncommutative case. The physical reason
behind this was explained at the end of section 4.1. We remark also that the
constructions we have presented in this section do not account for all possible
gauge theories on the noncommutative torus. In two and three dimensions,
generic $U(N)$ bundles on tori admit connections of vanishing $SU(N)$ curvature
(i.e. with constant curvature $[ D_i, D_j]=f_{ij}\,\id_N$, like the ones we
have considered above)~\cite{connesbook}. However, for $D\geq4$, even in the
commutative case not all bundles admit constant curvature
connections~\cite{thooftflux}. The connections that we have considered in this
section correspond to BPS states in the gauge theory~\cite{cds,bpscurv}.
Noncommutative gauge theories on ${\bf T}^4$ with non-constant $SU(N)$ flux
have been studied in~\cite{kkklly}.

\newsection{Duality in Noncommutative Yang-Mills Theory}

In this section we will derive a remarkable equivalence relation on the space
of noncommutative Yang-Mills
theories~\cite{lls1,cds,amns2,morita,schwarzmorita}. This is a special type of
geometrical symmetry which relates two apparently distinct ``spaces'' in
noncommutative geometry. It was originally introduced in the mathematics
literature as a resolution to certain paradoxes that arise in the context of
the reconstruction of topological spaces from $C^*$-algebras. Let us give a
very simple example of this equivalence. Given any manifold $M$, consider the
two non-isomorphic associative algebras
\bea
{\cal A}&=&C(M) \ , \nn\\{\cal A}'&=&C(M)\otimes \mat(N,\complex) \ ,
\label{AAprime}\eea
where $C(M)$ is the space of smooth complex-valued functions on $M$ and
$\mat(N,\complex)$ is the finite-dimensional algebra of $N\times N$
complex-valued matrices. At the level of topology, a topological space may be
completely characterized by the algebra of continuous complex-valued functions
defined on it, because one may reconstruct the topology given the continuity
requirement of all functions on it. The algebra $\cal A$ is commutative, and if
we didn't know that it was a space of functions on $M$, but rather only knew
its algebraic properties, then we could still associate the manifold $M$ to it.
That this is possible is the content of the {\it Gel'fand-Naimark theorem}
which provides a one-to-one correspondence between the category of commutative
$C^*$-algebras and the category of Hausdorff topological
spaces~\cite{connesbook}. Given any {\it commutative} algebra $\cal A$, we may
formally construct a topological space $M$ for which $\cal A$ is naturally
isomorphic to the space of functions $C(M)$. The {\it Gel'fand transform} which
accomplishes this identifies points of the space with the characters (i.e. the
multiplicative linear functionals) of the algebra $\alg$~\cite{connesbook}. In
the case of a commutative algebra, all irreducible representations are
one-dimensional, and the space of characters coincides with the space of
irreducible representations. We will return to these points in section~8.

On the other hand, the algebra ${\cal A}'$ is the space of $N\times N$
matrix-valued functions on the manifold $M$, which is noncommutative. The
definition of the Gel'fand transform, which is used to reconstruct a space from
an algebra, becomes ambiguous for noncommutative algebras, and it is not
possible to formally reconstruct the space $M$ in this case. In particular, the
spaces of characters and irreducible representations of the algebra no longer
coincide. So if we were only given the algebra ${\cal A}'$, due to its
noncommutativity we would have no way of knowing that it is canonically
associated with a manifold. But clearly we would like to do so, because ${\cal
A}'$ is still a space of fields which are defined over some configuration
manifold, in the classical sense of the word. The only difference now is that
the fields have isospin degrees of freedom associated to them. The ambiguity
that arises in defining a point is removed by the realization that the
different points which are associated via the Gel'fand transform are smeared
over an $N$ dimensional sphere~\cite{lsrev}, and are related to each other by
global rotations in the isospin space. The algebra ${\cal A}'$ thereby
certainly captures the topological characteristics of the manifold
$M$.\footnote{\baselineskip=12pt These statements can be made more precise by
using the formalism of spectral triples in noncommutative
geometry~\cite{connesbook},\cite{fg}--\cite{lls1},\cite{lsrev}. The Riemannian
geometry of a manifold $M$ can be reconstructed from the operator algebraic
spectral data associated with the quantum mechanics of the free geodesic motion
of a test particle on $M$. In this context, $\cal A$ coincides with the algebra
of observables of the quantum theory. The algebra ${\cal A}'$, on the other
hand, coincides with the algebra of observables of a test particle moving on
$M$ which has some internal degrees of freedom.}

This paradox is resolved by the realization that the space $\mat(N,\complex)$
has only {\it one} irreducible representation as a $C^*$-algebra, namely its
defining representation. To capture the feature that both algebras
(\ref{AAprime}) describe the same space $M$, one says that they are {\it Morita
equivalent}. In general, two algebras are Morita equivalent if they become
isomorphic upon tensoring them with the algebra of compact operators on some
Hilbert space (heuristically, this is the algebra $\mat(N,\complex)$ for ``$N$
sufficiently large'' -- see section~8)~\cite{connesbook}. Morita equivalent
spaces share many common geometrical characteristics, for example they have the
same K-theory and cyclic homology. But gauge theories, or more precisely vector
bundles, defined over them can be very different. For instance, the algebra
${\cal A}$ in (\ref{AAprime}), being commutative, possesses only a $U(1)$
unitary subgroup of functions, while ${\cal A}'$ has a $U(N)$ unitary subgroup.
Therefore, under the Morita relation, a $U(1)$ gauge theory becomes equivalent
to a $U(N)$ gauge theory. This equivalence at the level of vector bundles
follows from the stability of the corresponding K-theory groups (which
characterize the cohomology of vector bundles over a space) under the Morita
transformation~\cite{connesbook}, and indeed gauge theories over Morita
equivalent spaces are canonically related. In this section we will see some
specific instances of this natural relation.

In the following we will present a field theoretical derivation of the Morita
equivalence between Yang-Mills theories on noncommutative tori. We will see
that this equivalence can be interpreted as a stringy $T$-duality symmetry of
noncommutative Yang-Mills theory~\cite{lls1,schwarzmorita}, which implies
certain remarkable symmetries of Matrix theory
compactifications~\cite{cds,ans,morita,schwarzmorita}. Indeed, we shall find an
explicit relationship with the $T$-duality symmetry of toroidally compactified
open strings~\cite{Tduality}. Another application will be to give a
quantitative explanation of the form of the perturbative gauge theory results
that we discussed in section~4.3. There we saw, for example, that the one-loop
renormalization of $U(1)$ noncommutative gauge theory is identical on $\real^4$
to that of ordinary large $N$ Yang-Mills theory (after a suitable rescaling of
the Yang-Mills coupling constant $g$). This is a strong indication that the
geometrical Morita equivalence property of noncommutative geometry, which holds
at the classical level, does indeed persist in regularized perturbation theory.
Therefore, Morita equivalence, along with the Eguchi-Kawai reduction of large
$N$ gauge theories which will be discussed in section~7, lends a natural
explanation of these coincidences. It also yields a quantitative explanation
for the deep relationship that exists between large $N$ reduced models (such as
the IKKT matrix model which we discussed in section 1.2) and noncommutative
Yang-Mills theory~\cite{amns2}. We will return to this latter point in the next
section.

\subsection{Morita Equivalence}

In this subsection we will demonstrate in some detail how to solve the twisted
boundary conditions (\ref{adjointsectionop}) and show that in this way we
naturally arrive at a physically equivalent, dual noncommutative Yang-Mills
gauge theory. We will see in the next subsection that this notion of duality is
identical to that of $T$-duality for toroidally compactified open strings in
background supergravity fields.

\subsubsection{Irreducible Representations of Twist Eaters}

We first need to digress briefly and describe the representation theory of the
Weyl-'t~Hooft algebra (\ref{WeyltHooftalg})~\cite{twistirrep}. The irreducible
representations of this algebra are called {\it twist-eating solutions} and for
any even dimensionality $D=2d$ they may be constructed as follows. The lattice
${\cal L}$ which generates the torus as the quotient space ${\bf
T}^D=\real^D/{\cal L}$ has automorphism group $SL(D,\zed)$ which becomes the
modular group of ${\bf T}^D$. Using this discrete geometrical symmetry of the
spacetime, we can rotate the 't~Hooft matrix $Q\mapsto S^\top QS$, $S\in
SL(D,\zed)$, into a canonical skew-diagonal form~\cite{igusa}
\beq
Q=\pmatrix{0&-q_1& & & \cr q_1&0& & & \cr & &\ddots& & \cr
& & &0&-q_d\cr & & &q_d&0\cr} \ .
\label{Qdiag}\eeq
Given the $d$ independent fluxes $q_\alpha\in\zed$, we introduce the two
relatively prime sets of $d$ integers
\bea
N_\alpha&=&\frac N{{\rm gcd}(N,q_\alpha)} \ , \nn\\
q_\alpha'&=&\frac{q_\alpha}{{\rm gcd}(N,q_\alpha)} \ ,
\label{3integers}\eeq
where gcd denotes the greatest common divisor. We then assume that
there exists an integer $N_0\in\zed_+$ which is a divisor of
the rank of the gauge group and the product of the $d$ integers $N_\alpha$ in
(\ref{3integers}),
\beq
N=N_0\,(N_1\cdots N_d) \ .
\label{NN0decomp}\eeq
The integer $N_1\cdots N_d$ is the dimension of the irreducible representation
of the Weyl-'t~Hooft algebra. The requirement (\ref{NN0decomp}) is a necessary
and sufficient condition for the existence of $D$ independent twist eating
solutions $\Gamma_a$~\cite{twistirrep}. It is a condition which must be met by
the geometrical parameters of the given constant curvature bundle.

The matrices $\Gamma_a$ may then be defined on the $SU(N)$ subgroup
$SU(N_1)\otimes\cdots\otimes SU(N_d)\otimes SU(N_0)$ as
\bea
\Gamma_{2\alpha-1}&=&\id_{N_1}\otimes\cdots\otimes V_{N_\alpha}
\otimes\cdots\otimes\id_{N_d}\otimes\id_{N_0} \ , \nn\\
\Gamma_{2\alpha}&=&\id_{N_1}\otimes\cdots\otimes
\left(W_{N_\alpha}\right)^{q_\alpha'}
\otimes\cdots\otimes\id_{N_d}\otimes\id_{N_0}
\label{Gammadefs}\eeq
for $\alpha=1,\dots,d$, where $V_N$ and $W_N$ are the $SU(N)$ shift and clock
matrices
\bea
V_N&=&\pmatrix{0&1& & &0\cr &0&1& & \cr & &\ddots&\ddots& \cr
 & & &\ddots&1\cr1& & & &0\cr} \ , \nn\\W_N&=&\pmatrix{1& & & & \cr
&\e^{2\pi i/N}& & &\cr & &\e^{4\pi i/N}& & \cr
 & & &\ddots& \cr & & & &\e^{2\pi i(N-1)/N}\cr}
\label{VNWNdef}\eea
which obey the commutation relations
\beq
V_NW_N=\e^{2\pi i/N}\,W_NV_N \ .
\label{VWalg}\eeq
The twist eaters (\ref{Gammadefs}) commute with the $SU(N_0)$ subgroup of
$SU(N)$ which is generated by the matrices of the form
$\id_{N_1}\otimes\cdots\otimes\id_{N_d}\otimes Z_0$ with $Z_0\in SU(N_0)$. Note
that
$(\Gamma_{2\alpha-1})^{N_\alpha}=(\Gamma_{2\alpha})^{N_\alpha}=\id_{N_\alpha}$
for each $\alpha=1,\dots,d$.

Since the integers $N_\alpha$ and $q_\alpha'$ in (\ref{3integers}) are
relatively prime, there exist integers $a_\alpha,b_\alpha$ such that
\beq
a_\alpha N_\alpha+b_\alpha q_\alpha'=1
\label{aibidef}\eeq
for each $\alpha=1,\dots,d$. In the basis (\ref{Qdiag}) where $Q$ is
skew-diagonal, we then introduce the four integral $D\times D$ matrices
\bea
A=\pmatrix{a_1&0& & & \cr0& a_1& & & \cr & &\ddots& & \cr
& & & a_d&0\cr & & &0& a_d\cr}~~~~~~&,&~~~~~~B=\pmatrix{0&-b_1& & & \cr
b_1&0& & & \cr & &\ddots& & \cr & & &0&-b_d\cr & & &b_d&0\cr} \ , \nn\\
Q'=\pmatrix{0&-q_1'& & & \cr q_1'&0& & & \cr & &\ddots& & \cr
& & &0&-q_d'\cr & & & q_d'&0\cr}~~~~~~&,&~~~~~~N'=\pmatrix{N_1&0& & &\cr
0&N_1& & & \cr & &\ddots& & \cr & & &N_d&0\cr & & &0&N_d\cr} \ . \nn\\& &
\label{4matrices}\eea
We may then use the $SL(D,\zed)$ modular symmetry of the torus to rotate the
matrix $Q$ back to general form, $Q\mapsto S^\top QS$, and similarly rotate the
four matrices in (\ref{4matrices}) as $A\mapsto(S')^{-1}AS$,
$B\mapsto(S')^{-1}B(S^{-1})^\top$, $Q'\mapsto S^\top Q'S'$, and $N'\mapsto
S^{-1}N'S'$, with $S,S'\in SL(D,\zed)$. The geometrical significance of the
extra $SL(D,\zed)$ matrix $S'$ is that it will parametrize the automorphism
group of the dual noncommutative torus that we shall obtain. As we will discuss
in the next subsection, such an interpolation between two dual spaces is a
general characteristic of the Morita transformation. In these general forms we
can write
\beq
Q=N\,Q'N'^{-1} \ ,
\label{QQ'rel}\eeq
and from (\ref{aibidef}) it follows in general that
\beq
AN'+BQ'=\id_D \ .
\label{ANBQrel}\eeq
Because of (\ref{ANBQrel}) and the antisymmetry of the matrices $AB^\top$ and
$Q'^\top N'$ we have the block matrix identity
\beq
\pmatrix{A&B\cr Q'^\top&-N'^\top\cr}\pmatrix{0&\id_D\cr\id_D&0\cr}
\pmatrix{A&B\cr Q'^\top&-N'^\top\cr}^\top=\pmatrix{0&\id_D\cr\id_D&0\cr}
\label{blockid}\eeq
which is equivalent to the statement that
\beq
\pmatrix{A&B\cr Q'^\top&-N'^\top\cr}\in SO(D,D;\zed)
\label{SODDrel}\eeq
with respect to the canonical basis of $\real^{D,D}$.

\subsubsection{Solving Twisted Boundary Conditions}

We are now ready to describe how to solve
(\ref{adjointsectionop})~\cite{amns2,morita}. For this, we make two key
observations. First of all, the twisted boundary conditions are solved for
gauge fields on $\real^D$ which are only afterwards regarded as functions on a
torus, so that the corresponding Weyl operators should likewise be thought of
as originating in this way. This is important because the solutions to
(\ref{adjointsectionop}) do not actually live on the original noncommutative
torus. Secondly, for any pair of relatively prime integers $N,q$ the set
$\{(V_N)^j\,(W_N)^{qj'}~|~j,j'\in\zed_N\}$ spans the $N^2$ dimensional complex
linear vector space $gl(N,\complex)$. From the construction above, it follows
that we may expand the Weyl operator $\weyl[{\cal A}]_i$ in an $SU(N_1\cdots
N_d)\otimes SU(N_0)$ subgroup of $SU(N)$ and leave a $U(N_0)$ sector of the
original gauge group corresponding to the subgroup of matrices which commute
with the twist eaters.

We may therefore write down the expansion
\beq
\weyl[{\cal A}]_i=\int d^Dk~\e^{ik_i\hat x^i}
\otimes\sum_{\vec j~{\rm mod}\,N'}~\prod_{a=1}^D(\Gamma_a)^{j^a}
\otimes\tilde a_i(k,\vec j\,)
\label{WeylcalAexp}\eeq
where $\tilde a_i(k,\vec j\,)$ is an $N_0\times N_0$ matrix-valued function
which is periodic in $\vec j$, $\tilde a_i(k,j^a)=\tilde a_i(k,j^a+N'^a_{~b})$
for each $a,b=1,\dots,D$. By applying the constraint (\ref{adjointsectionop})
to (\ref{WeylcalAexp}) using (\ref{WeyltHooftalg}) and (\ref{WeylOmega}), and
by equating the expansion coefficients on both sides of
(\ref{adjointsectionop}), we find that the functions $\tilde a_i(k,\vec j\,)$
vanish unless
\beq
\frac1{2\pi}\,k_i\left(N'^{-1}\right)^b_{~a}\left(\beta^{-1}\right)^i_{~b}
-j^b\,Q_{bc}'\left(N'^{-1}\right)^c_{~a}=n_a
\label{tildeavanish}\eeq
for some $n_a\in\zed$, where we have used (\ref{QQ'rel}) and introduced the
$D\times D$ matrix
\beq
\beta=\frac1{(\Sigma+\theta\,\alpha^\top)N'}=\frac1{(\id_D+\theta\,F)
\Sigma}\,\Bigl(2\,\id_D+\theta\,F\Bigr)\,\frac{N'^{-1}}2 \ .
\label{betadef}\eeq
Given the matrices $A$ and $B$ constructed above which satisfy (\ref{ANBQrel}),
we may then solve (\ref{tildeavanish}) by setting $n_a=A^b_{~a}\,m_b$ and
$j^a=B^{ba}\,m_b$ for some $m_a\in\zed$. Since
$\prod_a(\Gamma_a)^{N'^a_{~b}}=\id_N$ for each $b=1,\dots,D$, it follows that
for any given set of $D$ integers $m_a$ and fixed Fourier momentum $\vec k$,
this solution for $(\vec n,\vec j\,)$ is unique ${\rm mod}\,N'$. By
substituting this solution into (\ref{tildeavanish}), we may then solve for the
Fourier momenta $\vec k$ as $k_i=2\pi\beta^a_{~i}\,m_a$ and replace the
integration in (\ref{WeylcalAexp}) by a summation over all $\vec m\in\zed^D$.

What we have shown is that the most general solution to the constraint
(\ref{adjointsection}) takes the form
\beq
\weyl[{\cal A}]_i=\sum_{\vec m\in\zeds^D}\,\prod_{a=1}^D\left(\hat Z'^a
\right)^{m_a}\,\prod_{a<b}\e^{\pi i\,m_a\,\Theta'^{ab}\,m_b}\otimes\tilde
a_i(\vec m) \ ,
\label{generaltwistsol}\eeq
where $\tilde a_i(\vec m)=\tilde a_i(2\pi\beta\vec m,B\vec m)$ are $N_0\times
N_0$ matrix-valued Fourier coefficients which by Hermiticity obey $\tilde
a_i(-\vec m)=\tilde a_i(\vec m)^\dagger$. The operators
\beq
\hat Z'^a=\e^{2\pi i\beta^a_{~i}\hat x^i}\otimes\prod_{b=1}^D
(\Gamma_b)^{B^{ab}}
\label{hatZprime}\eeq
obey the commutation relations
\bea
\hat Z'^a\,\hat Z'^b&=&\e^{-2\pi i\,\Theta'^{ab}}\,\hat Z'^b\,\hat Z'^a \ ,
\label{Zprimealg}\\\left[\hat D_i\,,\,\hat Z'^a\right]&=&2\pi i\,
\left(\Sigma'^{-1}\right)_i^{~a}\,\hat Z'^a \ ,
\label{nablaZprime}\eea
where $\hat D_i$ is the covariant derivation (\ref{hatnabla}),
and\footnote{\baselineskip=12pt Note that the transformation
$\Theta\mapsto\Theta'$ is only well-defined on those $\Theta$ for which
$\Theta\,Q'-N'$ is an invertible matrix. Such $\Theta$'s span a dense subspace
of the whole space of antisymmetric $D\times D$ real-valued matrices.}
\bea
\Theta'&=&\frac1{\Theta\,Q'-N'}\,\Bigl(A\,\Theta+B\Bigr)^\top\label{Thetaprime}
\ , \\\Sigma'&=&\Sigma\,(\Theta\,Q'-N') \ .
\label{Sigmaprime}\eea
The commutation relations (\ref{Zprimealg},\ref{nablaZprime}) are of the same
form as the defining ones (\ref{noncommtorus},\ref{derivWeyl}) of the original
noncommutative torus. The operators (\ref{hatZprime}) thereby define a Weyl
basis which generates a new, dual noncommutative torus with deformation matrix
(\ref{Thetaprime}) and period matrix (\ref{Sigmaprime}). The canonical
coordinates $x'^i$ on this new torus may be used to define a new basis
$\hat\Delta'(x')$ for the mapping between spacetime fields and Weyl operators,
and they are obtained by formally choosing a rotation $x\mapsto x'$ in which
\beq
\left[\hat D_i\,,\,\hat\Delta'(x')\right]=-\partial_i'\,\hat\Delta'(x') \ ,
\label{nablaDeltaprime}\eeq
where $\partial_i'=\partial/\partial x'^i$. All the information about the
topological charges of the original gauge theory are now transfered into the
new noncommutativity parameters (\ref{Thetaprime}), and the new basis is given
as
\beq
\hat\Delta'(x')=\frac1{|\det\Sigma'|}\,\sum_{\vec m\in\zeds^D}\,\prod_{a=1}^D
\left(\hat Z'^a\right)^{m_a}~\prod_{a<b}\e^{-\pi i\,m_a\,\Theta'^{ab}
\,m_b}~\e^{-2\pi i\,(\Sigma'^{-1})_i^{~a}\,m_ax'^i} \ ,
\label{Deltaprime}\eeq
analogously to (\ref{Deltatorus}). Note that the commutation relations
(\ref{nablaZprime}) are tantamount to representing the covariant derivations
through
\beq
\e^{\Sigma^i_{~a}\hat D_i}=\e^{2\pi\Sigma^i_{~a}\hat\partial_i}~
\e^{i\alpha_{ai}\hat x^i}\otimes\Gamma_a\otimes\prod_{b=1}^D\left(\hat Z'^b
\right)^{Q'_{ba}} \ .
\label{nablaZrep}\eeq

We may now rewrite the expansion (\ref{generaltwistsol}) using the new basis
(\ref{Deltaprime}) which leads to the Weyl quantization
\beq
\weyl[{\cal A}]_i=\int d^Dx'~\hat\Delta'(x')\otimes{\cal A}_i'(x') \ ,
\label{WeylcalAprime}\eeq
where ${\cal A}_i'(x')$ is by construction a {\it single-valued} $U(N_0)$ gauge
field on the dual noncommutative torus. The remaining rank $N/N_0$ of the
original $U(N)$ gauge theory has now been absorbed into the new Weyl basis
(\ref{hatZprime}). The operator trace $\Tr'$ satisfying
$\Tr'\,\hat\Delta'(x')=1$ may be computed in terms of the original trace Tr as
\beq
\Tr'\otimes\tr^{~}_{N_0}=\frac{N_0}N\,\left|\frac{\det\Sigma'}
{\det\Sigma}\right|\,\Tr\otimes\tr^{~}_N \ .
\label{traceprimes}\eeq
Using (\ref{Deltaprime})--(\ref{traceprimes}), we find that the noncommutative
Yang-Mills action (\ref{SYMop}) when expanded in this new basis of Weyl
operators becomes
\beq
S_{\rm YM}=-\frac1{4g'^2}\,\int d^Dx'~\tr^{~}_{N_0}\Bigl({\cal F}_{ij}'(x')
\star'{\cal F}'^{ij}(x')\Bigr) \ ,
\label{dualSYM}\eeq
where
\beq
{\cal F}_{ij}'=\partial_i'{\cal A}_j'-\partial_j'{\cal A}_i'-i
\left({\cal A}_i'\star'{\cal A}_j'-{\cal A}_j'\star'{\cal A}_i'\right)
\label{calFprime}\eeq
and $\star'$ denotes the new Groenewold-Moyal product defined using the
deformation parameter $\theta'=\Sigma'\,\Theta'\,\Sigma'^\top/2\pi$ instead of
$\theta$. The new Yang-Mills coupling constant in (\ref{dualSYM}) is given by
\beq
g'=g~\sqrt{\Bigl|\det(\Theta\,Q'-N')\Bigr|} \ .
\label{gprime}\eeq
The exact equivalence between the two forms (\ref{SYMmulti}) and
(\ref{dualSYM}) of the noncommutative Yang-Mills action is the duality that we
are looking for. It shows that a noncommutative $U(N)$ Yang-Mills theory with
magnetic flux (\ref{fQrel}) (and hence multi-valued gauge fields) is equivalent
to a $U(N_0)$ noncommutative Yang-Mills theory with deformation parameter
transformed according to (\ref{Thetaprime}), no magnetic flux (and hence
single-valued gauge fields), and reduced gauge group rank $N_0$ defined by
(\ref{NN0decomp}). This duality is known as Morita equivalence of
noncommutative gauge theories and its basic transformation rules are summarized
for convenience in table~\ref{Morita}.

\begin{table}
\begin{center}
\begin{tabular}{|c||c|c|}
\hline & Original Gauge Theory & Dual Gauge Theory\\
\hline\hline Magnetic Flux & $Q$ & $0$\\
\hline Gauge Group & $U(N)$ & $U(N_0)$\\
\hline Noncommutativity & $\Theta$ & $(\Theta\,Q'-N')^{-1}(A\,\Theta+B)^\top$\\
\hline Periods & $\Sigma$ & $\Sigma\,(\Theta\,Q'-N')$\\
\hline Coupling Constant & $g$ & $g\,\Bigl|\det(\Theta\,Q'-N')\Bigl|^{1/2}$
\\ \hline
\end{tabular}
\end{center}
\caption{\baselineskip=12pt {\it Basic Morita equivalence of noncommutative
gauge theories on $D$ dimensional tori. The integer $N/N_0$ is the dimension of
the irreducible representation of the Weyl-'t~Hooft algebra
$\Gamma_a\Gamma_b=\e^{2\pi iQ_{ab}}\,\Gamma_b\Gamma_a$ in $D$ dimensions.}}
\label{Morita}\end{table}

\subsection{Applications}

Let us now make a series of remarks concerning the duality that we have found
in the previous subsection:
\begin{itemize}
\item{Modulo a straightforward conjugation of the transformation matrix in
(\ref{SODDrel}), the map $\Theta\mapsto\Theta'$ is the standard $SO(D,D;\zed)$
transformation that relates Morita equivalent noncommutative tori. In fact, it
is a theorem of noncommutative geometry that two noncommutative tori are Morita
equivalent if and only if their noncommutativity parameters are related in this
way~\cite{schwarzrieffel}. This statement holds also when the target gauge
bundle has non-vanishing topological charge, as the equivalence may then be
realized by the composition of two of the sort that we have described in the
previous subsection. In general, the transformation rule for the background
fluxes $f_{ij}$ is given by (see (\ref{fQrel}) and
(\ref{QQ'rel}))~\cite{schwarzmorita}
\beq
f'=(\Theta\,Q'-N')^\top\,f\,(\Theta\,Q'-N')+2\pi\,Q'\,(\Theta\,Q'-N') \ .
\label{fprime}\eeq
We will see below how the transformation (\ref{fprime}) may be explicitly
obtained. From the point of view of the derivation given in the previous
subsection, Morita equivalence may be simply regarded as a change of basis
$\hat\Delta(x)\mapsto\hat\Delta'(x')$ for the mapping between operators and
fields in Weyl quantization.}
\item{From the transformation rule (\ref{Sigmaprime}) for the period matrix of
the torus, we find that the dual metric $G'=\Sigma'^\top\,\Sigma'$ is given by
\beq
G'=(\Theta\,Q'-N')^\top\,G\,(\Theta\,Q'-N') \ .
\label{Gprime}\eeq
The transformations (\ref{Thetaprime}) and (\ref{Gprime}) are recognized as
those of the $B$-field and the open string metric under the $SO(D,D;\zed)$
target space duality group of the torus, acting on the open string
parameters~\cite{Tduality,sw}.\footnote{\baselineskip=12pt Note that for open
strings, which do not wind around the cycles of ${\bf T}^D$, the mapping is
linear in the metric and, unlike the closed string metric, there is no
transformation which maps $G\mapsto G^{-1}$.} This similarity holds in the
usual decoupling limit $\alpha'\to0$ and modulo the conjugation we mentioned
above. It is also only true modulo the normalization of the operator trace Tr
(which determines the transformation rule for the Yang-Mills coupling constant
$g$), a point that we shall return to in the next subsection. It is possible to
work out the transformation rules for higher exterior powers of the
noncommutative field strength (see below) and show that they transform in a
spinor representation of $SO(D,D;\zed)$~\cite{ans,schwarzmorita}. This is
because a differential form of even degree can be identified with a bi-spinor
of the rotation group $SO(D)$ in $D$ dimensions, while a spinor of $SO(D,D)$
can be identified with a bi-spinor of its $SO(D)$ subgroup. It simply reflects
the fact that the target space duality group acts on D-brane charges (or more
precisely on Ramond-Ramond potentials) in a spinor representation of the group
$SO(D,D;\zed)$~\cite{Uduality}. Notice, however, that the $T$-duality
transformations along a single direction of ${\bf T}^D$ are absent in the
present formalism (they are in fact elements of $O(D,D;\zed)$), because such a
map takes Type IIA strings to Type IIB strings (and vice versa) and is
therefore not a symmetry of the corresponding gauge theory. Nevertheless, it is
a remarkable feature of noncommutative Yang-Mills theory that a stringy
symmetry such as $T$-duality acts at a field theoretical level, rather than
mixing the noncommutative gauge field modes with string winding states and
other stringy excitations. This makes noncommutative Yang-Mills theory a very
powerful description of the low-energy effective dynamics of strings, in
contrast to ordinary Yang-Mills theory which is not invariant under
$T$-duality~\cite{sw}.}
\item{The Morita transformation has several very interesting special cases. For
$N_0=1$ (so that $N$ corresponds to the dimension of the irreducible twist
eating solution), the non-abelian nature of the gauge theory is completely
absorbed into the noncommutativity $\Theta'$ of spacetime. All the internal
matrix structure of the gauge fields is absorbed by the Weyl operators $\hat
Z'^a$ in this case. This is true even for $\Theta=0$, so that an ordinary
$U(N)$ gauge theory is equivalent to a noncommutative gauge theory with $U(1)$
gauge group. We can therefore transform an ordinary non-abelian gauge theory
into a gauge theory with an abelian gauge group, at the cost of making the
spacetime noncommutative. On the other hand, the original and dual ranks can be
made equal within the present framework only when there is no background,
$Q=0$, in which case $N=N_0$ (see (\ref{3integers})). When $\Theta=0$, we see
in fact that the dual $\Theta'$ is rational-valued, and we find that
noncommutative Yang-Mills theory with rational-valued deformation parameters is
dual to ordinary Yang-Mills theory with 't~Hooft flux. But these dualities
between commutative and noncommutative gauge theories are not the whole story.
The various theories should be properly understood as being members of a
hierarchy of models~\cite{hierarchy}, in which the noncommutative description
is the physically significant one in the infrared regime as a local field
theory of the light degrees of freedom, even though the theory is equivalent by
duality to ordinary Yang-Mills theory. This is due to the extra infrared
degrees of freedom that noncommutative field theories contain, as we discussed
in section 3, and it is evident in the dual supergravity descriptions of
noncommutative Yang-Mills theory~\cite{wilsonsugra}. When $\Theta$ is
irrational-valued, there is no commutative dual. But this remarkable duality
does allow one to interpolate continuously, through noncommutative gauge
theories, between two ordinary Yang-Mills theories with gauge groups of
different rank and appropriate background magnetic fluxes~\cite{hierarchy}.}
\item{It has been shown that the one-loop ultraviolet structure of
noncommutative Yang-Mills theory on ${\bf T}^D$ is the same as that on
$\real^D$~\cite{krajwulk,sheikhren}. Given the duality between commutative and
noncommutative Yang-Mills theories that we have discussed above, we now have a
precise explanation for the equivalence of the one-loop renormalizations in the
two types of theories that were discussed in section~4.3. The reason why the
large $N$ limit is relevant for this equivalence will become clear in section~7
when we examine the Eguchi-Kawai reduction.}
\end{itemize}

\subsubsection{Other Transformation Rules}

We will now briefly describe the transformation properties of the other quantum
numbers of noncommutative Yang-Mills theory under the Morita map. We can
formulate these in a collective form by using the noncommutative Chern
character which was introduced in section 5.2. The new noncommutative
Yang-Mills theory over the dual torus determines a vector bundle $E'$ which may
likewise be classified topologically by the cohomology class
\beq
{\rm ch}_{\Theta'}(E')=\Tr'\otimes\tr^{~}_{N_0}\,\exp
\frac{\weyl[{\cal F}']}{2\pi} \ .
\label{Cherndual}\eeq
There is a simple and elegant relation between (\ref{Cherndual}) and the Chern
character (\ref{chThetamuE}) of the original noncommutative gauge
theory~\cite{schwarzmorita}. For this, we recall that the new curvature
two-form ${\cal F}'$ in (\ref{Cherndual}) is obtained from the original one as
a shift by the constant background flux, i.e. ${\cal F}'=F-f$ (compare
(\ref{SYMmulti}) and (\ref{SYMsingle})). Taking into account the change
(\ref{traceprimes}) in the normalization of the trace, we may then write
\beq
{\rm ch}_{\Theta'}(E')=\frac{N_0}N\,\Bigl|\det(\Theta\,Q'-N')\Bigr|~
\exp\left(-\frac1{2\pi}\,f_{ab}\,\rho^a\rho^b\right)~{\rm ch}_\Theta(E) \ ,
\label{chprimerel}\eeq
where $f_{ab}=\Sigma^i_{~a}\,f_{ij}\,\Sigma^j_{~b}$ is the noncommutative
curvature (\ref{fQrel}) of the corresponding frame bundle.

For example, in two dimensions the formula (\ref{chprimerel}) gives
\beq
{\rm ch}_{\Theta'}(E')=\frac{{\rm gcd}(N,Q)}{N^2}\,\Bigl(N-Q\,\Theta\Bigr)^3
\ ,
\label{chprime2D}\eeq
consistent with the fact that the magnetic flux vanishes in the target theory.
This is also consistent with the way that D-brane charges transform under
$T$-duality (even in the case $\Theta=0$)~\cite{Uduality}. Similar formulas can
be worked out for gauge theories in higher dimensions. For the cases $f'\neq0$
when the target gauge bundle has non-vanishing magnetic flux, the
transformation rule (\ref{fprime}) now follows from (\ref{chThetamuE}) and
(\ref{chprimerel}).

Finally, let us comment on the transformation of observables of noncommutative
Yang-Mills theory~\cite{amns2}, i.e. the noncommutative Wilson lines of section
4.2. In the target theory, where there are no large star-gauge transformations,
the observables ${\cal O}'(C_v)$ associated with an arbitrary oriented contour
$C_v$ can be constructed using the relations (\ref{partransport}),
(\ref{stargaugeobs}) and (\ref{vtorus}), and replacing all un-primed quantities
with primed ones. In the original theory, however, we have to be a bit more
careful because the gauge fields are multi-valued functions on the torus and
transform according to the twisted boundary conditions (\ref{twistedbc}). The
corresponding parallel transport operator (\ref{partransport}) is likewise
multi-valued and obeys the boundary conditions
\beq
{\cal U}(x+\Sigma^i_{~a}\,\hat\imath;C_v)=\Omega_a(x)\star{\cal U}(x;C_v)
\star\Omega_a(x+v)^\dagger \ .
\label{calUtwist}\eeq
To construct a single-valued observable ${\cal O}(C_v)$, we use a path-ordered
star-exponential of the background abelian gauge field (\ref{aifixed}) to
absorb the global gauge transformation in (\ref{calUtwist}). We then arrive at
the observable
\beq
{\cal O}(C_v)=\int d^Dx~\tr^{~}_N\left[\,{\cal U}(x;C_v)\star~{\rm P}~
\exp_\star\left(i\int\limits_{C_v}d\xi^i~a_i(x+\xi)\right)^\dagger\,\right]
\star\e^{ik_i(v)x^i} \ ,
\label{calOmulti}\eeq
which can be shown~\cite{amns2} to be equivalent to those of the target
noncommutative gauge theory under the Morita map by using Weyl quantization and
the change of basis $\hat\Delta(x)\mapsto\hat\Delta'(x')$ of the previous
subsection.

As an explicit example of this equivalence, let us start with a commutative
Yang-Mills theory, $\Theta=0$, with topologically non-trivial gauge fields.
Fixing a loop $C_{\vec n}$ which winds $n^a$ times around the $a$-th cycle of
${\bf T}^D$, the integrand of the observable (\ref{calOmulti}) is then the
usual gauge-invariant Polyakov line
\beq
{\cal P}(x;C_{\vec n})=\tr^{~}_N\left[\,
{\rm P}~\exp\left(i\int\limits_{C_{\vec n}}d\xi^i
{}~A_i(x+\xi)\right)~\prod_{a=1}^D(\Gamma_a)^{n^a}~\e^{2\pi i\,
(\Sigma^{-1})_i^{~a}\,\delta_{ab}\,n^bx^i}\right] \ .
\label{Polyakovline}\eeq
We have used the fact, discussed at the end of subsection 5.3, that in the
commutative case we may set $\alpha=0$ in (\ref{Omegagauge}) and maintain
global gauge invariance by using only the twist eaters $\Gamma_a$. In the
Morita equivalent theory, we have noncommutativity $\Theta'=N'^{-1}B^\top$ and
periods $\Sigma'=\Sigma\,N'$. As discussed above, in this case the complete
matrix structure of the gauge theory may be absorbed into the noncommutativity
of spacetime and the target theory has gauge group $U(1)$. The twist eaters in
(\ref{Polyakovline}) are therefore eaten up by the Morita transformation and
one is left with an open line observable (\ref{stargaugeobs}) with momentum
(\ref{linemomentum}), where the endpoint separation distance vector
$v^i=(N'^{-1})^a_{~b}\,\Sigma'^i_{~a}\,n^b$ in general does not wind around the
cycles of the dual torus. Therefore, we have the equivalence
\beq
{\cal O}'(C_{\vec n})=\int d^Dx~{\cal P}(x;C_{\vec n}) \ ,
\label{calOcalPrel}\eeq
and the Polyakov lines of ordinary Yang-Mills theory map to open noncommutative
Wilson lines under the Morita transformation~\cite{amns2}.

\subsection{Projective Modules}

Within the operator algebraic setting of noncommutative
geometry~\cite{connesbook}, there is a more precise notion of a vector bundle
over a noncommutative space, and along with it a more formal definition of
Morita equivalence. We shall not enter much into this technical definition, but
simply satisfy ourselves here with the stronger notion of Morita equivalence of
noncommutative gauge fields that we developed above. But let us give a brief
indication of the more formal definition, which will be exploited to some
extent in section~8. Consider an algebra ${\cal A}$ of Weyl operators. By a
{\it module} for $\cal A$ we will mean a separable Hilbert space $\cal H$ on
which ${\cal A}$ acts. We will use only right actions of the algebra and denote
them by $\psi\cdot\weyl[f]$ for $\psi\in\hil$ and $\weyl[f]\in\alg$. The action
is required to satisfy the condition
\beq
\Bigl(\psi\cdot\weyl[g]\Bigr)\cdot\weyl[f]=\psi\cdot\Bigl(\weyl[g]\,\weyl[f]
\Bigr)
\label{actionreq}\eeq
for $\psi\in\hil$ and $\weyl[f],\weyl[g]\in\alg$, so that a module generates an
explicit representation of the Weyl operators. For noncommutative algebras
there are also left modules of $\alg$, while in the commutative case there is
no distinction between the two types of actions. The module ${\cal H}$ is said
to be {\it projective} if it can be embedded as a direct summand of a freely
generated module, i.e. if there exists another ${\cal A}$-module $\tilde{\cal
H}$ such that ${\cal H}\oplus\tilde{\cal H}$ is a direct sum of copies of the
algebra ${\cal A}$ itself, completed to a Hilbert space in an appropriate inner
product. The latter space is trivially an $\cal A$-module, with action defined
by $\weyl[g]\cdot\weyl[f]=\weyl[g]\,\weyl[f]$ for $\weyl[f],\weyl[g]\in\alg$.
In this simplest case the defining condition (\ref{actionreq}) of a module is
equivalent to the associative law of the algebra $\alg$.

The space $\Gamma(E)$ of smooth sections of a vector bundle $E$ over a manifold
$M$ is naturally a projective $C(M)$-module, again with action defined by
$s\cdot f=fs$ for $f\in C(M)$ and $s\in\Gamma(E)$. The condition
(\ref{actionreq}) is a trivial consequence of the commutativity of the function
algebras, while Swan's theorem~\cite{connesbook}, i.e. that $E$ is a direct
summand of a trivial bundle, guarantees that this module is projective. In
fact, there is an analog of the Gel'fand-Naimark theorem for vector bundles
known as the {\it Serre-Swan theorem}~\cite{connesbook}, which asserts that
there is a one-to-one correspondence between the category of smooth vector
bundles over a manifold $M$ and the category of finitely generated projective
$C(M)$-modules. Therefore, we may formally define a vector bundle over a
noncommutative space to be a representation space $\cal H$ for its Weyl
operator algebra.

The purpose of this subsection is to describe some of these modules explicitly
in the case of the noncommutative torus. Although the derivation given thus far
in this section is completely independent of any of these representations, as
we have indicated earlier there are many instances in which one would like to
have explicit representations of the Weyl operators. For example, we will see
that these modules arise naturally in Matrix theory and also as the Hilbert
spaces of physical states in open string quantization. This will enable us to
make a more precise identification of the Morita equivalence of noncommutative
Yang-Mills theory with the $T$-duality symmetry of string
theory~\cite{lls1,sw}. Furthermore, we will use this formalism in
section~8 to give a more precise description of noncommutative gauge
transformations in terms of matter fields in the fundamental representation of
the star-gauge group.

A more mathematical reason for wanting to study these modules is that it gives
a more concise definition of Morita equivalence~\cite{connesbook}. A Morita
equivalence of two algebras $\cal A$ and ${\cal A}'$ provides a natural
one-to-one correspondence between their projective modules. Precisely, it
provides {\it equivalence bi-modules} ${\cal M}$ and ${\cal M}'$ for ${\cal
A}\times{\cal A}'$ and ${\cal A}'\times{\cal A}$, respectively. The Hilbert
space $\cal M$ is simultaneously a right $\alg$-module and a left
$\alg'$-module, and vice versa for ${\cal M}'$, with the right and left actions
of $\alg$ and $\alg'$ commuting. Using $\cal M$ one can define a map from right
${\cal A}'$-modules to right $\cal A$-modules by ${\cal H}'\mapsto{\cal
M}\otimes_{{\cal A}'}{\cal H}'$, with inverse map ${\cal H}\mapsto{\cal
M}'\otimes_{\cal A}{\cal H}$. The algebra ${\cal A}'$ is the {\it commutant} of
$\cal A$ in the module $\cal M$, i.e. the set of operators on ${\cal M}$ which
commute with $\cal A$ is precisely ${\cal A}'$, and vice versa in ${\cal M}'$.
Together these algebras of Weyl operators act irreducibly on the Hilbert spaces
${\cal M}$ and ${\cal M}'$. The Morita equivalence derived at the field
theoretical level above then asserts that gauge theory over $\cal A$ in a
certain bi-module $\cal M$ is equivalent to gauge theory over ${\cal A}'$ in
another bi-module ${\cal M}'$. It may be checked explicitly that the Weyl
operators (\ref{Weylbasis}) commute with the dual Weyl basis
(\ref{hatZprime}). We will now give some illustrative examples to see
how this more abstract notion of Morita equivalence works in
practise.

A particularly simple equivalence bi-module is provided by taking $\cal M$ to
be the Hilbert space of square-integrable functions on the torus ${\bf T}^D$.
We may then represent the Weyl algebra (\ref{noncommtorus}) of the
noncommutative torus by the operators
\beq
\hat Z^a=\exp\left[2\pi i\left(\left(\Sigma^{-1}\right)_i^{~a}\,x^i+i\,
\Sigma^i_{~b}\,\Lambda^{ab}\,\frac\partial{\partial x^i}\right)\right]
\label{Weylsquarerep}\eeq
acting on $\cal M$, where $\Lambda$ is any constant, real-valued $D\times D$
matrix whose antisymmetric part is given by
\beq
\Lambda-\Lambda^\top=\Theta \ .
\label{Lambdaantisym}\eeq
In the case of a gauge bundle of vanishing topological charge, $Q=0$, we may
take $B=0$ and $A=N'=\id_D$ in order to satisfy the relation (\ref{ANBQrel}).
{}From (\ref{Thetaprime}) it then follows that the dual noncommutativity
parameter is $\Theta'=-\Theta$, and so the dual Weyl algebra (\ref{Zprimealg})
may be represented on $\cal M$ by taking
\beq
\hat Z'^a=\exp\left[2\pi i\left(\left(\Sigma^{-1}\right)_i^{~a}\,x^i+i\,
\Sigma^i_{~b}\,\Lambda^{ba}\,\frac\partial{\partial x^i}\right)\right] \ .
\label{Weylprimesquarerep}\eeq
It is easy to see that the set of operators (\ref{Weylprimesquarerep}) generate
the commutant of the set (\ref{Weylsquarerep}) in $\cal M$. The appropriate
linear derivations may also be represented on $\cal M$ as
\beq
\hat\partial_i=\hat D_i=\frac\partial{\partial x^i} \ .
\label{derivsquarerep}\eeq

The situation for gauge bundles of non-vanishing magnetic flux is somewhat more
complicated. In this case, a convenient representation of the algebra of the
noncommutative torus is provided by the {\it fundamental sections} of the given
gauge bundle. These are the $\complex^N$ vector-valued functions $\psi(x)$ on
$\real^D$ which transform under large gauge transformations in the fundamental
representation of the star-gauge group,
\beq
\psi(x+\Sigma^j_{~a}\,\hat\jmath)=\Omega_a(x)\star\psi(x) \ .
\label{fundsection}\eeq
We shall now solve the twisted boundary conditions (\ref{fundsection}) and
thereby explicitly construct the module corresponding to the Hilbert space of
sections of the corresponding fundamental bundle. For illustration we will work
only in $D=2$ spacetime dimensions. The equation can be truncated to such a
form always by picking Darboux coordinates in which the noncommutativity
parameter assumes its canonical form (\ref{thetacan}). In addition, we will
work on a square torus ${\bf T}^2$ of unit size, i.e. we take
$\Sigma^j_{~a}=\delta^j_{~a}$, and make the gauge choice
\beq
\Omega_1(x)=\e^{2\pi iqx^2/N}\otimes(W_N)^q~~~~~~,~~~~~~
\Omega_2(x)=1\otimes V_N
\label{T2gaugechoice}\eeq
with relatively prime positive integers $q$ and $N$, where $W_N$ and $V_N$ are
the $SU(N)$ clock and shift operators (\ref{VNWNdef}).

Given the fundamental section $\psi(x)$ obeying (\ref{fundsection}), we
introduce the section
\beq
\Psi(x)\equiv\psi_\mu\Bigl(x-(\mu-1)\,\hat2\Bigr)=\sum_{\nu=1}^N\Bigl[
(V_N)^{-(\mu-1)}\Bigr]_{\mu\nu}\,\psi_\nu(x) \ .
\label{newsection}\eeq
By using the explicit representation (\ref{VNWNdef}) we find that
\beq
\Psi(x)=\psi_1(x)
\label{varphipsi1}\eeq
is independent of the vector index $\mu=1,\dots,N$ of the fundamental sections.
Furthermore, since $(V_N)^N=\id_N$, the field
(\ref{newsection},\ref{varphipsi1}) is a periodic function of period $N$,
\beq
\Psi(x+N\,\hat2)=\psi_\mu\Bigl(x+(N-\mu+1)\,\hat2\Bigr)=\psi_1(x)=\Psi(x) \ .
\label{varphiper}\eeq

As for the other boundary condition in (\ref{fundsection}), by using
(\ref{T2gaugechoice}) and (\ref{VNWNdef}) we find
\beq
\Psi(x+\hat1)=\e^{2\pi iq(x^2+1)/N}\star\Psi(x) \ .
\label{otherbc}\eeq
The general solution to (\ref{otherbc}) may be written as
\beq
\Psi(x)=\NO\exp_\star\left(\frac qN\,\Bigl(x^2+1\Bigr)\,,\,2\pi ix^1
\right)\NO\star\varphi(x) \ ,
\label{gentwistsoln}\eeq
where $\varphi(x)$ is a periodic function, $\varphi(x+\hat1)=\varphi(x)$, and,
for any two functions $f(x)$ and $g(x)$ whose Moyal bracket $f\star g-g\star f$
is a constant function on ${\bf T}^2$, the normal-ordered exponential function
is defined by
\beq
\NO\exp_\star(f,g)\NO=\frac1{1-f\star g+g\star f}\,\sum_{n=0}^\infty
\frac1{n!}\,f^{\star n}\star g^{\star n}
\label{expNO}\eeq
with $f^{\star n}=f\star\cdots\star f$ ($n$ times). The function (\ref{expNO})
reduces to the ordinary exponential function $\e^{fg}$ in the commutative limit
and generically it shares similar properties,
\bea
\NO\exp_\star(f,g)\NO\star\NO\exp_\star(-g,f)\NO&=&1 \ , \nn\\
\NO\exp_\star(f+c,g)\NO&=&\NO\exp_\star(f,g)\NO\star\exp_\star(cg) \ , \nn\\
\NO\exp_\star(f,g+c)\NO&=&\exp_\star(cf)\star\NO\exp_\star(f,g)\NO \ ,
\label{expNOprops}\eea
where $c$ is any constant function on the torus. The periodic field
$\varphi(x)$ may be expanded in a Fourier series
\beq
\varphi(x)=\sum_{n=-\infty}^\infty\e^{2\pi inx^1}\star\varphi_n(x^2) \ ,
\label{varphiFourier}\eeq
and, using the properties (\ref{expNOprops}) of the normal-ordered exponential
function, we arrive at the solution
\beq
\Psi(x)=\sum_{n=-\infty}^\infty\NO\exp_\star\left(\frac qN\,
\Bigl(x^2+1\Bigr)+n\,,\,2\pi ix^1\right)\NO\star\varphi_n(x^2) \ .
\label{Psisoln}\eeq

Let us now rewrite the series (\ref{Psisoln}) in terms of the decomposition
$n=qm+j$ with $m\in\zed$ and $j=1,\dots,q$ as
\beq
\Psi(x)=\sum_{m=-\infty}^\infty~\sum_{j=1}^q\NO\exp_\star\left(\frac qN\,
\Bigl(x^2+1\Bigr)+qm+j\,,\,2\pi ix^1\right)\NO\star\varphi_{m,j}(x^2) \ ,
\label{Psisolnrewrite}\eeq
where $\varphi_{m,j}(x^2)=\varphi_{qm+j}(x^2)$. The periodicity property
(\ref{varphiper}) implies $\varphi_{m-1,j}(x^2+N)=\varphi_{m,j}(x^2)$, and so
by induction it follows that
\beq
\varphi_{m,j}(x^2)=\varphi_{0,j}(x^2+Nm) \ .
\label{varphimj}\eeq
Therefore, by inverting the definition (\ref{newsection}), we arrive finally at
the general expression for the fundamental sections in the form
\bea
\psi_\mu(x)&=&\sum_{m=-\infty}^\infty~\sum_{j=1}^q\NO\exp_\star\left(\frac qN
\,\Bigl(x^2+\mu+Nm\Bigr)+j\,,\,2\pi ix^1\right)\NO\nn\\&&\star\,\chi\left(x^2+
\mu+Nm+\frac Nq\,j\,,\,j\right) \ ,
\label{psimufinal}\eea
where the functions $\chi(s,j)=\varphi_{0,j}(s-1-\frac Nq\,j)$ are defined on
the whole of the domain $\real\times\zed_q$ and are only restricted by the
requirement that they be Schwartz functions of $s\in\real$. They form a basis
of vectors in the Hilbert space of fundamental sections of the given gauge
bundle parametrized by the rank $N$ and magnetic flux $q$.

All operators of the algebra of functions on the noncommutative torus may now
be represented on the basis $\chi(s,j)$. In particular, the actions of the
covariant derivatives (\ref{hatnabla}) and the dual Weyl basis
(\ref{hatZprime}) on the operators $\weyl[\psi]_\mu$ induce their
representations on these basis functions. After some tedious algebra, we arrive
at the explicit representations
\bea
\hat D_1=-ifs~~~~~~&,&~~~~~~\hat D_2=\frac\partial{\partial s} \ ,
\nn\\\hat Z'^1=U^1\otimes(V_q)^a~~~~~~&,&~~~~~~\hat Z'^2=U^2\otimes W_q \ ,
\label{chisjreps}\eea
where the integer $a$ is defined by (\ref{aibidef}). The $q\times q$ shift and
clock matrices in (\ref{chisjreps}) act on the vector indices $j\in\zed_q$ of
the functions $\chi(s,j)$, while the operators $U^a$, $a=1,2$ act as shift and
clock type operators on the continuous indices $s\in\real$ by
\beq
U^1\chi(s,j)=\chi\left(s-\frac1q\,,\,j\right)~~~~~~,~~~~~~
U^2\chi(s,j)=\e^{2\pi is/(N-q\,\Theta)}\,\chi(s,j)
\label{Uaclockshift}\eeq
and thereby generate the algebra
\beq
U^1\,U^2=\e^{-2\pi i/q(N-q\,\Theta)}\,U^2\,U^1 \ .
\label{Uaalg}\eeq
It is straightforward to verify that the operators (\ref{chisjreps}) yield a
representation of the commutation relations
(\ref{Zprimealg},\ref{nablaZprime}). These are the irreducible modules ${\cal
H}_{N,q}$ over the noncommutative torus that were used in~\cite{cds} in the
context of matrix theory compactifications. Other representations corresponding
to the standard form of $T$-duality, mapping a gauge field into the position of
a D-string on the dual noncommutative torus, may also be
constructed~\cite{bpscurv,morzum}. The constant curvature modules over a
four-dimensional noncommutative torus are explicitly constructed
in~\cite{kimkimlee}. In the general case the irreducible modules correspond to
linear spaces of Schwartz functions on $\real^p\times\zed^q\times\Gamma$, where
$2p+q=D$ and $\Gamma$ is a finite abelian group~\cite{cds,rieffelhigher}. Such
representations of the noncommutative torus are known as {\it Heisenberg
modules}. The expression (\ref{psimufinal}) shows that a Heisenberg module may
be regarded as a deformation of the space of sections of a vector bundle over
the ordinary, commutative torus~\cite{bpscurv}.

\subsubsection{String Theoretical Interpretation}

\begin{figure}[htb]
\epsfxsize=3in
\bigskip
\centerline{\epsffile{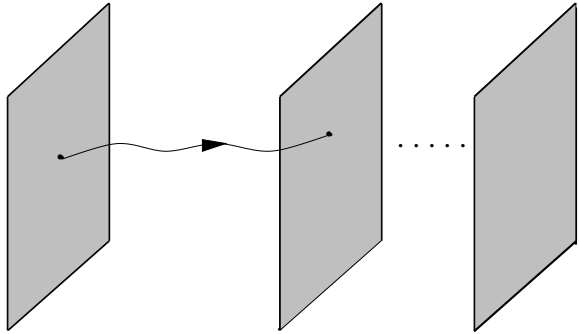}}
\caption{\baselineskip=12pt {\it An open string (wavy line) stretching from a
single D2-brane (shaded region) to a cluster of $N$ D2-branes carrying $q$
units of D0-brane charge. Quantization of the point particle at the left end of
the string produces a Hilbert space $\hil=\hil_{1,0}$ while quantization at the
right end yields $\hil'=\hil_{N,q}$.}}
\bigskip
\label{stringmodules}\end{figure}

The Heisenberg modules described above admit an elegant interpretation in the
context of the quantization of open strings in external
$B$-fields~\cite{cds,sw}. Consider an open string with one endpoint terminating
on a D2-brane, and the other one on a configuration of $N$ coincident D2-branes
with $q$ units of D0-brane charge which is equivalent to $q$ units of magnetic
vortex flux~\cite{douglaswz} (see section~5.2). The situation is depicted
schematically in fig.~\ref{stringmodules}. Let us consider first the simplest
case whereby the open string stretches between a pair of D2-branes. In the
Seiberg-Witten scaling limit considered earlier (see section~1.3), the
topological open string $\sigma$-model action (\ref{SSigmabdry}) in the case
that the worldsheet $\Sigma$ is an infinite strip is given by
\beq
S_B=-\frac i2\,\int dt~B_{ij}\,x^i\,\frac{dx^j}{dt}+\frac i2\,\int dt~B_{ij}\,
\tilde x^i\,\frac{d\tilde x^j}{dt} \ ,
\label{SBstrip}\eeq
where $x^i$ and $\tilde x^i$ denote the values of the string fields at opposite
boundaries of the strip. Canonical quantization of the action (\ref{SBstrip})
yields the quantum commutators (\ref{noncommalg}) and $[\hat{\tilde
x}^i,\hat{\tilde x}^j]=-i\,\theta^{ij}$. However, one has to remember that the
particles described by the configurations $x^i$ and $\tilde x^i$ are connected
together by a string. The contribution to the energy from the bulk kinetic term
$g_{ij}\,\partial_a x^i\,\partial_a x^j$ in (\ref{SSigmabulk}) is minimized by
a string which is a geodesic from $x^i$ to $\tilde x^i$. In the decoupling
$\alpha'\to0$ limit, the fluctuations about this minimum have infinite energy,
and we may thereby identify the classical phase space of the theory
(\ref{SBstrip}) as consisting of a pair of points $x^i$ and $\tilde x^i$ along
with a geodesic line connecting them. This leads to the parametrization
\beq
x^i=y^i+\frac12\,s^i~~~~~~,~~~~~~\tilde x^i=y^i-\frac12\,s^i \ ,
\label{xyspar}\eeq
where $y\in{\bf T}^2$ is the midpoint of the geodesic joining $x$ and $\tilde
x$, and the coordinate $s\in\real^2$ keeps track of how many times the geodesic
wraps around the cycles of the torus.

Canonical quantization is then tantamount to taking $\hat y^i$ to be
multiplication operators by $y^i$ and $\hat s^i$ the canonical momentum
operators
\beq
\hat s^i=i\,\theta^{ij}\,\frac\partial{\partial y^j} \ .
\label{sicanmom}\eeq
The physical Hilbert space ${\cal M}$ of open string ground states thus
consists of functions on an ordinary torus ${\bf T}^2$ with coordinates $y$.
The algebras of functions at the left and right endpoints of the open string
are thereby generated by operators of the type (\ref{Weylsquarerep}) and
(\ref{Weylprimesquarerep}), respectively (c.f. (\ref{Fourierseries})). In the
general case, the map $\Theta\mapsto\Theta'$ in (\ref{Thetaprime}) represents
the Morita equivalence between the Heisenberg modules $\hil_{1,0}$ and
$\hil_{N,q}$, and it coincides with the $T$-duality transformation that maps a
configuration of (D2,D0) brane charges $(1,0)$ to a brane cluster of charges
$(N,q)$. The latter collection of D2-branes supports a $U(N)$ Chan-Paton gauge
bundle with connection of constant curvature $q/N$. The Hilbert space
$\hil_{N,q}$ may then be constructed explicitly by quantizing open strings on
an $N^2$-fold cover of ${\bf T}^2$ that end on a cluster with brane charges
$(1,Nq)$, which can be easily described by the modular transformation $B\mapsto
B+NQ$ of the action in (\ref{SBstrip}) governing the dynamics of $\tilde
x\in{\bf T}^2$, and then orbifolding by the action of the discrete group
$\zed_N\times\zed_N$~\cite{sw}. Generally, the different algebras obtained by
quantizing open strings with different boundary conditions, in the low-energy
limit with fixed open string parameters, are all Morita equivalent~\cite{sw}.
The equivalence bi-modules are generated by the open string tachyon states with
a given boundary condition on the left and another one on the right. That the
actions of a pair of algebras commute with each other in these modules follows
from the fact that they act at opposite ends of the open strings. That they act
irreducibly on the bi-modules follows from the fact that they generate the full
algebra of observables in the quantum mechanics of the open string ground
states, a point we shall return to in section~8. It is for this latter reason
in fact that the modules above are refered to as Heisenberg modules.

\newsection{Matrix Models of Noncommutative Yang-Mills Theory}

As we discussed in section~1, noncommutative gauge theories in string theory
originally appeared through the large $N$ limits of matrix models. There is in
fact a very deep relationship between matrix models and noncommutative
Yang-Mills theory that we shall now spend some time analysing. This will be
particularly useful for our analysis in the next section. Moreover, it
demonstrates the existence of a very natural non-perturbative regularization of
noncommutative gauge theories which has no counterpart in ordinary quantum
field theory. This in itself proves that these models exist as well-defined
field theories, even beyond perturbation theory. Henceforth we shall focus our
attention on the simplest instance of structure group rank $N=1$. That the
ensuing conclusions hold as well without loss of generality for the generic
cases will become clearer in section~8.3. For the remainder of this paper, we
shall also assume for simplicity that the spacetime dimensionality $D=2d$ is
even and that $\theta^{ij}$ is non-degenerate.

\subsection{Twisted Reduced Models}

The remarkable aspect of noncommutative gauge theory that we shall build on in
this section is the fact that the derivatives $\partial_i$ (or
$\hat\partial_i$) can be completely absorbed into the noncommutative gauge
fields. There is no analog of the following manipulation in ordinary,
commutative Yang-Mills theory. For this, we introduce the {\it covariant
coordinates}~\cite{covcoord}
\beq
C_i=\left(\theta^{-1}\right)_{ij}\,x^j+A_i
\label{covcoord}\eeq
which, on using the representation (\ref{xderiv}), are seen to transform
covariantly under the noncommutative gauge transformations (\ref{stargaugetr}),
\beq
C_i~\longmapsto~g\star C_i\star g^\dagger \ .
\label{Citransf}\eeq
In this sense, the operators (\ref{covcoord}) may be thought of as the gauge
covariant momentum operators that one introduces in the quantum mechanics of a
charged particle in a background magnetic field (c.f. section~1.3). Indeed,
they are completely analogous to the covariant derivative operators
(\ref{hatnabla}) introduced in the case of a constant background flux. Their
remarkable property though in the present context is that the entire
noncommutative gauge theory may be expressed in terms of them. The star-gauge
covariant derivatives $\nabla_i$, defined by (\ref{covderiv}), are given as the
star-commutators
\beq
\nabla_if=i\,f\star C_i-i\,C_i\star f \ ,
\label{nablaCi}\eeq
while the field strength tensor (\ref{starfieldstrength}) is the sum
\beq
F_{ij}=-i\,\Bigl(C_i\star C_j-C_j\star C_i\Bigr)+\left(\theta^{-1}
\right)_{ij} \ .
\label{fieldstrengthCi}\eeq
Notice that the operators (\ref{covcoord}) are essentially elements of the {\it
abstract}, deformed algebra of functions, and so from (\ref{fieldstrengthCi})
we see that spacetime derivatives have completely disappeared in this rewriting
from the action (\ref{starYMaction}).

Passing to the Weyl representation $\hat C_i=\weyl[C]_i$, the noncommutative
Yang-Mills action (\ref{WeylYMaction}) may then be written as
\beq
S_{\rm YM}=-\frac1{4g^2}\,\Tr\,\sum_{i\neq j}\left(-i\,\left[\hat C_i\,,\,
\hat C_j\right]+\left(\theta^{-1}\right)_{ij}\right)^2 \ .
\label{YMactionCi}\eeq
The classical vacua of the gauge theory, i.e. the flat noncommutative gauge
fields with $F_{ij}=0$, are in the representation (\ref{YMactionCi})
interpreted as the momenta which obey the commutation relations $[\hat C_i,\hat
C_j]=-i\,(\theta^{-1})_{ij}$. The remarkable feature of the action
(\ref{YMactionCi}) is that it is just an infinite dimensional matrix model
action, as the fields $\hat C_i$ are formally space independent, i.e. it is a
large $N$ version of the matrix model action (\ref{IIBbosaction}). Such a
theory is known as a {\it reduced model}, becomes it formally derives from the
dimensional reduction of gauge theory by taking all fields to be independent of
the spacetime coordinates. In fact, one could start from the action
(\ref{YMactionCi}), expand the infinite matrices $\hat C_i$ as in
(\ref{covcoord}) with a noncommuting background (\ref{noncommalg}), and thereby
{\it derive} noncommutative gauge theory from a large $N$ matrix model. Such a
matrix expansion about a non-trivial background is the way that noncommutative
Yang-Mills theory is obtained from the large $N$ limit of the IIB matrix model
(\ref{IIBbosaction})~\cite{aiikkt}. The spacetime dependence appears from
expanding around a classical vacuum, but initially it is hidden in the
infinitely many degrees of freedom of the large $N$ matrices $\hat C_i$. This
is in fact the basis of the original appearence of noncommutative gauge theory
from string theory~\cite{cds}.

This intimate connection with reduced models is just a special instance of the
Eguchi-Kawai reduction of multi-colour field theories~\cite{EK} which was
argued long ago to reproduce the physics of ordinary, large $N$ Yang-Mills
theory. The addition of the constants $(\theta^{-1})_{ij}$ in
(\ref{YMactionCi}) removes what would otherwise be an infinite constant and
corresponds to a ``twist'' in the reduced model~\cite{twistedEK}. It is
required in order that the reduced model be equivalent to the 't~Hooft
limit~\cite{thooft} of large $N$ quantum field theory on continuum spacetime,
precisely it restores a certain symmetry of the theory that is otherwise broken
in the loop expansion of the model. The fact that noncommutative gauge theories
are deeply connected to matrix models implies some rather surprising aspects of
them and their gauge groups that we shall now proceed to explore.

\subsection{Finite-Dimensional Representations}

The twisted reduced model (\ref{YMactionCi}) is intrinsically
infinite-dimensional, because its classical equations of motion admit solutions
only over an infinite-dimensional Hilbert space, as is usual for
Heisenberg-type commutation relations (see the next section). This simply
indicates that we must specify derivatives for it somehow. We can now ask
whether there exists a finite dimensional, $N\times N$ matrix model version of
(\ref{YMactionCi}) which reproduces noncommutative Yang-Mills theory in the
large $N$ limit, as these are the types of models in string theory from which
one starts from. Of course, any operator representation of noncommutative gauge
theory is formally a matrix model, but we are really seeking a
finite-dimensional version which can regulate the continuum quantum field
theory at a non-perturbative level. Such a matrix model does indeed
exist~\cite{amns1,amns2}. Detailed reviews of these constructions are given
in~\cite{reducedrev}. Related work can also be found in~\cite{reducedother}.

\subsubsection{The Twisted Eguchi-Kawai Model}

A regulated, $N\times N$ matrix model formulation of noncommutative gauge
theory is provided by the twisted Eguchi-Kawai model~\cite{twistedEK}. The
action is
\beq
S_{\rm TEK}=-\frac1{4g^2}\,\sum_{i\neq j}\zeta_{ij}^*\,\tr^{~}_N
\left(V_i\,V_j\,V_i^\dagger\,V_j^\dagger\right) \ ,
\label{STEK}\eeq
where $V_i$, $i=1,\dots,D$, are $N\times N$ unitary matrices and the
$\zed_N$-valued twist factors are given by
\beq
\zeta_{ij}=\e^{2\pi iQ_{ij}/N} \ ,
\label{twists}\eeq
with $Q$ an antisymmetric $D\times D$ integral matrix. This action is the
natural non-perturbative analog of the infinite-dimensional matrix model
(\ref{YMactionCi}). By identifying $V_i=\e^{i\epsilon\hat C_i}$, with
$\epsilon$ a dimensionful lattice spacing, the action (\ref{YMactionCi}) is
obtained in the continuum limit of the twisted Eguchi-Kawai model corresponding
to $\epsilon\to0$, $N\to\infty$, and
\beq
\left(\theta^{-1}\right)_{ij}=\frac{2\pi Q_{ij}}{N\epsilon^2} \ .
\label{latticethetainv}\eeq
Identifications of this sort are of course well-known. It is the basis of
Weyl's finite version of quantum mechanics~\cite{weylbook} which follows from
the simple observation that while the Heisenberg commutation relations do not
admit any finite dimensional representations, their exponentiated form in terms
of unitary operators do in some special instances.

The unitary matrix model (\ref{STEK}) originates from the ordinary, Wilson
lattice gauge theory version~\cite{wilson} of the commutative counterpart of
the torus model with background 't~Hooft flux that we studied in section~5.3.
The {\it commutative} action is given by
\beq
S_{\rm W}=-\frac1{4g^2}\,\sum_x\,\sum_{i\neq j}\tr^{~}_N
\left[U_i(x)\,U_j(x+\epsilon\,\hat\imath)\,U_i(x+\epsilon\,\hat\jmath)^\dagger
\,U_j(x)^\dagger\right] \ ,
\label{Wilsontwisted}\eeq
where the sum over $x$ runs through sites on a periodic hypercubic lattice, and
the gauge fields $U_i(x)$ are $N\times N$ unitary matrices on the links of the
lattice. As in section~5.3, we assume that the gauge fields are multi-valued
functions around the periods of the lattice. They thereby satisfy
exponentiated, discrete versions of the twisted boundary conditions
(\ref{twistedbc}). However, we recall from the discussion at the end of
section~5.3 that in the commutative case we may choose $\alpha=0$ in the
transition functions (\ref{Omegagauge}), so that the $\Omega_i$ are constant
and given by the twist eating solutions $\Gamma_i$. Now let us dimensionally
reduce the action (\ref{Wilsontwisted}) to the point $x=0$. Then, up to an
irrelevant dimensionless volume factor, the reduced model describes a
one-plaquette version of lattice gauge theory with multi-valued gauge fields.
We can use this multi-valuedness to generate the gauge fields at the other
corners of the plaquette from the unitary matrix $U_i\equiv U_i(0)$ via the
twisted boundary conditions
\beq
U_i(\epsilon\,\hat\jmath)=\Gamma_j\,U_i\,\Gamma_j^\dagger \ .
\label{redtwistedbc}\eeq
Substituting (\ref{redtwistedbc}) into the reduced action induced from
(\ref{Wilsontwisted}) yields
\beq
S_{\rm W}^{\rm red}=-\frac1{4g^2}\,\sum_{i\neq j}\tr_N^{~}
\left(U_i\,\Gamma_i\,U_j\,\Gamma_i^\dagger\,\Gamma_j\,U_i^\dagger\,
\Gamma_j^\dagger\,U_j^\dagger\right) \ .
\label{SWred}\eeq
By using the Weyl-'t~Hooft commutation relations (\ref{WeyltHooftalg}) and
defining the unitary matrices
\beq
V_i=U_i\,\Gamma_i \ ,
\label{ViUidef}\eeq
the action (\ref{SWred}) reduces to (\ref{STEK}).

As we mentioned in the previous subsection, the twisted Eguchi-Kawai model was
originally used as a matrix model which is equivalent to Yang-Mills gauge
theory in the large $N$ limit. However, for finite $N$ the model admits another
interpretation, which leads to a complete justification of the result that
noncommutative gauge theory is equivalent to all orders of perturbation theory
to a twisted large $N$ reduced model, namely the IIB matrix model
(\ref{IIBbosaction}) with D-brane backgrounds~\cite{aiikkt}. For this, we will
assume, for simplicity, that the rank $N$ of the unitary matrices is an odd
integer. Note that the action (\ref{STEK}) possesses the global $U(N)$ gauge
symmetry
\beq
V_i~\longmapsto~\Omega\,V_i\,\Omega^\dagger~~~~~~,~~~~~~\Omega\in U(N) \ ,
\label{TEKgaugesym}\eeq
and a $U(1)^D$ center symmetry
\beq
V_i~\longmapsto~\e^{i\alpha_i}\,V_i~~~~~~,~~~~~~\alpha_i\in\real \ .
\label{centersym}\eeq
The vacuum configuration $V_i^{(0)}$ of the theory is given, up to a $U(N)$
gauge transformation (\ref{TEKgaugesym}), by the twist-eating solutions for
$SU(N)$,
\beq
V_i^{(0)}=\Gamma_i \ .
\label{TEKvac}\eeq
We will also restrict the twist matrices and rank to be of the form
\beq
Q_{ij}=2L^{d-1}\,\varepsilon_{ij}~~~~~~,~~~~~~N=L^d \ ,
\label{QijNrestr}\eeq
where $L$ is an odd integer and $\varepsilon$ is the $D\times D$ skew-diagonal
matrix defined by
\beq
\varepsilon=\pmatrix{0&-1\cr1&0\cr}\otimes\id_d \ .
\label{varepsilondef}\eeq
Then, following the general construction of section~6.1, we have $N_\alpha=L$,
$q_\alpha'=1$ for each $\alpha=1,\dots,d$, and $N_0=1$. It follows that the
$SU(N)$ twist eaters are all constructed from $L\times L$ clock and shift
matrices. They therefore satisfy $(\Gamma_i)^L=\id_N$. Note that the constraint
(\ref{aibidef}) may then be satisfied by taking $a_\alpha=0$, $b_\alpha=1$ for
each $\alpha=1,\dots,d$ without loss of generality.

\subsubsection{The Matrix-Field Correspondence}

The relationship between the unitary matrix model (\ref{STEK}) and
noncommutative gauge theory comes about because there is a very natural
finite-dimensional version of the Weyl-Wigner correspondence. Let us introduce
the $N\times N$ unitary, unimodular matrices
\beq
J_k=\prod_{i=1}^D\,(\Gamma_i)^{k_i}\,\prod_{j<i}\e^{\pi i Q_{ij}k_ik_j/N}
\label{Jkdef}\eeq
defined for integer-valued vectors $k$. The phase factor is included in
(\ref{Jkdef}) to symmetrically order the product of twist eaters. Since
$(\Gamma_i)^L=\id_N$, these matrices have the periodicity properties
\beq
J_{L-k}=J_{-k}=J_k^\dagger \ ,
\label{Jkperiod}\eeq
and they obey the algebraic relations
\beq
J_k\,J_q=\prod_{i=1}^D\,\prod_{j=1}^D\e^{\pi ik_iQ_{ij}q_j/N}~J_{k+q} \ .
\label{Jkcommrels}\eeq
The $J_k$'s have the same formal algebraic properties as the plane wave Weyl
basis $\e^{ik_i\hat x^i}$ for the continuum noncommutative field theory on the
torus. The basic operators were defined in (\ref{Weylbasis}) and are the
analogs of the twist eaters $\Gamma_i$ in the present case. Owing to the
property (\ref{Jkperiod}), there are only $N^2$ independent matrices. As we
will see, the integers $k_i$ label momenta on a periodic lattice which are
restricted to a Brillouin zone $k\in\zed_L^D$.

The matrices (\ref{Jkdef}) obey the orthonormality and completeness relations
\bea
\frac1N\,\tr^{~}_N\left(J_k\,J_q^\dagger\right)&=&\delta_{k,q\,({\rm mod}
\,L)} \ , \nn\\
\frac1N\,\sum_{k\in\zeds_L^D}\,\left(J_k\right)_{\mu\nu}\,\left(J_k
\right)_{\lambda\rho}&=&\delta_{\mu\rho}\,\delta_{\nu\lambda} \ .
\label{orthocompleteJk}\eea
They thereby form the {\it Weyl basis} for the linear space $gl(N,\complex)$ of
$N\times N$ complex matrices~\cite{weylbook,ffz}. In particular, the
fluctuation modes $U_i$ in (\ref{ViUidef}) about the classical vacuum
configuration (\ref{TEKvac}) can be expanded as
\beq
U_i=\frac1{N^2}\,\sum_{k\in\zeds_L^D}U_i(k)~J_k~~~~~~,~~~~~~
U_i(k)=N\,\tr^{~}_N\left(U_i\,J_k^\dagger\right) \ ,
\label{Uikdef}\eeq
where the c-numbers $U_i(k)$ may be interpreted as Fourier coefficients for the
expansion of a lattice field on the discrete torus. These $N^2$ momentum space
coefficients now describe the dynamical degrees of freedom in the twisted
Eguchi-Kawai model. The underlying discrete, noncommutative space described by
these matrices is sometimes called a {\it fuzzy torus}.

We can now make a discrete Fourier transformation to define lattice fields on a
discrete torus. In complete analogy with the continuum formalism, we define
these via the $N\times N$ matrix fields
\beq
\Delta(x)=\frac1{N^2}\,\sum_{k\in\zeds_L^D}J_k~\e^{-2\pi ik_ix^i/\ell} \ ,
\label{Deltafinite}\eeq
where
\beq
\ell=\epsilon\,L
\label{elldef}\eeq
is the dimensionful extent of the hypercubic lattice with $N^2=L^D$ sites
$x^i$. Because of the relations (\ref{Jkperiod}), the matrices $\Delta(x)$ are
Hermitian and periodic in $x^i$ with period $\ell$. This means that the
underlying lattice is a discrete torus. Since the algebraic relations satisfied
by the matrices $J_k$ are completely analogous to their continuum counterparts,
the matrices (\ref{Deltafinite}) have the same formal properties as the
continuum ones (\ref{Deltatorus}) and thereby yield an invertible map between
$N\times N$ matrices and lattice fields. In particular, they obey the relations
\bea
\tr^{~}_N\Bigl(J_k\,\Delta(x)\Bigr)&=&\frac1N~\e^{2\pi ik_ix^i/\ell} \ , \nn\\
\frac1N\,\sum_x\,\Delta(x)_{\mu\nu}\,\Delta(x)_{\lambda\rho}&=&\delta_{\mu\rho}
\,\delta_{\nu\lambda} \ , \nn\\\frac1N\,\tr^{~}_N\Bigl(\Delta(x)\,
\Delta(y)\Bigr)&=&N^2\,\delta_{x,y\,({\rm mod}\,\ell)} \ .
\label{Deltafiniterels}\eea
A lattice field may then be associated to the unitary matrix $U_i$ by the
Fourier series
\beq
{\cal U}_i(x)\equiv\frac1{N^2}\,
\sum_{k\in\zeds_L^D}U_i(k)~\e^{2\pi ik_ix^i/\ell}=\frac1N\,
\tr^{~}_N\Bigl(U_i\,\Delta(x)\Bigr) \ .
\label{calUix}\eeq

Because of the relation (\ref{QijNrestr}), the field ${\cal U}_i(x)$, which
depends on $L^D$ spacetime points, describes the same $N^2$ degrees of freedom
as the $N\times N$ unitary matrix $U_i$. In fact, as we will see below, the
matrix trace $\tr^{~}_N$ can be substituted by a summation $\sum_x$ over
lattice points, in complete analogy again with the continuum property
(\ref{Tracedef}). However, while the matrices $J_k$ are unitary as in
(\ref{Jkperiod}), their linear combination in (\ref{Uikdef}) need not be
unitary in general (a linear combination of unitary matrices does not
necessarily stay in the group $U(N)$). Instead, the unitarity condition
$U_i\,U_i^\dagger=U_i^\dagger\,U_i=\id_N$ on the matrices
\beq
U_i=\frac1{N^2}\,\sum_x\,{\cal U}_i(x)\,\Delta(x)
\label{UiDelta}\eeq
reads
\beq
{\cal U}_i(x)\star{\cal U}_i(x)^*={\cal U}_i(x)^*\star{\cal U}_i(x)=1 \ ,
\label{latticestarunitary}\eeq
where the {\it lattice star-product} is defined by
\bea
{\cal F}(x)\star{\cal G}(x)&\equiv&\frac1N\,\tr^{~}_N\Bigl(F\,
G\,\Delta(x)\Bigr)\nn\\&=&\frac1{N^2}\,\sum_y\,\sum_z{\cal F}(x+y)\,
{\cal G}(x+z)~\e^{2i(\theta^{-1})_{ij}y^iz^j}
\label{latticestarproddef}\eeq
with (dimensionful) noncommutativity parameter
\beq
\theta_{ij}=\frac{\epsilon^2L}\pi\,\varepsilon_{ij} \ .
\label{latticetheta}\eeq
The star-product (\ref{latticestarproddef}) reduces to the Fourier integral
kernel representation (\ref{Weylprodcoord}) in the continuum limit
$\epsilon\to0$. It is a proper discretized, finite-dimensional form of the
continuum Groenewold-Moyal star-product, with which it shares the same
algebraic properties (with spacetime integrals replaced by lattice sums).

\subsubsection{Discrete Noncommutative Yang-Mills Theory}

We are finally ready to interpret the twisted Eguchi-Kawai model in terms of
noncommutative gauge theory. For this, we substitute (\ref{ViUidef}) and the
completeness relation
\beq
\frac1{N^2}\,\sum_x\,\Delta(x)=\id_N
\label{completeness}\eeq
into the action (\ref{STEK}) to write
\beq
S_{\rm TEK}=-\frac1{4g^2N^2}\,\sum_x\,\sum_{i\neq j}\tr_N^{~}\left[U_i\,
\Bigl(\Gamma_i\,U_j\,\Gamma_i^\dagger\Bigr)\,\Bigl(\Gamma_j\,U_i^\dagger
\,\Gamma_j^\dagger\Bigr)\,U_j^\dagger\,\Delta(x)\right] \ ,
\label{STEKrewrite}\eeq
where we have used the Weyl-'t~Hooft algebra (\ref{WeyltHooftalg}) to rearrange
the twist eaters in (\ref{STEKrewrite}). The key observation now is that the
matrices $\Gamma_i$ act as lattice shift operators in this picture, i.e. they
are discrete derivatives $\e^{\epsilon\hat\partial_i}$. Using (\ref{Jkdef}),
(\ref{Deltafinite}) and (\ref{WeyltHooftalg}) we may easily compute
\beq
\Gamma_i\,\Delta(x)\,\Gamma_i^\dagger=\Delta(x-\epsilon\,\hat\imath) \ ,
\label{GammaDeltashift}\eeq
from which it follows that shifts of the lattice gauge fields may be
represented as
\beq
{\cal U}_i(x+\epsilon\,\hat\jmath)=\frac1N\,\tr^{~}_N\Bigl(\Gamma_j\,U_i\,
\Gamma_j^\dagger\,\Delta(x)\Bigr) \ .
\label{Uishift}\eeq
Using (\ref{calUix}), (\ref{latticestarproddef}) and (\ref{Uishift}), the
action (\ref{STEKrewrite}) finally becomes
\beq
S_{\rm TEK}=-\frac1{4\lambda^2}\,\sum_x\,\sum_{i\neq j}\,{\cal U}_i(x)\star
{\cal U}_j(x+\epsilon\,\hat\imath)\star{\cal U}_i(x+\epsilon\,\hat\jmath)^*
\star{\cal U}_j(x)^* \ ,
\label{latticeNCYM}\eeq
where
\beq
\lambda=\sqrt{g^2N}
\label{thooftcoupling}\eeq
is the 't~Hooft coupling constant.

Thus the twisted Eguchi-Kawai model (\ref{STEK}) can be rewritten exactly as
the noncommutative $U(1)$ lattice gauge theory (\ref{latticeNCYM}). Using the
matrix-field correspondence established above, we see that the $U(N)$
invariance (\ref{TEKgaugesym}) of the unitary matrix model translates into the
local star-gauge symmetry of the lattice model,
\beq
{\cal U}_i(x)~\longmapsto~g(x)\star{\cal U}_i(x)\star g(x+\epsilon\,
\hat\imath)^* \ ,
\label{latticestarsym}\eeq
where $g(x)$ is a star-unitary lattice field, $g(x)\star g(x)^*=g(x)^*\star
g(x)=1$.  The lattice gauge theory (\ref{latticeNCYM}) reduces to the Wilson
plaquette model (\ref{Wilsontwisted}) in the commutative limit $\theta\to0$. It
is in this sense that the twisted Eguchi-Kawai model can be interpreted as
noncommutative $U(1)$ Yang-Mills theory (on a periodic lattice). In particular,
the matrix model provides a non-perturbative regularization of the field
theory, and all results derived in this setting will be completely rigorous.
For example, the integration measure for the path integral of the twisted
Eguchi-Kawai model is determined in terms of the invariant Haar measures
$[dU_i]$ for the unitary Lie group $U(N)$, which are invariant under the gauge
transformations (\ref{TEKgaugesym}). Using the above correspondence it
determines the Feynman measure for path integration in the noncommutative gauge
theory (\ref{latticeNCYM}) which is invariant under the lattice star-gauge
transformations (\ref{latticestarsym}) as
\beq
\prod_x\,\prod_{i=1}^D\,\Bigl[d\,{\cal U}_i(x)\Bigr]=\prod_{k\in\zeds_L^D}\,
\prod_{i=1}^D\,\Bigl[dU_i(k)\Bigr]=\prod_{i=1}^D\,[dU_i] \ .
\label{latticemeasure}\eeq
The simplicity in writing down the quantum theory here, as compared to the
continuum case, is a consequence of the mapping of the $N\times N$ matrix
degrees of freedom into a lattice of size $N^2=L^D$ and with $U(1)$ fields. In
particular, we can identify the star-gauge symmetry group of the $U(1)$
noncommutative gauge theory (\ref{latticeNCYM}) with the symmetry group
$U(L^d)$ of the unitary matrix model. These facts will all be instrumental in
the analysis of the next section.

We will conclude our discussion of the matrix model formulations of
noncommutative Yang-Mills theories with a number of remarks concerning the
above construction:
\begin{itemize}
\item{There are two sorts of continuum limits that the lattice gauge theory
(\ref{latticeNCYM}) admits. If we take the limit $N=L^d\to\infty$ first for
finite lattice spacing $\epsilon$, followed by the continuum limit
$\epsilon\to0$, then from (\ref{latticetheta}) it follows that
$\theta\to\infty$ and hence only planar Feynman diagrams survive
(c.f.~(\ref{maxplanar})). This is the usual way to reproduce the 't~Hooft limit
of ordinary large $N$ Yang-Mills theory on continuum spacetime from twisted
reduced models~\cite{twistedEK}. Alternatively, we can take the continuum limit
of the reduced model by keeping the noncommutativity $\theta$ fixed in the
correlated limit $N\to\infty$, $\epsilon\to0$, with $L\epsilon^2$ finite. The
extent of the lattice (\ref{elldef}) in this limit is $\ell\sim\sqrt
L=N^{1/D}\to\infty$, and we recover noncommutative gauge theory on flat,
infinite space $\real^D$, as in the large $N$ limit of the IIB matrix
model~\cite{aiikkt}. In both types of large $N$ limits the Yang-Mills coupling
constant $g$ must be tuned to be a function of $\epsilon$, in order that the
't~Hooft coupling constant (\ref{thooftcoupling}) be finite in the limit. The
case of finite $N$ corresponds to the noncommutative version
(\ref{latticeNCYM}) of Wilson lattice gauge theory, which is a non-perturbative
lattice regularization of the continuum noncommutative Yang-Mills theory. We
have thereby formulated a well-defined finite-dimensional matrix model
representation of noncommutative Yang-Mills theory on $\real^D$. Among other
things, this relationship completes the explanation of the remarkable
coincidence of the perturbative beta-functions in the planar commutative and
noncommutative gauge theories that we discussed in section~4.3.}
\item{Given that the twisted Eguchi-Kawai model is derived as the dimensional
reduction of the Wilson lattice gauge theory (\ref{Wilsontwisted}) with
background 't~Hooft flux, it would appear that we have also derived a
relationship between this latter, commutative lattice gauge theory and the
noncommutative lattice gauge theory (\ref{latticeNCYM}) which has single-valued
fields. The two theories are in fact Morita equivalent~\cite{amns2}.
Noncommutative lattice gauge theory is always Morita equivalent to a
commutative lattice gauge theory, because the finite dimensionality of the
representation of the noncommutative algebra of functions on the lattice
necessitates a rational-valued dimensionless deformation parameter $\Theta$.
This establishes that the phenomenon of Morita equivalence of noncommutative
gauge theories holds in and beyond regulated perturbation theory.}
\item{It is possible to generalize the construction of this subsection to
induce a noncommutative $U(r)$ lattice gauge theory with rank $r>1$. For this,
we take the unitary matrices of the twisted Eguchi-Kawai model to live in a
direct product group $U(r)\otimes U(N)$. The trace over the $U(N)$ indices is
treated as before and transformed into a sum over lattice points. The remaining
indices are left unaltered and become the non-abelian colour indices of the
resulting field theory. This is tantamount to choosing a more general
background flux $Q$ for which $N_0>1$, as we did in the derivation of Morita
equivalence in section~6.}
\item{The noncommutativity parameter (\ref{latticetheta}) can be written as
$\theta=\ell\epsilon/\pi$, so that finite noncommutativity requires keeping the
quantity $\ell\epsilon$ fixed in the continuum limit. In fact, finite
noncommutativity in the lattice formulation necessarily implies a finite size
$\ell=\pi\theta/\epsilon$ of the spacetime~\cite{amns2}. As the lattice spacing
$\epsilon$ is an ultraviolet cutoff for the dynamics of the field theory and
$\ell$ serves as an infrared cutoff, this is just a non-perturbative
manifestation of the UV/IR mixing phenomenon in noncommutative quantum field
theory that we unravelled in section~3.3. It is very explicitly evident in the
discrete formalism that the two limits $\ell\to\infty$ (giving the
noncommutative planar limit) and $\epsilon\to0$ (giving the commutative limit)
do not commute. This gives a very direct interpretation of this novel property
of noncommutative field theories, which in the present case occurs at a {\it
kinematical} level. The reason for this is the very drastic regularization
provided by the matrix model.}
\item{It is possible to modify the above construction and arrive at a continuum
gauge theory on a noncommutative {\it torus}. Within the present framework this
is not possible, because from (\ref{elldef}) and (\ref{latticetheta}) it
follows that it is not possible to take a large $N$ continuum limit which keeps
both the size and noncommutativity of the spacetime finite. One can, however,
repeat the above construction by introducing two more integers $n$ and $m$ with
$L=nm$, and modifying the map (\ref{Deltafinite}) to~\cite{amns1}
\beq
\Delta_n(x)=\frac1{N^2}\,\sum_{k\in\zeds_m^D}\left(J_k\right)^n~
\e^{-2\pi ik_ix^i/\epsilon m} \ .
\label{Deltan}\eeq
The $N\times N$ matrix-valued lattice fields (\ref{Deltan}) provide a
one-to-one Weyl-Wigner correspondence between lattice fields and elements of
$gl(N,\complex)$ which commute with the matrices
\beq
\Omega_i=\prod_{j=1}^D\left(\Gamma_j\right)^{m\varepsilon_{ij}} \ .
\label{Omegai}\eeq
The resulting noncommutative lattice gauge theories thereby follow from the
{\it constrained} twisted Eguchi-Kawai model (\ref{STEK}) obtained by
restricting its unitary matrices (\ref{ViUidef}) to those which obey the
constraints $U_i\,\Omega_j=\Omega_j\,U_i$ for each $i,j=1,\dots,D$. Since
$(\Gamma_i)^L=\id_N$, for $n=1$ the matrices (\ref{Omegai}) are trivial and we
recover the previous construction, with $\Delta(x)=\Delta_1(x)$. For $n>1$ the
dimensionless noncommutativity parameter is $\Theta=n/m$ and it is possible
keep the noncommutativity of spacetime finite as $N\to\infty$ even for finite
extent $\epsilon m$. Thus the resulting continuum noncommutative field theory
lives on a torus. The geometrical meaning of the constraints is that they
enforce the compactification of the matrices of the unitary matrix model on a
$d$-dimensional torus~\cite{amns1}. Indeed, they are just equivalent to unitary
versions of the {\it quotient conditions}~\cite{cds} for toroidal
compactifications of the IIB matrix model (\ref{IIBbosaction}). Unlike this
Hermitian matrix model, the quotient conditions for the unitary matrix model
admit finite-dimensional solutions. The resulting solutions also have a natural
interpretation in terms of Morita equivalences of noncommutative
tori~\cite{amns1}. It is also possible to view such correspondences between
matrix models and lattice field theories by using only Morita
equivalence~\cite{amns2}, without the quotient conditions, but with more
complicated twist matrices $Q$. Such generalizations also allow the
construction of noncommutative field theories with the most general deformation
parameters $\theta_{ij}$. We should point out, however, that the noncommutative
lattice gauge theories which originate from twisted reduced models are not the
only ones that can be constructed~\cite{amns2}.}
\end{itemize}

\newsection{Geometry and Topology of Star-Gauge Transformations}

In this final section we will take a look at the structure of the group of
star-gauge transformations in noncommutative Yang-Mills theory. Gauge
symmetries in the noncommutative case are very different from their commutative
counterparts, because they involve an intriguing mixing between spacetime and
internal, $U(N)$ symmetries. This mixing was responsible for the duality that
we described in the previous section, in that the spacetime degrees of freedom
were able to absorb some of the colour degrees of freedom of the gauge fields.
It is evident in fact immediately from the operatorial form of the
noncommutative Yang-Mills action (\ref{WeylYMaction}), which shows how the
spacetime and $U(N)$ traces are interlocked and cannot be separated from one
another. We recall that for this reason it was difficult to construct local,
star-gauge invariant observables.  Another aspect of noncommutative gauge
theory which is intimately tied to the mixing between spacetime and gauge
degrees of freedom was its connection with matrix models, which followed from
the observation that all derivatives, and hence all spacetime dependence, can
be completely absorbed into the noncommutative gauge fields. This is
particularly transparent in the discrete representations of the previous
section, whereby the $N\times N$ matrix degrees of freedom are in a one-to-one
correspondence with $N^2$ spacetime lattice points.

In the remainder of this paper we will pose a very elementary question: What is
the gauge symmetry group of noncommutative Yang-Mills theory on flat infinite
space? In our attempt to formulate an answer to this question, we will be
guided by two main themes, with the aim of clarifying the structure of the
local and global star-gauge symmetry group:
\begin{itemize}
\item{$\underline{\rm Geometry:}$ We have seen in section~4.2 that spacetime
translations can be regarded to a certain extent (to be discussed below) as
star-gauge transformations. The only other theory with such a property is
general relativity. This feature has been used to suggest~\cite{grsuggest} that
general coordinate transformations may be realized as genuine gauge symmetries
of noncommutative Yang-Mills theory. In fact, via certain dimensional reduction
techniques~\cite{langmannsz}, the translational symmetry can be gauged to
induce a field theory which contains as special limits some gauge models of
gravitation. The most compelling evidence however has been via the
strong-coupling dual supergravity description of maximally supersymmetric
noncommutative Yang-Mills theory in four dimensions, in which it is possible to
identify the Newtonian gravitational force law~\cite{iikkgrav}. Other
indications include the observation that the unitary group of a closed string
vertex operator algebra contains generic reparametrizations of the spacetime
coordinates~\cite{ls1,lls1} (see~\cite{ls2} for an explicit description of this
group), the identification of the one-loop long-ranged potential particular to
noncommutative Yang-Mills theory with the gravitational interaction in Type IIB
superstring theory~\cite{aiikkt,ikkiib}, and the couplings of noncommutative
gauge fields to massless closed string modes in flat
space~\cite{sugracoupling}. However, we will see in this section that this
assertion is incorrect, in that noncommutative gauge transformations can only
realize a certain subgroup of the diffeomorphism group of
spacetime~\cite{lsz}.}
\item{$\underline{\rm Topology:}$ We are also interested in global properties
of noncommutative gauge theories, and the global gauge group is an important
object for the topological classification of solitons, among other things. The
star-gauge group is infinite-dimensional and has been identified previously as
both the infinite unitary group $U(\infty)$ and also as the group $U(\hil)$ of
unitary operators on a separable Hilbert space $\hil$. These two groups are
very different, and in fact both of these proposals for the star-gauge group
are incorrect. The former group can never be identified with a space of
functions (but rather only completions of it can), while the latter group is,
as we will discuss, contractible and so it doesn't possess the interesting
topological characteristics that its commutative counterpart has which leads to
effects like anomalies and topological solitons.}
\end{itemize}

In this section we will clarify some of these misconceptions, and also
illustrate some very precise mathematical aspects of the star-gauge symmetry
group of noncommutative Yang-Mills theory. This will bring out some more of the
deeper operator-algebraic formalism of noncommutative geometry, already
unleashed at the end of section~6. Throughout this section we will denote the
algebra of Schwartz functions on $\real^D\to\complex$ equipped with the
star-product (\ref{starproddef}) by $\alg_\theta$. Its commutative limit of
ordinary functions will be denoted $\alg_0=C(\real^D)$. The representation of
the algebra $\alg_\theta$ by operators on a Hilbert space $\hil$ will be
denoted $\alg_\theta(\hil)\subset{\rm End}(\hil)$. The treatment of
this section will mostly employ an operator formalism with Heisenberg
commutation relations, mostly ignoring the star product
structure. However, within the framework of deformation quantization,
the structure of star-gauge symmetries is rather well understood
through the formalism of deformed vector bundles~\cite{BurWald}, along
the lines of what we described in section~6.3. Along these lines an
important structure which arises is a cocycle condition on the gauge
transformations, which is treated implicitly in the operator
framework. While the following material is somewhat more mathematical
than that of previous sections, it introduces some more fundamental
techniques and ideas of noncommutative geometry. Most of the material
of this section follows closely~\cite{lsz}, where more technical
details may be found.

\subsection{Star-Gauge Symmetries Revisited}

We will begin by defining more precisely the notion of gauge symmetry in the
noncommuting setting, and will proceed throughout this subsection in a somewhat
abstract fashion. In the commutative case, gauge fields arise through covariant
derivatives which specify parallel transport along the fibers of a given vector
bundle over $\real^D$. In turn, as we saw in section~6.3, a vector bundle is
completely characterized by its (Hilbert) space of sections. The commutative
algebra $\alg_0$ acts naturally on this space, so that the sections form an
$\alg_0$-module. Thus the noncommutative analog of a vector bundle over
noncommutative $\real^D$ is an $\alg_\theta$-module $\hil$, whose vectors
$\psi$ will be interpreted as fundamental matter fields, i.e. fields which
transform under the fundamental representation of the gauge symmetry group.
These representations will be particularly important for the explicit
identification of the star-gauge symmetry group. We then seek a covariant
derivative $\nabla_i$ such that $\nabla_i\psi$ is a matter field in the same
representation as $\psi$. This is the algebraic version of the parallel
transport condition.

A canonical choice of projective module is the Hilbert space of
square-integrable fundamental matter fields,
\beq
\hil_{\rm m}=L^2(\real^D)\otimes\complex^N \ .
\label{hilmdef}\eeq
{}From a geometric standpoint and also to analyse properly the gauge
symmetries, we need to investigate the reducibility of this representation of
the algebra. To this end, let us begin by analysing the commutative case. We
may define an action of $\alg_0$ on $\hil_{\rm m}$ by
\beq
\psi\cdot f=f\,\psi
\label{commhilmaction}\eeq
with $f\in\alg_0$ and $\psi\in\hil_{\rm m}$. The defining condition of a
module, $(\psi\cdot g)\cdot f=\psi\cdot(f\,g)$, is a trivial consequence of the
commutativity of pointwise multiplication of functions. We can then decompose
the space (\ref{hilmdef}) into irreducible components with respect to this
action,
\beq
\hil_{\rm m}=\int\limits_x\!\!\!\!\!\!\!~\!\!\ominus~\delta_x
\otimes\complex^N \ ,
\label{hilmdecomp}\eeq
where $\delta_x:\alg_0\to\complex$ is the evaluation functional at
$x\in\real^D$ defined by
\beq
\delta_x(f)=\int d^Dy~\delta^D(x-y)\,f(y) \ .
\label{deltaxf}\eeq
By approximating the delta-function by functions of $\alg_0$, we may view the
functional $\delta_x(f)=f(x)$ as a character of the algebra $\alg_0$, and also
as a one-dimensional unitary irreducible representation of $\alg_0$ on
$\hil_{\rm m}$ via pointwise multiplication of functions,
\beq
\delta_x(f)\cdot\psi=f(x)\,\psi \ .
\label{deltaxfpsi}\eeq
We see that the points $x\in\real^D$ are formally ``reconstructed'' from the
unitary irreducible representations (or equivalently the characters) of the
commutative algebra $\alg_0$. As discussed at the beginning of section~6, this
is the geometric basis of the Gel'fand-Naimark theorem and the association of
topological spaces to commutative $C^*$-algebras.

Let us now consider the noncommutative case. We may define a right action of
$\alg_\theta$ on $\hil_{\rm m}$ by
\beq
\psi\cdot f=\psi\star f \ .
\label{nchilmaction}\eeq
The requisite condition $(\psi\cdot g)\cdot f=\psi\cdot(g\star f)$ follows from
associativity of the star-product. Such fields $\psi$ are, as we will soon see,
naturally interpreted as fundamental matter fields. For a left action
$f\cdot\psi=f\star\psi$, the $\psi$'s would instead be thought of as
anti-fundamental matter fields. Again, this action defines a reducible
representation. To see this, let us rotate coordinates to the Darboux basis
(\ref{thetacan}), in which the coordinate operators $\hat x^i$ split into $d$
mutually commuting blocks in each of which the Heisenberg commutation relations
\beq
\left[\hat x^{2\alpha-1}\,,\,\hat x^{2\alpha}\right]=i\,\vartheta_\alpha
{}~~~~~~,~~~~~~\alpha=1,\dots,d
\label{Heisenberg}\eeq
hold. By the Stone-von~Neumann theorem~\cite{rieffelstone}, the Lie algebra
(\ref{Heisenberg}) has a {\it unique} irreducible representation, the Hilbert
space of quantum mechanics, i.e. the Schr\"odinger representation $\hil_{\rm
q}=L^2(\real^d)$ seen as functions of the coordinates $x^{2\alpha}$,
$\alpha=1,\dots,d$. From this fact it is evident that the Hilbert space
(\ref{hilmdef}) is reducible.

The Schr\"odinger $\alg_\theta$-module $\hil_{\rm q}$ is a separable Hilbert
space, i.e. it is countably infinite-dimensional, because it can be expressed
in terms of the usual Fock space of creation and annihilation operators.
Mathematically, this is the completion to the space of square-summable
sequences
\beq
\hil_{\rm q}\cong\ell^2(\zed_+^d)=\overline{
\bigoplus_{\vec n\in\zeds_+^d}\complex\,|\vec n\,\rangle} \ ,
\label{Fockspace}\eeq
where $|\vec n\,\rangle=|n_1,\dots,n_d\rangle$ is a multi-particle state, and
the Fock space creation and annihilation operators are defined by
\beq
\hat c_\alpha=\frac1{\sqrt{2|\vartheta_\alpha|}}\,\left(\hat x^1+i\,{\rm sgn}
(\vartheta_\alpha)\,\hat x^2
\right)~~~~~~,~~~~~~\hat c_\alpha^\dagger=\frac1{\sqrt{2|\vartheta_\alpha|}}\,
\left(\hat x^1-i\,{\rm sgn}(\vartheta_\alpha)\,\hat x^2\right)
\label{cranops}\eeq
with the non-vanishing commutation relations
\beq
\left[\hat c_\alpha\,,\,\hat c_\beta^\dagger\right]=\delta_{\alpha\beta} \ .
\label{crancommrels}\eeq
The vectors $|\vec n\,\rangle$ are then the simultaneous orthonormal
eigenstates of the $d$ number operators $\hat n_\alpha=\hat
c_\alpha^\dagger\hat c_\alpha$ with eigenvalue $n_\alpha\in\zed_+$, $\hat
n_\alpha|\vec n\,\rangle=n_\alpha|\vec n\,\rangle$, and the actions of the
operators (\ref{cranops}) on this basis are defined by
\beq
\hat c_\alpha|\vec n\,\rangle=\sqrt{n_\alpha}\,|\vec n-1_\alpha\rangle
{}~~~~~~,~~~~~~\hat c_\alpha^\dagger|\vec n\,\rangle=\sqrt{n_\alpha+1}\,
|\vec n+1_\alpha\rangle \ ,
\label{cranaction}\eeq
with $1_\alpha$ the integer vector whose components are
$(1_\alpha)_\beta=\delta_{\alpha\beta}$. The Hilbert space (\ref{Fockspace}) is
projective as a right $\alg_\theta$-module. To see this, we use (\ref{cranops})
to expand the Weyl operators (\ref{Weylopdef}) over the Fock space
(\ref{Fockspace}), and for each fixed integer vector $\vec n_0\in\zed_+^d$
consider the operator $\hat p_{\vec n_0}=|\vec n_0\rangle\langle\vec n_0|$. It
is the orthogonal projection onto the one-dimensional subspace of $\hil_{\rm
q}$ spanned by the vector $|\vec n_0\rangle$. In the Weyl representation of the
trivial rank-$N$ $\alg_\theta$-module $\hil_{\rm m}$, we may write the
orthogonal decomposition $\hil_{\rm m}=\hat p_{\vec n_0}\hil_{\rm
m}\oplus(\id_{\hil_{\rm q}}-\hat p_{\vec n_0})\hil_{\rm m}$. Under the
correspondence $\langle\vec n|\leftrightarrow|\vec n_0\rangle\langle\vec n|$,
we have the natural isomorphism $\hat p_{\vec n_0}\hil_{\rm m}\cong\hil_{\rm
q}$ as right $\alg_\theta$-modules, and hence $\hil_{\rm q}$ is projective.

By the stronger Mackey form of the Stone-von~Neumann
theorem~\cite{rieffelstone} it follows that any $\alg_\theta$-module $\hil$ is
a direct sum of Fock modules. In particular, by iterating the preceding
arguments it follows that the Hilbert space (\ref{hilmdef}) contains infinitely
many copies of the Schr\"odinger representation, so that as
$\alg_\theta$-modules there is a natural isomorphism
\beq
\hil_{\rm m}=\bigoplus_{n=0}^\infty\hil_{\rm q}\otimes\complex^N \ .
\label{hilmhilqrel}\eeq
In this sense, Fock space $\hil_{\rm q}$ is the analog of a single point in the
noncommutative space. It also follows that any fundamental matter field may be
expanded with respect to this decomposition as
\beq
\weyl[\psi]=\sum_{\vec n\in\zeds_+^d}\,\left(\sum_{\vec m\in\zeds_+^d}
\psi_{\vec n,\vec m}\,|\vec m\rangle\right)\langle\vec n\,| \ ,
\label{psidecompFock}\eeq
where $\psi_{\vec n,\vec m}\in\complex^N$ and the states $\langle\vec n\,|$
label the points (Fock representations) on the noncommutative space. Note that
the superposition $\sum_{\vec m}\psi_{\vec n,\vec m}\,|\vec m\rangle$ is an
element of $\hil_{\rm q}$ and so is the analog of a field with support at only
one point in spacetime. The field (\ref{psidecompFock}) is in fact an element
of the Weyl representation of the algebra $\alg_\theta$ on the Hilbert space
(\ref{Fockspace}), defined by $\psi\cdot f=\weyl[\psi]\,\weyl[f]$. This simply
means that we are working in the defining (or fundamental) representation of
$\alg_\theta$ and it is completely analogous to the commutative case
(\ref{hilmdecomp},\ref{deltaxf}). This superposition carries information about
the infinite dimensional unitary gauge symmetry represented on the module
$\hil_{\rm q}$. Notice also that the irreducibility of the Fock module
automatically implies that all of the algebras $\alg_\theta$ for different
deformation parameters $\theta$ are Morita equivalent. This is in marked
contrast to the noncommutative torus, which possesses a non-trivial topological
structure.

In analogy to the commutative case, we may then consider the gauge
transformations
\beq
\psi(x)~\longmapsto~g(x)\star\psi(x) \ .
\label{psigaugetransf}\eeq
It commutes with the right action of $\alg_\theta$ on $\hil_{\rm m}$ and
thereby preserves the representation of the algebra. It also preserves the
$L^2$-norm of the matter field $\psi$ provided that the gauge function $g(x)$
satisfies the star-unitarity condition (\ref{starunitary}). This implies that
$g$ is an element of the group $U(N,\alg_\theta)=U(\mat(N,\alg_\theta))$ of
unitary elements of the algebra
$\mat(N,\alg_\theta)=\alg_\theta\otimes\mat(N,\complex)$ of $N\times N$
matrices with entries in the algebra $\alg_\theta$. Strictly speaking, however,
the algebra $\alg_\theta$ of Schwartz functions has no unit and so it is
necessary to define unitary elements of the algebra $\alg_\theta\oplus\complex$
obtained from $\alg_\theta$ by adjoining an identity element. Geometrically,
this extension corresponds to studying functions on the topological (but not
metric) one-point compactification of $\real^D$. While geometrically such a
compactification can have dramatic effects on the topological properties of the
field theory, it is perfectly harmless at the algebraic level. We shall always
implicitly assume such a unital extension, and discuss some of its properties
further in section~8.4. The group $U(N,\alg_\theta)$ is the {\it gauge symmetry
group} that we shall study in this section.

We now introduce a covariant derivative as the anti-Hermitian operator
$\nabla_i:\hil_{\rm m}\to\hil_{\rm m}$ defined by
\beq
\nabla_i(\psi)=\partial_i\psi-i\,A_i\star\psi \ ,
\label{nablaipsi}\eeq
where $A_i$ is a Hermitian element of the algebra $\mat(N,\alg_\theta)$, i.e. a
gauge field. This operator has the properties we need. Since the derivative
$\partial_i$ satisfies the Leibnitz rule with respect to the star-product,
\beq
\partial_i(f\star g)=(\partial_if)\star g+f\star(\partial_ig) \ ,
\label{starLeibnitz}\eeq
it follows that $\nabla_i$ satisfies a right Leibnitz rule,
\beq
\nabla_i(\psi\star f)=\psi\star(\partial_if)+\nabla_i(\psi)\star f \ .
\label{nablaiLeibnitz}\eeq
This ensures that $\nabla_i(\psi)$ lies in the same representation of the
algebra $\alg_\theta$ as the matter field $\psi$, as desired. In particular,
$\nabla_i(\psi)$ should transform in the same way as $\psi$ under the gauge
transformations (\ref{psigaugetransf}), which fixes the gauge transformation
rule $\nabla_i\mapsto\nabla_i^g$, where
\beq
\nabla_i^g(\psi)=g\star\nabla_i\left(g^\dagger\star\psi\right) \ .
\label{nablaigdef}\eeq
By using (\ref{nablaipsi}) we then find that the covariant transformation law
(\ref{nablaigdef}) is equivalent to the usual star-gauge transformation law
(\ref{stargaugetr}), and also that the noncommutative field strength tensor
(\ref{starfieldstrength}) is given by the star-commutator
\beq
F_{ij}=i\,\nabla_i\star\nabla_j-i\,\nabla_j\star\nabla_i \ .
\label{Fijnabla}\eeq
It follows that the curvature tensor $F_{ij}\in\mat(N,\alg_\theta)$ commutes
with the action of $\alg_\theta$ on the Hilbert space $\hil_{\rm m}$, and so it
lies in the corresponding commutant of the algebra representation, i.e.
$F_{ij}\in{\rm End}_{\alg_\theta}(\hil_{\rm m})$.

The derivation presented in this subsection thereby brings us back to the
models of noncommutative Yang-Mills theory that were described in section~4. In
particular, the Weyl representation (\ref{WeylYMaction}) gives a rewriting of
noncommutative gauge theory as {\it ordinary} Yang-Mills theory (on a
noncommutative space) with local fields and with the extended, infinite
dimensional gauge symmetry group $U(N,\alg_\theta)$. This point of view has
proven fruitful for analysing the renormalization properties of noncommutative
Yang-Mills theory~\cite{sheikhren}, and it shows rather explicitly the
transmutation of $U(N)$ colour degrees of freedom into spacetime degrees of
freedom along the noncommutative directions. We remark also that the
fundamental matter fields $\psi(x)$ induce {\it local} star-gauge invariant
observables of noncommutative gauge theory through the density operators
$\rho(x)=\psi(x)^\dagger\star\psi(x)$.

\subsection{Inner Automorphisms}

The discussion of the previous subsection emphasizes, among other things, the
point that gauge transformations correspond to the {\it inner automorphisms}
$f\mapsto g\star f\star g^\dagger$ of the algebra $\mat(N,\alg_\theta)$. These
transformations form the group
\beq
{\rm Inn}(N,\alg_\theta)=\Bigl\{\,\imath_g~\Bigm|~\imath_g(f)=g\star f\star
g^\dagger~,~f\in\mat(N,\alg_\theta)~,~g\in U(N,\alg_\theta)\Bigr\} \ .
\label{InnNalg}\eeq
They rotate the algebra elements and correspond to internal fluctuations of the
spacetime geometry in a sense that we will now describe. In general, the group
(\ref{InnNalg}) is a proper, normal subgroup of the automorphism group ${\rm
Aut}(N,\alg_\theta)$, the group of transformations which preserve the algebra
$\mat(N,\alg_\theta)$. The remaining automorphisms are called {\it outer
automorphisms} and together they form an exact sequence of groups,
\beq
\id^{~}_{\mats(N,\alg_\theta)}~\longrightarrow~{\rm Inn}(N,\alg_\theta)~
\longrightarrow~{\rm Aut}(N,\alg_\theta)~\longrightarrow~
{\rm Out}(N,\alg_\theta)~\longrightarrow~\id^{~}_{\mats(N,\alg_\theta)} \ .
\label{exactseq}\eeq
Equivalently, the group ${\rm Aut}(N,\alg_\theta)$ is the semi-direct product
of ${\rm Inn}(N,\alg_\theta)$ by the natural action of ${\rm
Out}(N,\alg_\theta)$ on the elements $\imath_g\in{\rm Inn}(N,\alg_\theta)$.

To get some feel for this somewhat abstract characterization, let us again turn
to the commutative limit $\alg_0$. Then ${\rm Inn}(N,\alg_0)$ is the group of
$U(N)$ gauge transformations on $\real^D\to U(N)$, while ${\rm Out}(N,\alg_0)$
is naturally isomorphic to the group ${\rm Diff}(\real^D)$ of diffeomorphisms
of $\real^D$~\cite{connesdiffeo}. Given a smooth function
$\phi:\real^D\to\real^D$, there is a natural automorphism
$\alpha_\phi:\alg_0\to\alg_0$ defined by
\beq
\alpha_\phi(f)=f\circ\phi^{-1}~~~~~~,~~~~~~f\in\alg_0 \ .
\label{autodiffeo}\eeq
If we now represent, as in the previous subsection, the algebra $\alg_0$ on the
Hilbert space $\hil_{\rm m}$ of fundamental matter fields, then evidently all
inner automorphisms are given via conjugation by unitary operators on
$\hil_{\rm m}$. The same property is in fact true of the outer automorphisms.
Given $\phi\in{\rm Diff}(\real^D)$, we may define a unitary operator $\hat
g_\phi$ on $\hil_{\rm m}$ by
\beq
\hat g_\phi\,\psi(x)=\left|\frac{\partial\phi}{\partial x}\right|^{1/2}\,
\psi\left(\phi^{-1}x\right) \ .
\label{gphipsi}\eeq
Thus, in the commutative case the group ${\rm Aut}(N,\alg_0)$ may be modelled
on the group $U(\hil_{\rm m})$ of unitary endomorphisms of the Hilbert space
(\ref{hilmdef}), and the unitary group of $\alg_0$ coincides with the ordinary
$N\times N$ unitary group $U(N)$ of Yang-Mills theory.

In the noncommutative case $\theta\neq0$, we will soon see that it is also true
that the automorphism group ${\rm Aut}(N,\alg_\theta)$ is related to the group
$U(\hil)$ of unitary operators on some Hilbert space $\hil$. However, $U(\hil)$
is not the right candidate for the gauge symmetry group of noncommutative
Yang-Mills theory. The problem is that for any {\it separable} Hilbert space
$\hil$ (one with a countably infinite basis), it is a fundamental
fact, known as Kuiper's theorem~\cite{kuiper}, that $U(\hil)$ is
contractible, i.e. as a manifold, all closed loops on $U(\hil)$ can be
continuously contracted to a point. In particular, all of its homotopy
groups are trivial,
\beq
\pi_n\Bigl(U(\hil)\Bigr)=0 \ ,
\label{pinUhil}\eeq
and we would thereby lose all of the nice topology residing in noncommutative
gauge theory and the ensuing topological configurations like solitons,
instantons and D-branes, to name but a few. Furthermore, many such topological
quantities should be stable under algebra deformations, i.e. they should be
preserved in the commutative limit. The lesson to be learned here is that not
all automorphisms of the algebra (or unitary endomorphisms of a Hilbert space)
generate gauge transformations, but rather only the inner automorphisms do. We
shall soon see what the appropriate gauge group is. For the remainder of this
subsection we will describe some more basic aspects of the group of (inner)
automorphisms of the noncommutative algebra $\alg_\theta$.

\subsubsection{The Tomita Involution}

A lot of what we have described in this section thus far has been based on the
representation of $\alg_\theta$ on some Hilbert space $\hil$, and indeed this
will be important for the remainder of our discussion. The automorphism group
${\rm Aut}(N,\alg_\theta)$ may then be computed via its lift to this Hilbert
space as~\cite{lsz,connesdiffeo}
\beq
{\rm Aut}_\hil(N,\alg_\theta)=\left\{\,\hat g\in U(\hil)~\left|~\hat g\,J=J
\,\hat g~,~\imath_{\hat g}\in{\rm Aut}\Bigl(N\,,\,\alg_\theta(\hil)\Bigr)
\right.\right\} \ .
\label{Authilalg}\eeq
The operator $J$ is called the {\it Tomita involution} and it induces a
bi-module structure for the given representation of the algebra $\alg_\theta$.
It is inserted in the definition (\ref{Authilalg}) because the structure of the
automorphism group of $\alg_\theta$ shouldn't depend on whether the algebra
acts on $\hil$ from the right or left. The operator $J$ is formally the
anti-linear, self-adjoint unitary isometry of $\hil$ such that
$J\alg_\theta(\hil)J^{-1}=\alg_\theta'(\hil)$ is the commutant of the algebra
$\alg_\theta$ in the module $\hil$. If $\alg_\theta$ acts on $\hil$ from the
right (resp. left), then $J\hil$ is a left (resp. right) $\alg_\theta$-module.
{}From the Hilbert space lift (\ref{Authilalg}) we recover the automorphisms of
$\alg_\theta$ from the Wigner projection $\Pi:{\rm
Aut}_\hil(N,\alg_\theta)\to{\rm Aut}(N,\alg_\theta)$ defined by
\beq
\Pi(\hat g)\Bigl[f(x)\Bigr]=\Tr^{~}_\hil\Bigl(\hat g\,\weyl[f]\,\hat g^{-1}\,
\hat\Delta(x)\Bigr) \ ,
\label{Wignerproj}\eeq
where $\hat\Delta(x)$ is the map (\ref{Deltadef}) in the representation of
$\alg_\theta$ on $\hil$, and $\Tr^{~}_\hil$ denotes the trace over states of
the Hilbert space $\hil$. Clearly these constructions also hold true in the
restriction of the automorphism group of the algebra to gauge transformations.

A physical interpretation of the Tomita involution $J$ may be given as follows.
When $\hil=\hil_{\rm m}$ is the Hilbert space (\ref{hilmdef}), we define the
action of $J$ on fundamental matter fields $\psi$ by
\beq
J(\psi)=\psi^\dagger \ .
\label{Jpsi}\eeq
Thus in this case $J$ is a charge conjugation operator, and the commutant
$\alg_\theta'(\hil_{\rm m})=\alg_{-\theta}(\hil_{\rm m})$ is naturally
isomorphic to the algebra $\alg_\theta(\hil_{\rm m})$. In fact, it is simply
the algebra obtained from $\alg_\theta(\hil_{\rm m})$ by multiplying its
elements in the opposite order. In this case the symmetry operator $J$ has the
effect of enlarging the irreducible Fock module $\hil_{\rm q}$ to $\hil_{\rm
m}$~\cite{lsz}. This means that, as anticipated, the gauge symmetries of the
noncommutative space are only visible when the full set of ``points'' (Fock
representations) of the space are incorporated. This is necessary because
connections on the Fock module are trivial and induce, due to irreducibility,
only the gauge group $U(1)$~\cite{gn}.

This induced bi-module structure also arises naturally within the context of
open string quantization in background $B$-fields and D-branes, as described in
section~6.3.1. We quantize the point particle at an endpoint of an open string
to produce a Hilbert space $\hil$, upon which the algebra $\alg_\theta$ acts.
This is depicted as in fig.~\ref{stringmodules}, with the {\it same}
configurations of D-branes at the opposite ends of the string, i.e. in the
usual Seiberg-Witten scaling limit, in which the string oscillations can be
neglected, we impose identical boundary conditions at both endpoints of an open
string. This yields the bi-module ${\cal M}=\hil\otimes\hil^\vee$, where
$\hil^\vee$ is the conjugate $\alg_\theta$-module to $\hil$ corresponding to
the opposite orientations of a pair of Type II string endpoints. The Hilbert
space $\cal M$ is naturally an algebra which coincides with the algebra of
Hilbert-Schmidt operators on $\cal H$ that represent the joining of open string
endpoints (see section~8.4 below)~\cite{witten2,wittentachyon}. As explained in
section~6.3.1, we may naturally identify the Hilbert spaces $\hil=\hil_{\rm q}$
and $\hil^\vee=J\hil$. Then the condition involving the real structure $J$ in
(\ref{Authilalg}) simply reflects the fact that a lifted gauge transformation
from the worldvolume field theory (or more generally a lifted algebra
automorphism) should preserve the actions of $\alg_\theta$ at opposite ends of
the open string. In fact, $J$ may be thought of as a worldsheet parity
operator, mapping Type IIB D-branes onto Type I D-branes and the associated
orientifold planes. This real structure can be thereby used to construct
non-unitary noncommutative gauge groups~\cite{sospstar}, as indicated at the
end of section~4.1.

\subsubsection{Geometrical Aspects}

{}From a very heuristic point of view, in the simplest instance of $U(1)$ gauge
symmetry, the automorphism group ${\rm Aut}(\alg_\theta)\equiv{\rm
Aut}(1,\alg_\theta)$ lies ``somewhere'' in between that of the commutative
algebra of functions $\alg_0$ and a finite-dimensional matrix algebra
$\mat(N,\complex)$. In the former case there are no inner
automorphisms,\footnote{\baselineskip=12pt More precisely, the inner
automorphisms correspond to abelian, $U(1)$ gauge transformations.} so that all
automorphisms are outer automorphisms and generate spacetime coordinate
transformations,
\beq
{\rm Inn}(\alg_0)=\{\id\}~~~~~~,~~~~~~{\rm Out}(\alg_0)={\rm Diff}(\real^D) \ .
\label{autoalg0}\eeq
On the other hand, all automorphisms of the algebra $\mat(N,\complex)$ can be
represented via rotations by $N\times N$ unitary matrices, so that
\beq
{\rm Inn}\Bigl(\mat(N,\complex)\Bigr)=U(N)~~~~~~,~~~~~~
{\rm Out}\Bigl(\mat(N,\complex)\Bigr)=\{\id\} \ .
\label{automatN}\eeq
For the algebra $\alg_\theta$ there is a non-trivial mixing between the two
structures (\ref{autoalg0}) and (\ref{automatN}). Unlike the commutative case,
it is no longer true that the group $U(N,\alg_\theta)$ is the product of a
function algebra and a finite-dimensional Lie group, and so much richer
geometric and algebraic structures will emerge. In what follows we will attempt
to make this mixing between spacetime and internal degrees of freedom more
precise.

We have already seen an example of this mixing in section~4.2. Namely, the
star-unitary plane waves $g_v(x)=\e^{ik_i(v)x^i}$, with momentum given by
(\ref{linemomentum}), determine inner automorphisms of the algebra
$\alg_\theta$ which generate the translations (\ref{startransl}) of functions
by the constant vectors $v\in\real^D$. The corresponding gauge transformation
(\ref{stargaugetr}) is given by
\beq
A_i(x)~\longmapsto~A_i(x+v)-k_i(v) \ .
\label{gaugetransl}\eeq
The overall constant shift of the gauge field in (\ref{gaugetransl}) drops out
of the field strength (\ref{starfieldstrength}) and has no physical effect in
flat, infinite space, i.e. it corresponds to a global symmetry transformation
of the field theory. In this simple instance we thereby find that the
noncommutative gauge group contains spacetime translations. This seemingly
remarkable conclusion must, however, be taken in appropriate context. The plane
waves $g_v(x)$ are not Schwartz functions, because they only oscillate very
rapidly at infinity in $\real^D$. They can of course be approximated by
Schwartz functions, in a distributional sense, and for many applications this
would suffice to deduce that they generate gauge symmetries of noncommutative
Yang-Mills theory. For other applications, such as those involving
noncommutative solitons whereby the details of the asymptotic, topological
configurations of the fields are crucial, this conclusion is not entirely
valid.

It is natural to ask if this construction can be repeated for more general,
non-constant functions $v^i=v^i(x)$ on $\real^D$. These will produce more
general spacetime transformations, which we may wish to compare with
diffeomorphisms of $\real^D$. Let us examine, at an infinitesimal level, the
expansion of the Moyal commutator bracket in powers of $\theta$. Using
(\ref{starcomm}) we find
\beq
\delta_\phi f\equiv i\,\phi\star f-i\,f\star\phi=\{\phi,f\}^{~}_\theta
+O(\partial^2\phi\,\partial^2f) \ ,
\label{deltaphif}\eeq
where
\beq
\{\phi,f\}^{~}_\theta=\theta^{ij}\,\partial_i\phi\,\partial_jf
\label{Poissonbracket}\eeq
is the Poisson bracket based on the symplectic form $\theta$ of $\real^D$. From
(\ref{deltaphif}) it follows that, to leading orders in the deformation
parameter $\theta$ (equivalently for slowly-varying fields), the noncommutative
gauge group coincides with the group of canonical transformations which
preserve the symplectic structure $\theta$. These diffeomorphisms form the {\it
symplectomorphism group} ${\rm Diff}_\theta(\real^D)$ of $\real^D$, and we see
that $U(\alg_\theta)\sim{\rm Diff}_\theta(\real^D)$ in the limit $\theta\to0$.
This limit is analogous to the classical limit in quantum mechanics, and in
this truncation the noncommutative fields can be treated as ordinary functions
rather than operators. The group ${\rm Diff}_\theta(\real^D)$ is the natural
symmetry group of membranes, dynamical systems, and hydrodynamic
systems~\cite{membrane}, and in this limit the noncommutative Yang-Mills action
reduces to the corresponding bosonic membrane actions. Moreover, as we
indicated in section~2.2, this is the starting point for the deformation
quantization description of the theory~\cite{bffls,kontsevich} which can be
carried out over any Poisson manifold.

Although the higher-derivative terms in (\ref{deltaphif}) modify this
interpretation of noncommutative gauge transformations, we will see that the
general symplectomorphism nature of the spacetime symmetries induced by the
star-gauge symmetry will always be the same~\cite{lsz}. We shall now proceed to
try to understand better this unification of spacetime and gauge symmetries. At
the same time we will also attempt to clarify more precisely in what sense the
matrix degrees of freedom are deformed into spacetime ones, such that some
outer automorphisms of the commutative algebra $\alg_0$ become inner
automorphisms of the noncommutative algebra $\alg_\theta$ and thereby generate
genuine gauge symmetries of noncommutative Yang-Mills theory.

\subsubsection{Violations of Lorentz Invariance}

We will first digress momentarily to make some quick remarks on the Lorentz
transformation properties of noncommutative gauge theories. For this, we first
notice that the global translational symmetry above naturally generalizes to
other symplectic diffeomorphisms. For instance, for $D=2$ we can define the
star-unitary function
\beq
g_\alpha(x)=\sqrt{1+\alpha^2\theta^2}~\e^{i\alpha|x|^2}
\label{galphadef}\eeq
where $\theta=\theta^{12}$ and $\alpha$ is a real parameter. It generates the
inner automorphism
\beq
g_\alpha(x)\star f(x)\star g_\alpha(x)^\dagger=f(x_\alpha) \ ,
\label{galphainner}\eeq
where
\beq
\pmatrix{x_\alpha^1\cr x_\alpha^2\cr}=\pmatrix{\cos\gamma&\sin\gamma\cr
-\sin\gamma&\cos\gamma\cr}\pmatrix{x^1\cr x^2\cr}
\label{xrot}\eeq
is a rotation in the plane through angle
\beq
\gamma=\arctan(\alpha\theta) \ .
\label{gammaalphatheta}\eeq
Thus, as mentioned already in section~4.2, it is possible to realize spacetime
rotations via noncommutative gauge transformations. A similar property holds
for global, discrete symmetries. For instance, the star-unitary function
\beq
g_{\rm p}(x)=\pi^{D/2}\,{\rm Pfaff}(\theta)\,\delta^D(x)
\label{gpdef}\eeq
generates the parity reflection
\beq
g_{\rm p}(x)\star f(x)\star g_{\rm p}(x)^\dagger=f(-x) \ .
\label{parity}\eeq

Using these geometrical properties we can now make some general remarks
concerning the Lorentz invariance of the theory, which is superficially broken
by the presence of the tensor $\theta^{ij}$ in the spacetime commutation
relations (\ref{noncommalg}). For the so-called ``observer'' Lorentz
transformations of the theory, rotations or boosts of an observer inertial
reference frame leaves the physics unchanged because a unitary transformation
of the matrix $\theta^{ij}$ can be gauged away by a star-gauge transformation,
as we have just shown above. However, the theory is not invariant under
``particle'' Lorentz transformations which correspond to rotations or boosts of
localized field configurations within a fixed observer frame. Such
transformations leave the noncommutativity parameters unchanged, and lead to
spontaneous Lorentz violation because $\theta^{ij}$ provides a directionality
to spacetime in any fixed inertial frame. One can thereby compare
noncommutative gauge theory in four dimensions with Lorentz-violating
extensions of the standard model. Many terms in such extensions can be
eliminated because noncommutative field theories are $CPT$
invariant~\cite{cptsheikh}. Comparisons with the QED sector yield bounds of the
order~\cite{chklo}
\beq
\|\theta\|<(10\,{\rm TeV})^{-2} \ ,
\label{thetabounds}\eeq
which follow from an analysis of atomic clock-comparison experiments or by
comparison with standard QED processes.

\subsection{Universal Gauge Symmetry}

A striking consequence of the connection between noncommutative gauge theories
and twisted reduced models that was described in section~7.1 is the
universality of the noncommutative gauge group~\cite{gn,universal}, in a sense
that we shall now explain. Recall that the classical vacua of the action
(\ref{YMactionCi}) determined the noncommuting momenta $[\hat C_i,\hat
C_j]=-i\,(\theta^{-1})_{ij}$ with the {\it opposite} deformation parameter
$-\theta$. This rewriting is therefore naturally associated with the bi-module
structure based on the (equivalent) noncommutative function algebras
$\alg_\theta$ and $\alg_{-\theta}$, and it is the basis for the relationship
between the commutative and noncommutative descriptions of the same field
theory~\cite{covcoord}. In particular, the covariant coordinate operators
(\ref{covcoord}) may be interpreted as gauge fields appropriate to the
splitting of covariant derivatives on a generic $\alg_\theta$-module into a
free part $\hil_{\rm m}$ (represented by the gauge field $A_i$) and a twisted
part of constant magnetic flux corresponding to copies of the Fock module
$\hil_{\rm q}$~\cite{gn,SchwarzFock}. In the D-brane picture, the
corresponding global minima are identified with the closed string vacuum
possessing no open string excitations.

By using global Euclidean invariance, we may rotate to Darboux coordinates and
consider the noncommutative gauge theory (\ref{YMactionCi}) independently in
each $2\times2$ skew-block of the deformation matrix (\ref{thetacan}). In this
subsection we will therefore consider only the simplest case of $D=2$ spacetime
dimensions. The general case can be easily obtained by stitching the
independent blocks together again by means of an $SO(D)$ transformation. We
will therefore consider the noncommutative Yang-Mills action
\beq
S_{\rm YM}=-\frac1{g^2}\,\Tr\left(\left[\hat C_{z}
\,,\,\hat C_{\overline z}\right]+\frac1{2\theta}\right)^2 \ ,
\label{YMCiDarboux}\eeq
where
\beq
\hat C_{z}=\frac12\,\left(\hat C_1+i\,\hat C_2
\right)~~~~~~,~~~~~~\hat C_{\overline z}=\frac12\,\left(
\hat C_1-i\,\hat C_2\right) \ .
\label{CzCbarz}\eeq
We will restrict the quantum field theory to those field configurations which
have finite action. This requires the field strength $F_{ij}$ to vanish almost
everywhere and corresponds to the classical vacua of the theory. It will enable
us to evaluate the corresponding Feynman path integral in a semi-classical
approximation. From (\ref{YMCiDarboux}) it follows that this is equivalent to
the conditions
\beq
\left[\hat C_{\overline z}\,,\,\hat C_{z}\right]=\frac1{2\theta} \ .
\label{vacuaalpha}\eeq
In a particular $\alg_\theta$-module $\hil$, we require that the equations
(\ref{vacuaalpha}) hold for all but a finite number of matrix elements of the
operators (corresponding to a set of measure zero in field space).

The algebraic conditions (\ref{vacuaalpha}) are simply the Heisenberg
commutation relations that we encountered in section~8.1. By the
Stone-von~Neumann-Mackey theorem, we know that (up to unitary equivalence) its
unique unitary irreducible representation is the Schr\"odinger representation
on Fock space (\ref{Fockspace}). For $\theta>0$ it is given by
\beq
\hat C_{z}^{(1)}=-\frac{\hat c^\dagger}{\sqrt{2\theta}}
{}~~~~~~,~~~~~~\hat C_{\overline z}^{(1)}=-\frac{\hat c}{\sqrt{2\theta}} \ ,
\label{Cz1}\eeq
where $\hat c^\dagger$ and $\hat c$ are the Fock space creation and
annihilation operators (\ref{cranops}) for $d=1$. For $\theta<0$ one should
interchange $\hat c^\dagger$ and $\hat c$ in (\ref{Cz1}). The most general
solution to the classical equations of motion (\ref{vacuaalpha}) is given by a
countable direct sum of Fock modules. We may label the solution space by an
integer $N\geq1$ which corresponds to a representation by operators $\hat
C_{z}^{(N)},\hat C_{\overline z}^{(N)}$ acting on the separable Hilbert space
\beq
\bigoplus_{\mu=0}^{N-1}\hil_{\rm q}\cong\hil_{\rm q}\otimes
\complex^N\equiv\hil_{\rm q}^{(N)} \ .
\label{hilNdef}\eeq
There is a (non-canonical) isomorphism $\hil_{\rm q}^{(N)}\cong\hil_{\rm q}$
which can be used to rewrite the noncommutative gauge theory
(\ref{YMCiDarboux}), evaluated (without loss of generality) in the
Schr\"odinger $\alg_\theta$-module $\hil_{\rm q}$, in terms of field
configurations in the sector labelled by $N$. This follows from the {\it
Hilbert hotel argument} which regroups the Fock space states $|n\rangle$,
$n=0,1,2,\dots$, into the basis vectors $|p,\mu\rangle$, $p=0,1,2,\dots$,
$\mu=0,1,\dots,N-1$, of $\hil_{\rm q}^{(N)}$ as
\beq
|n\rangle=|pN+\mu\rangle\equiv|p,\mu\rangle \ ,
\label{Fockregroup}\eeq
where the index $p$ labels the infinite-dimensional, Fock space component of
$\hil_{\rm q}^{(N)}$ and $\mu$ indexes the finite-dimensional part
$\complex^N$. In this basis we may write the vacuum state, for $\theta>0$, up
to a star-gauge transformation as
\bea
\hat C_z^{(N)}&=&-\frac{\hat c^\dagger}{\sqrt{2\theta}}\otimes\id_N
{}~=~-\frac1{\sqrt{2\theta}}\,\sum_{p=0}^\infty\,\sum_{\mu=0}^{N-1}\sqrt p\,
|p,\mu\rangle\langle p-1,\mu| \ , \nn\\\hat C_{\overline z}^{(N)}&=&
-\frac{\hat c}{\sqrt{2\theta}}\otimes\id_N~=~-\frac1{\sqrt{2\theta}}\,
\sum_{p=0}^\infty\,\sum_{\mu=0}^{N-1}\sqrt{p+1}\,|p,\mu\rangle
\langle p+1,\mu| \ ,
\label{genvacN}\eea
and similarly for $\theta<0$.

The vacuum configuration (\ref{genvacN}) has two types of unitary gauge
symmetries. There is the infinite-dimensional $U(\hil_{\rm q})$ symmetry acting
on the Fock space labels, under which even the original action
(\ref{YMCiDarboux}) is invariant. There is also a $U(N)$ symmetry acting by
finite-dimensional rotations of the $\mu$ labels. The integer $N$ labels the
gauge inequivalent vacua of the noncommutative gauge theory and can be given as
the {\it analytical index}
\beq
N=\Tr^{~}_{\hil_{\rm q}^{(N)}}\left[\hat C_z^{(N)}\,,\,
\hat C_{\overline z}^{(N)}
\right]=\dim\ker\hat C_z^{(N)}\hat C_{\overline z}^{(N)}\ ,
\label{Nindex}\eeq
which counts the difference between the number of zero eigenvalues of the
operators $\hat C_z^{(N)}\hat C_{\overline z}^{(N)}$ and $\hat C_{\overline
z}^{(N)}\hat C_z^{(N)}$ whose non-zero eigenvalues all coincide. Here one must
remember that we are dealing with infinite-dimensional operators, so that
generally $\Tr[\hat C_{\overline z},\hat C_z]\neq0$. The quantity
(\ref{Nindex}) is thereby a topological invariant which detects the
differential operators that are hidden in the fields $C_i$. In particular, $N$
cannot be changed by any local gauge transformation. In fact, it is this
quantity that identifies sectors with a higher-dimensional interpretation in
what is naively a zero-dimensional theory (i.e. a matrix model).

Any operator on Fock space which transforms covariantly under star-gauge
transformations may now be re-expressed, via the isomorphism
(\ref{Fockregroup}), in terms of matrices transforming in the adjoint
representation of the finite-dimensional unitary group $U(N)$. For example, for
the gauge field strengths $\hat F\equiv\weyl[F]_{\overline z z}$ we may write
\bea
\hat F&=&\sum_{m=0}^\infty\,\sum_{n=0}^\infty
\,\left(\hat F\right)_{mn}\,|m\rangle\langle n|
\nn\\&=&\sum_{p=0}^\infty\,\sum_{q=0}^\infty\,\sum_{\mu=0}^{N-1}\,
\sum_{\nu=0}^{N-1}\,\left(\hat F^{(N)}
\right)_{pq}^{\mu\nu}\,|p,\mu\rangle\langle q,\nu|\nn\\&=&\sum_{p=0}^\infty
\,\sum_{q=0}^\infty\,
\sum_{a=1}^{N^2}\,\left(\hat F^{(N)a}\right)_{pq}\,\sum_{\mu=0}^{N-1}
\,\sum_{\nu=0}^{N-1}\left(t_a\right)^{\mu\nu}\,|p,\mu\rangle\langle q,\nu| \ ,
\label{hatCiUN}\eea
where $t_a$ are the generators of $U(N)$ in the fundamental representation. As
in section~8.1, the Fock indices $p,q$ label both the dependence of the fields
on the ``coordinates'' of the noncommutative space and the internal star-gauge
symmetry. But now there are new indices $\mu$ on the fields, representing a
hidden internal $U(N)$ gauge symmetry in the given topological vacuum sector
labelled by $N$. This is quite remarkable, given that we started with only
$U(1)$ gauge fields.

Let us now examine how the quantum field theory decomposes according to the
vacuum configurations that we have found. Evidently, any path in field space
which connects different vacua has infinite action, and so the quantum theory
constructed about any one of these vacua will not mix with any of the others.
Evaluating the corresponding path integral as a sum over each of the classical
vacuum field configurations thereby splits it into a sum of partition functions
for each $U(N)$ theory,
\beq
Z=\int\!\!\!\int
\frac{DC_z~DC_{\overline z}}{{\rm vol}\Bigl(U(\hil_{\rm q})\Bigr)}~
\e^{-S_{\rm YM}}=\sum_{N=1}^\infty Z_N \ .
\label{partsplit}\eeq
In the sector labelled by $N$, we can expand the quantum theory described by
the partition function $Z_N$ around the classical vacuum configuration
(\ref{genvacN}) as
\beq
\hat C_i=\hat C_i^{(N)}+\hat A_i^{(N)} \ .
\label{vacexpand}\eeq
By using (\ref{hatCiUN}) we may then write the noncommutative Yang-Mills action
in the topological sector $N$ which defines the partition function $Z_N$ for
$\theta>0$ as
\bea
S_{\rm YM}&=&-\frac1{2g^2}\,\Tr^{~}_{\hil_{\rm q}}\hat F^2\nn\\&=&-\frac1{2g^2}
\,\Tr^{~}_{\hil_{\rm q}^{(N)}}\left(\hat F^{(N)}\right)^2\nn\\
&=&-\frac1{2g^2}\,
\Tr^{~}_{\hil_{\rm q}}\otimes\tr^{~}_N\left(\frac1{\sqrt{2\theta}}\,\Biggl\{
\left[\hat c\otimes\id_N\,,\,\hat A_z\right]+\left[\hat c^\dagger\otimes\id_N
\,,\,\hat A_{\overline z}\right]\Biggr\}+\left[\hat A_{\overline z}\,,\,
\hat A_z\right]\right)^2 \ , \nn\\&&
\label{SYMsectorN}\eea
where $(\hat A_i)_{\mu\nu}$ is an $N\times N$ matrix-valued operator on
ordinary Fock space $\hil_{\rm q}$ which is defined by its matrix elements in
taking the traces in (\ref{SYMsectorN}),
\beq
\langle p|\left(\hat A_i\right)_{\mu\nu}|q\rangle=
\langle p,\mu|\hat A_i^{(N)}|q,\nu\rangle \ .
\label{matrixeltmap}\eeq
Like the coordinate operators $\hat x^i$, the oscillator operators $\hat c$ and
$\hat c^\dagger$ generate derivatives of fields in the Weyl representation,
owing to the commutation relations (\ref{crancommrels}),
\beq
\left[\hat c\,,\,\weyl[f]\right]=\sqrt{\frac{|\theta|}2}~
\weyl[\partial_{\overline
z}f]~~~~~~,~~~~~~\left[\hat c^\dagger\,,\,\weyl[f]\right]=-
\sqrt{\frac{|\theta|}2}~\weyl[\partial_zf] \ ,
\label{cderivs}\eeq
where $z,\overline{z}=x^1\pm i\,{\rm sgn}(\theta)\,x^2$. It follows that the
field strength squared which appears in the argument of the trace in
(\ref{SYMsectorN}) is just the standard Weyl representation of that for
noncommutative $U(N)$ gauge theory, and the action (\ref{SYMsectorN}) is of
precisely the same form as~(\ref{WeylYMaction}).

We thus conclude that $U(1)$ noncommutative Yang-Mills theory contains
noncommutative $U(N)$ gauge theory for {\it all} values of $N$. The rank $N$ of
the gauge group emerges as a superselection parameter, labelling separate,
star-gauge inequivalent vacuum sectors of the original quantum Hilbert space.
Therefore, noncommutative Yang-Mills theory is a {\it universal gauge theory},
containing all Yang-Mills theories (including the noncommutative ones).
Universal gauge theories had been sought some time ago~\cite{rajeev} in terms
of models based on a gauge symmetry group $U(\infty)$ defined through a
sequence of embeddings of $U(N)$ structure groups,
\beq
U(1)~\subset~U(2)~\subset~\dots~\subset~U(N)~\subset~U(N+1)~\subset~
\dots~\subset~U(\infty) \ .
\label{UNsequence}\eeq
In this sequence the unitary group $U(N)$ is viewed as consisting of operators
on an infinite-dimensional Hilbert space which are equal to the identity
operator except in an $N\times N$ submatrix. The $N\to\infty$ limit then
defines the {\it inductive limit} $U(\infty)$, which is the group of all
unitary operators that differ from the identity by a finite-rank operator.
However, the gauge group $U(\infty)$ is rather difficult to deal with, in
particular within the setting of noncommutative geometry. An appropriate
enlargement to unitary groups within the Schatten ideals proves to be
manageable, and in some instances even leads to an exactly solvable gauge
theory~\cite{rajeev}. We shall see in the next section that these same gauge
groups are the appropriate ones for noncommutative Yang-Mills theory, although
they arise much more naturally and for different reasons.

\subsection{Large $N$ Limits}

The basic symmetry underlying the remarkable construction of the previous
subsection is some version of the infinite unitary group $U(\infty)$. We are
thus posed with the problem of determining which version, i.e. how to embed
finite-rank structure groups starting from $U(N)$ to end up with $U(\infty)$.
Different ways of embedding lead to very different inductive limits. In
particular, to make contact with the function algebras we are dealing with, an
appropriate completion is required. The appearence of this gauge group and the
ensuing universal gauge symmetry was a consequence of the rewriting of
noncommutative Yang-Mills theory as the infinite-dimensional reduced model
(\ref{YMactionCi}). However, to understand better the sequence of embeddings
(\ref{UNsequence}) from finite-dimensional symmetry groups to
infinite-dimensional ones, we should first appeal to the {\it finite}
dimensional matrix models underlying noncommutative gauge theory that were
described in section~7.2. This will help with the understanding of how the
appropriate large $N$ limit should be taken above. In these matrix models, the
group of inner automorphisms may be written down in a very precise closed form,
and this will help us understand the structure of the noncommutative gauge
symmetry group $U(\alg_\theta)$.

\subsubsection{Algebraic Description}

We will begin with an algebraic formulation of the star-gauge symmetry group,
as it is somewhat more straightforward to describe. From the matrix model
formulation of noncommutative gauge theory that we presented in section~7.2, it
is clear that the gauge group is intimately related to the infinite unitary
group $U(\infty)$. But this is also apparent from the representation of the
algebra $\alg_\theta$ on the irreducible Fock space (\ref{Fockspace}). The
natural algebra acting on $\hil_{\rm q}$ consists of $d$ copies
$\mat(\infty,\complex)\oplus\cdots\oplus\mat(\infty,\complex)
\cong\mat(\infty,\complex)$ of the algebra of finite-rank operators
\beq
\mat(\infty,\complex)=\bigcup_{N=1}^\infty\mat(N,\complex) \ ,
\label{matinftydef}\eeq
which is defined with respect to the natural system of embeddings of finite
dimensional matrix algebras,
\bea
\mat(N,\complex)&\hookrightarrow&\mat(N+1,\complex)\nn\\
M&\longmapsto&\pmatrix{M&0\cr0&0\cr} \ .
\label{matNembeddings}\eea
It is important to note that (\ref{matinftydef}) consists of arbitrarily large
but {\it finite} matrices.

\begin{figure}[htb]
\epsfxsize=3in
\bigskip
\centerline{\epsffile{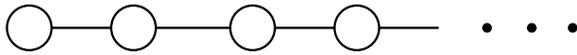}}
\caption{\baselineskip=12pt {\it The semi-infinite Dynkin diagram for the
infinite-dimensional Lie algebra $su(\infty)$ represented on Fock space. The
corresponding step operators for the Cartan basis are $E_{\vec n,\vec m}=|\vec
n\,\rangle\langle\vec m|$.}}
\bigskip
\label{dynkin}\end{figure}

At the finite level, the unitary group of $\mat(N,\complex)$ is of course just
the usual $N\times N$ unitary group $U(N)$, which coincides with the gauge
symmetry group of the twisted Eguchi-Kawai model (\ref{STEK}), or equivalently
of the noncommutative lattice gauge theory (\ref{latticeNCYM}). The group
homomorphism $\imath:U(N)\to{\rm Inn}(N,\complex)$, $U\mapsto\imath_U^{~}$, is
injective and has kernel $\ker\imath=U(1)$ generated by the unit matrix
$\id_N$. The group of finite-dimensional inner automorphisms is thereby given
explicitly as
\beq
{\rm Inn}(N,\complex)=U(N)/U(1)=SU(N)/\zed_N \ ,
\label{InnNexplicit}\eeq
and the unitary group of the inductive limit (\ref{matinftydef}) is the
expected
\beq
U\Bigl(\mat(\infty,\complex)\Bigr)=U(\infty)=\bigcup_{N=1}^\infty U(N) \ .
\label{Uinftydef}\eeq
The corresponding semi-simple Lie algebra $su(\infty)$ is described by the
Dynkin diagram depicted in fig.~\ref{dynkin}~\cite{kac}.

We should remember, however, that the Fock module (\ref{Fockspace}) is defined
with an appropriate completion, emphasizing the fact that it derives from a
space of {\it functions} on $\real^D$. To relate this construction to the
function algebras of interest, we must consider appropriate norm completions of
the algebra $\mat(\infty,\complex)$, and hence of the infinite unitary group
$U(\infty)$. At the level of Schwartz functions, the completions may be defined
in terms of $L^p$-norms for $1\leq p\leq\infty$. For $p=\infty$ this is defined
in (\ref{Linftynorm}), while for $p<\infty$ we define
\beq
\|f\|^{~}_p=\left(\,\int d^Dx~\Bigl|f(x)\Bigr|^p\,\right)^{1/p}
\label{Lpnorm}\eeq
on the space of $p$-integrable functions $f\in L^p(\real^D)$. The $L^p$-spaces
form a sequence of completions
\beq
L^1(\real^D)~\subset~L^2(\real^D)~\subset~\dots~\subset~L^\infty(\real^D) \ ,
\label{Lpincl}\eeq
where the Banach space $L^\infty(\real^D)$ contains the algebra of Schwartz
functions that we have been working with.

Under the large $N$ matrix model Weyl-Wigner correspondence, we should now
translate this structure into a statement about operators acting on the
irreducible Fock module $\hil_{\rm q}$. Recalling from (\ref{Tracedef}) that
spacetime integrals map onto traces of Weyl operators in ${\rm End}(\hil_{\rm
q})$, it follows that the operator algebras should be completed in the {\it
Schatten $p$-norms}. For $1\leq p<\infty$ these are defined by
\beq
\left\|\weyl[f]\right\|^{~}_p=\left(\Tr^{~}_{\hil_{\rm q}}\left(
\weyl[f]^\dagger\,\weyl[f]\right)^{p/2}\,\right)^{1/p}
\label{Schattenpnorm}\eeq
on the space of $p$-summable operators $\ell^p(\hil_{\rm q})$ on Fock space.
For $p=\infty$ it is given by the {\it operator norm}
\beq
\left\|\weyl[f]\right\|_\infty^{~}=\sup_{\langle\psi|\psi\rangle\leq1}\,
\Biggl(\left\langle\left.\weyl[f]\,\psi~\right|~\weyl[f]\,\psi\right
\rangle\Biggr)^{1/2}
\label{operatornorm}\eeq
defined on the algebra of {\it compact operators} $\ell^\infty(\hil_{\rm
q})\equiv{\cal K}(\hil_{\rm q})$. Starting from the algebra of finite-rank
operators (\ref{matinftydef}), there is also a sequence of completions
\beq
\mat(\infty,\complex)~\subset~\ell^1(\hil_{\rm q})~\subset~\ell^2(
\hil_{\rm q})~\subset~\dots~\subset~{\cal K}(\hil_{\rm q})
\label{lpincl}\eeq
in correspondence with the functional sequence (\ref{Lpincl}). In other words,
integrable functions correspond to trace-class operators, square-integrable
functions to Hilbert-Schmidt operators, and so on. Notice that there is no
functional analog of the finite-rank operators.

Of special interest to us is the algebra ${\cal K}(\hil_{\rm q})$ of compact
operators. A compact operator $\weyl[f]$ is one for which the sequence of
eigenvalues of the Hermitian operator $\weyl[f]^\dagger\,\weyl[f]$ tends to
zero. They are therefore the natural analogs of functions which fall-off at
infinity in $\real^D$. They are as close to finite-rank or bounded operators as
one can get under the Weyl-Wigner correspondence. For instance, like the
finite-dimensional matrix algebras $\mat(N,\complex)$, the defining
representation of ${\cal K}(\hil_{\rm q})$ on $\hil_{\rm q}$ is, up to unitary
equivalence, the only irreducible representation of the $C^*$-algebra of
compact quantum mechanical operators. In particular, they are the natural
analogs of Schwartz functions~\cite{rieffelstone}. Now the map
$\weyl[g]\mapsto\imath_{\weyl[g]}$ generates a continuous homomorphism (in the
operator norm topology) of the unitary group $U(\hil_{\rm q})$ of Fock space
onto the automorphism group ${\rm Aut}({\cal K}(\hil_{\rm q}))$. It has kernel
$U(1)$ consisting of phase multiples of the Fock space identity operator
$\id_{\hil_{\rm q}}$. This identifies the automorphism group as the group of
projective unitary automorphisms $PU(\hil_{\rm q})=U(\hil_{\rm q})/U(1)$ of the
Hilbert space~$\hil_{\rm q}$,
\beq
{\rm Aut}\Bigl({\cal K}(\hil_{\rm q})\Bigr)=PU(\hil_{\rm q}) \ .
\label{projunit}\eeq
This is the natural completion of the matrix model automorphism group
(\ref{InnNexplicit}), whereby the global center subgroup $\zed_N$ of $SU(N)$ is
replaced with the phase group $U(1)$ in the large $N$ limit of
(\ref{InnNexplicit}) leading to (\ref{projunit}). This illustrates explicitly
how the matrix model degrees of freedom are transmuted into spacetime degrees
of freedom.

But again, not all of these automorphisms are gauge transformations. We need to
consider the unitary subgroups of the spaces of operators which comprise the
sequence of completions (\ref{lpincl}). They themselves form a natural sequence
of completions starting from the group $U(\infty)$ of finite-rank unitary
operators on $\hil_{\rm q}$,
\beq
U(\infty)~\subset~U_1(\hil_{\rm q})~\subset~U_2(\hil_{\rm q})~\subset~\dots~
\subset~U_\infty(\hil_{\rm q}) \ .
\label{Uincl}\eeq
Putting these facts all together now identifies the star-gauge transformations
as the automorphisms (\ref{projunit}) generated by the compact unitaries
$U_\infty(\hil_{\rm q})$. Therefore, we arrive at a very nice physical
interpretation of the gauge symmetry group of noncommutative Yang-Mills theory.
Namely, it is the group of compact unitary operators on Fock space,
\beq
U(\alg_\theta)=U_\infty(\hil_{\rm q}) \ .
\label{UalgUinfty}\eeq

The result (\ref{UalgUinfty}) is the right answer for the gauge symmetry group.
The driving reasons for this are the properties of the infinite unitary group
(\ref{Uinftydef}). First of all, it contains all finite-rank structure groups
$U(N)$, so that the group (\ref{UalgUinfty}) has the appropriate universality
properties that we encountered in the previous section, and it in fact
coincides with the gauge groups used in~\cite{rajeev} to build models of
universal gauge theory. Secondly, the group $U(\infty)$ has homotopy groups
determined by Bott periodicity which are non-trivial in every odd dimension,
\beq
\pi_n\Bigl(U(\infty)\Bigr)=\left\{{\begin{array}{lll}
\zed~~~~&,&~~~~n=2k+1 \ , \\0~~~~&,&~~~~n=2k \ . \end{array}}\right.
\label{Uinftyhomotopy}\eeq
By Palais' theorem~\cite{palais}, all the unitary groups that appear in the
sequence (\ref{Uincl}) have the same homotopy type as $U(\infty)$, i.e. this is
a topological property that is preserved under the completions in ${\rm
End}(\hil_{\rm q})$. Therefore, unlike the full unitary group of Hilbert space
which has trivial homotopy (\ref{pinUhil}), the subgroup (\ref{UalgUinfty})
recovers the correct topological properties. Furthermore, the identification
(\ref{UalgUinfty}) agrees with the natural gauge orbit space that one should
integrate over in the Euclidean path integral formulation of the quantum gauge
theory~\cite{nph}. In the commutative case this would be the quotient of the
space of gauge field configurations on $\real^D$ by the group of gauge
transformations which are connected to the identity, i.e. which approach the
identity at infinity in $\real^D$. But this connectedness property is precisely
what is possessed by the compact unitaries $U_\infty(\hil_{\rm q})$ under the
Weyl-Wigner correspondence. This provides a direct relationship between the
topology of the gauge group $U_\infty(\hil_{\rm q})$ and that of the
configuration space of noncommutative gauge fields~\cite{nph}. The analytic
description (\ref{UalgUinfty}) of the star-gauge symmetry group is therefore
the correct one.

\subsubsection{Geometric Description}

In (\ref{UalgUinfty}) we have unveiled a very precise, analytic description of
the star-gauge symmetry group which illustrates clearly its topological
properties. What is less transparent in this formalism is its geometrical
characteristics. To determine these, we appeal once more to the
finite-dimensional matrix model representation of noncommutative gauge theory.
{}From (\ref{Jkcommrels}) it follows that the fundamental discrete generators
in the finite $N$ matrix model obey the commutation relations~\cite{ffz}
\beq
[J_k,J_q]=2i\,\sin\left(\frac{\pi i}N\,\sum_{i<j}k_i\,Q_{ij}\,q_j\right)~
J_{k+q} \ .
\label{Jksinebracket}\eeq
We now take the large $N$ limit of the relations (\ref{Jksinebracket}) in the
dynamical regime of momentum space whereby the discrete noncommutative fields
contain only small Fourier modes $k_i,q_j\ll\sqrt{N}$. This restricts the
Fourier momenta of the fields to lie in the interior of the Brillouin zone.
After an appropriate rescaling of the operators $J_k$ by $N$, the sine function
in (\ref{Jksinebracket}) can be expanded, resulting in the large $N$
commutation relations of the $W_\infty$ algebra
\beq
\left[J_k^{(\infty)}\,,\,J_q^{(\infty)}\right]=2\pi i\,k\wedge q~
J^{(\infty)}_{k+q} \ ,
\label{Jkinftycommrels}\eeq
with $k\wedge q=\sum_{i<j}k_i\,\varepsilon_{ij}\,q_j$.

The commutation relations (\ref{Jkinftycommrels}) coincide with the Lie algebra
of the vector fields
\beq
V_\phi=\theta^{ij}\,\partial_i\phi~\frac\partial{\partial x^j}
\label{VcalF}\eeq
for the functions $\phi(x)=\phi_k(x)=\e^{2\pi ik_ix^i/\ell}$ which constitute
the complete set of harmonics on a $D$-dimensional hypercubic torus. The
infinitesimal diffeomorphisms generated by the vector fields (\ref{VcalF}),
represented on functions as
\beq
f~\longmapsto~\delta_\phi f=V_\phi(f) \ ,
\label{deltaphifV}\eeq
generate canonical transformations of the spacetime coordinates, as in
(\ref{deltaphif},\ref{Poissonbracket}). In particular, they realize the
Poisson-Lie algebra
\beq
\left[V_\phi\,,\,V_{\phi'}\right]=V_{\{\phi\,,\,\phi'\}^{~}_\theta} \ .
\label{PoissonLiealg}\eeq
It follows that for smooth matrices $U_i$ whose low Fourier modes dominate the
expansion (\ref{Uikdef}), the commutator bracket of large $N$ noncommutative
fields can be substituted by the Poisson bracket. As we remarked in
section~8.2.2, this reduction is reminescent of the semi-classical
approximation of quantum mechanics. In this limit the Moyal and Poisson
brackets coincide, and the group of symplectomorphisms may be identified with
an appropriate completion of the infinite unitary group $U(\infty)$~\cite{ffz}.
Of course, this is no longer true for fields which
have high momentum modes, as we saw already in section~8.2.2. Given that the
Moyal bracket represents the commutator bracket of quantum mechanical
operators, we may thereby arrive at a geometrical characterization of the
star-gauge symmetry group. Namely, $U(\alg_\theta)$ is a {\it quantum
deformation} of the symplectomorphism group ${\rm Diff}_\theta(\real^D)$. There
are, however, many ways to see that this deformation must still consist of
operators which preserve the symplectic structure $\theta$~\cite{lsz}. For
instance, these are precisely the
transformations which preserve the Poisson bi-vector
$\overleftarrow{\partial_i}\,\theta^{ij}\,\overrightarrow{\partial_j}$ which
appears in the formula (\ref{starproddef}) for the star-product. Further
aspects of this deformation are described in~\cite{lsz,ChanLee}.

There are several subtleties with the geometrical description that we have just
presented. First of all, it is only a local description, because the
correspondence has been established only at the level of the Fourier basis for
a torus. It has also neglected the boundaries of the Brillouin zone in momentum
space, and subtleties associated with the periodicity $N$ of the lattice in the
large $N$ limit. This latter property is the reason why, for instance, the
$U(\infty)$ symmetry group associated with $\real^D$ (which has the
semi-infinite Dynkin diagram of fig.~\ref{dynkin}) is not the same as that on
${\bf T}^D$ (which has an infinite Dynkin diagram)~\cite{PopeStelle}. There are
many different algebras that can be obtained starting from $\mat(N,\complex)$
by taking inductive $N\to\infty$ limits with more complicated embeddings than
the simplest, canonical one in (\ref{matNembeddings}). Indeed, there are
infinitely many, non-mutually pairwise isomorphic versions of the infinite
dimensional Lie group $SU(\infty)$ which depend on the way that the large $N$
limit is taken~\cite{HoppeSchaller}. In particular, the Lie groups $U(\infty)$
and ${\rm Diff}_\theta(\real^D)$ are not isomorphic~\cite{bhss}, their
differences lying precisely in the high frequency components of the fields. The
proper way to relate the infinite matrix algebra to the algebra of functions on
the noncommutative torus is described in~\cite{lls2}. One can embed the algebra
$\alg_\theta$ in this case into the completion of an infinite dimensional
algebra of finite rank matrices. The nice geometric feature of this
construction is that the embedding algebra contains all Morita equivalent tori,
and hence the continuum noncommutative Yang-Mills theory is approximated by
gauge theories on discrete spaces that at the same time approximate all dual
field theories. This is the basis of the finite-dimensional Morita equivalences
constructed in~\cite{amns2}.

While the description of this subsection still leaves some imprecision as to
the precise nature of star-gauge symmetries in noncommutative quantum field
theory, we have at least captured some of the general geometrical features of
the mixing between spacetime and colour degrees of freedom. At this stage it
does not seem likely to capture complete diffeomorphism invariance using
star-gauge transformations alone. However, investigation of star-gauge
invariant operators, such as the open Wilson lines, could suggest how the
noncommutative gauge fields couple to gravity~\cite{sugracoupling}. Indeed, the
results obtained in this subsection are very natural in the D-brane
picture~\cite{covcoord,HKL}. As we explained at the end of section~5.2,
noncommutative field theories appear as a description of D-branes carrying a
uniform distribution of lower-dimensional brane charges. In this way a
noncommutative D-brane (i.e. one in a constant $B$-field) may be described as a
configuration of infinitely-many lower-dimensional D-branes. Then the usual
$U(1)$ gauge theory on the brane is represented as a $U_\infty(\hil_{\rm q})$
gauge symmetry in the lower-dimensional field theory corresponding to
diffeomorphisms which leave the volume of the brane invariant. A larger class
of diffeomorphisms may be obtained by considering the gauge theory operators
which couple to closed string states in the bulk of the D-brane,
\beq
\tilde{\cal C}(k)=\int d^Dx~{\rm P}\,\exp_\star\left[\,\int\limits_0^1dt~
\Bigl(i\,v^i\,A_i(x+vt)+y_a\,\phi^a(x+vt)\Bigr)\right]\star\e^{ik_ix^i} \ ,
\label{closedop}\eeq
where $\phi^a$ are the embedding coordinates of the D-brane in target space,
$v^i=k_j\,\theta^{ji}$ is the separation of the straight open Wilson line, and
$y_a=2\pi\alpha'\,k_a$ with $k_a$ the momentum in the directions transverse to
the D-brane.

\subsection*{Acknowledgments}

The author thanks A.~Armoni, C.-S.~Chu, J.~Erickson, J.~Gomis,
J.~Gracia-Bond\'{\i}a, D.R.T.~Jones, F.~Lizzi and
V.~Rivelles for very helpful comments on the manuscript. He would
especially like to thank the referee for the many suggestions and
remarks which have helped to greatly improve the manuscript. This work was
supported in part by an Advanced Fellowship from the Particle Physics and
Astronomy Research Council~(U.K.).

\end{document}